\documentclass[%
reprint,
superscriptaddress,
 amsmath,amssymb,
 aps,
]{revtex4-1}

\usepackage{mathtools}
\usepackage{graphicx}% Include figure files
\usepackage{dcolumn}% Align table columns on decimal point
\usepackage{bm}% bold math
\usepackage{amsmath}
\usepackage{amssymb}
\usepackage{amsthm}
\usepackage{graphicx}
\usepackage{algorithmic}
\usepackage{mathrsfs}
\usepackage{braket}
\usepackage{xcolor}
\usepackage{tikz}
\usepackage[roman]{complexity}
\usepackage[caption=false]{subfig}
 \usepackage{url}
 \usepackage[utf8]{inputenc}
\usepackage[T1]{fontenc}
\usepackage[clock]{ifsym}
\usepackage{comment}

\newtheorem{definition}{Definition}

\usepackage{pdfpages}
\usepackage{pgffor}
\makeatletter
\AtBeginDocument{\let\LS@rot\@undefined}
\makeatother

\usepackage[hyperindex]{hyperref}
\hypersetup{colorlinks,linkcolor={blue},citecolor={blue},urlcolor={black}} % 
\usepackage{cleveref}
\begin{document}

%%%%%%%%%%%%%%%%%%%%%%%%%%%%%%%%%%%%%%%%%%%%%%%%%%%%%%%%%%%%%%%%%%%%%%%%%%%%%%
\title{Data centers with quantum random access memory and quantum networks}% Force line breaks with \\

\author{Junyu Liu}
\email{junyuliu@uchicago.edu}
\affiliation{Pritzker School of Molecular Engineering, The University of Chicago, Chicago, IL 60637, USA}
\affiliation{Kadanoff Center for Theoretical Physics, The University of Chicago, Chicago, IL 60637, USA}
\affiliation{Chicago Quantum Exchange, The University of Chicago, Chicago, IL 60637, USA}
\affiliation{Institute for Quantum Information and Matter, California Institute of Technology, Pasadena, CA 91125, USA}
\affiliation{Walter Burke Institute for Theoretical Physics, California Institute of Technology, Pasadena, CA 91125, USA}

\author{Connor T.~Hann}
\email{connor.t.hann@gmail.com. This work was done prior to CTH joining the AWS Center for Quantum Computing.}
\affiliation{Pritzker School of Molecular Engineering, The University of Chicago, Chicago, IL 60637, USA}
\affiliation{Institute for Quantum Information and Matter, California Institute of Technology, Pasadena, CA 91125, USA}
\affiliation{AWS Center for Quantum Computing, Pasadena, CA 91125, USA}

\author{Liang Jiang}
\email{liang.jiang@uchicago.edu}
\affiliation{Pritzker School of Molecular Engineering, The University of Chicago, Chicago, IL 60637, USA}
\affiliation{Chicago Quantum Exchange, The University of Chicago, Chicago, IL 60637, USA}
\affiliation{AWS Center for Quantum Computing, Pasadena, CA 91125, USA}

\date{\today}% It is always \today, today,

\begin{abstract}
In this paper, we propose the Quantum Data Center (QDC), an architecture combining Quantum Random Access Memory (QRAM) and quantum networks. We give a precise definition of QDC, and discuss its possible realizations and extensions. We discuss applications of QDC in quantum computation, quantum communication, and quantum sensing, with a primary focus on QDC for $T$-gate resources, QDC for multi-party private quantum communication, and QDC for distributed sensing through data compression. We show that QDC will provide efficient, private, and fast services as a future version of data centers.

\end{abstract}
\maketitle
\textbf{\emph{Introduction}}. As a frontier subject of physics and computer science, quantum information science is currently a rapidly developing and highly valued research area, with wide applications in computation \cite{feynman2018simulating,shor1999polynomial,grover1997quantum,preskill2018quantum}, data science and machine learning \cite{wittek2014quantum,biamonte2017quantum}, communication \cite{gisin2002quantum,cleve1999share,hillery1999quantum,kimble2008quantum,caleffi2018quantum,muralidharan2014ultrafast,muralidharan2016optimal} and sensing \cite{degen2017quantum,giovannetti2006quantum,giovannetti2011advances}. In the near future, quantum computation may bring significant advantages to some specific algorithms; Quantum communication will strictly guarantee data security and privacy, boost transmission efficiency based on the laws of physics; Quantum sensing may boost the measurement precision significantly.

%However, the application of quantum information science in the real world may generate large amounts of data. Quantum data (in qubits) will carry the information, so we need some special units to handle quantum data as well as efficient methods to make the quantum data interact with the classical world. At the same time, the scale of classical data is getting bigger and bigger, and one could naturally consider storing and processing classical data in quantum devices with large Hilbert space dimensions. 
The generation, processing, and application of quantum data, and the treatment of those data together with their classical counterparts, are currently challenging theoretical and experimental problems in quantum science.

In this paper, we propose the idea of the so-called \emph{Quantum Data Center} (QDC), a unified concept referring to some specific quantum hardware that could efficiently deal with the quantum data, and would provide an efficient interface between classical data and quantum processors. The key component of the proposed QDC is a Quantum Random Access Memory (QRAM) \cite{giovannetti2008quantum,giovannetti2008architectures,hong2012robust,arunachalam2015robustness,hann2019hardware,di2020fault,paler2020parallelizing,hann2021resilience,connorthesis}, which is a device that allows a user to access multiple different elements in superposition from a database (which can be either classical or quantum). At minimum, a QDC consists of a QRAM coupled to a quantum network. 

We construct a theory of QDCs associated with original applications. We propose explicit constructions of examples including: QDCs as implementations of data-lookup oracles in fault-tolerant quantum computations; QDCs as mediators of so-called multi-party private quantum communications (defined below), which combines the Quantum Private Query (QPQ) \cite{giovannetti2008queries} and Quantum Secret Sharing \cite{cleve1999share,hillery1999quantum} protocols; and QDCs as quantum data compressors for distributed sensing applications. These three examples demonstrate that QDCs can provide significant advantages in the areas of quantum computing, quantum communication, and quantum sensing, respectively (see Figure \ref{fig:qdc_mean}), with all other technical details and extra examples in the Appendix.
\begin{figure}[h!]
\begin{center}  
\includegraphics[width=0.4\textwidth]{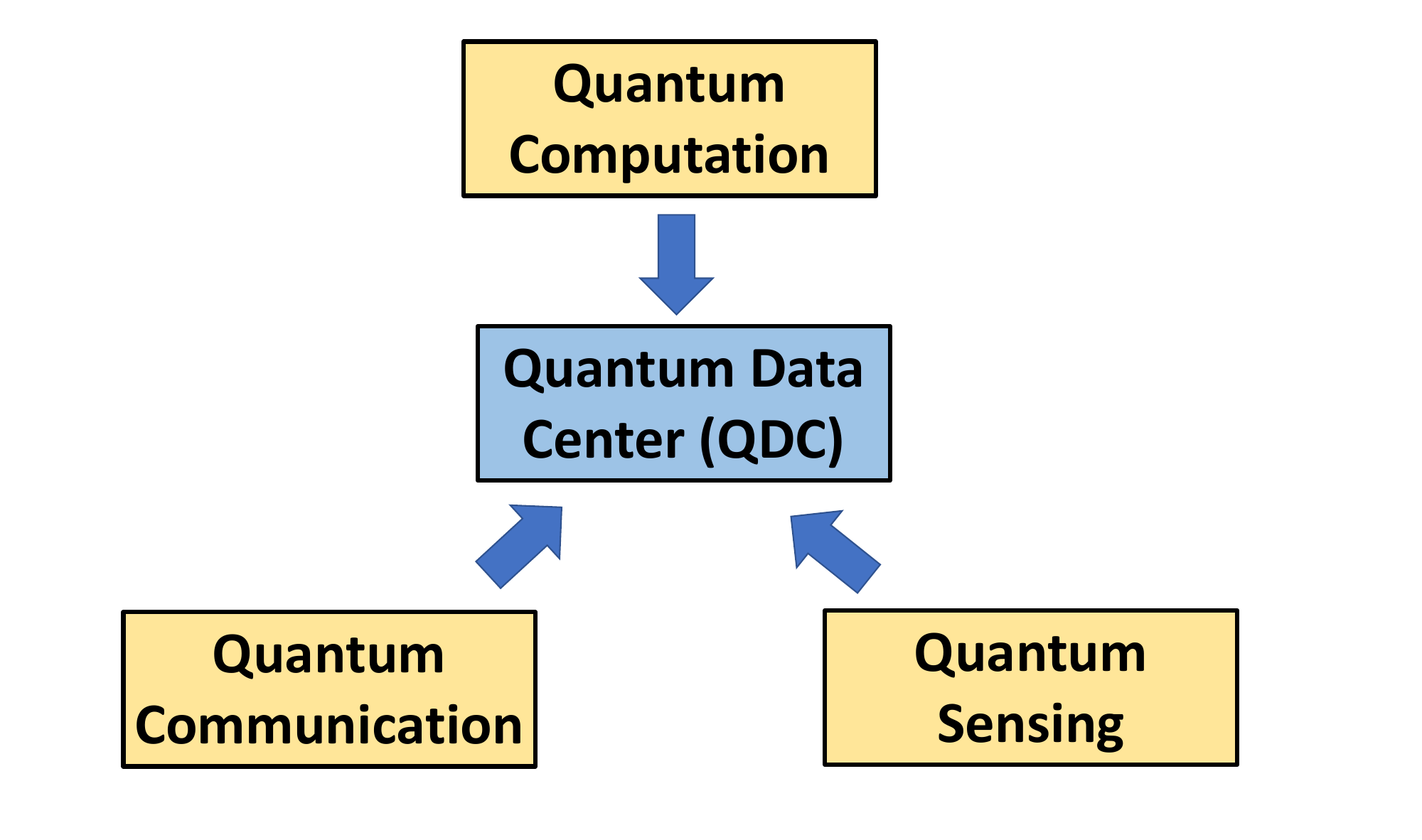}
\caption{Quantum data center (QDC) could potentially provide services about generation, processing, and application of quantum data, which could have wide applications in quantum computation, quantum communication, and quantum sensing. }
\label{fig:qdc_mean}
\end{center}  
\end{figure}  

\begin{figure}[h!]
\begin{center}  
\includegraphics[width=0.4\textwidth]{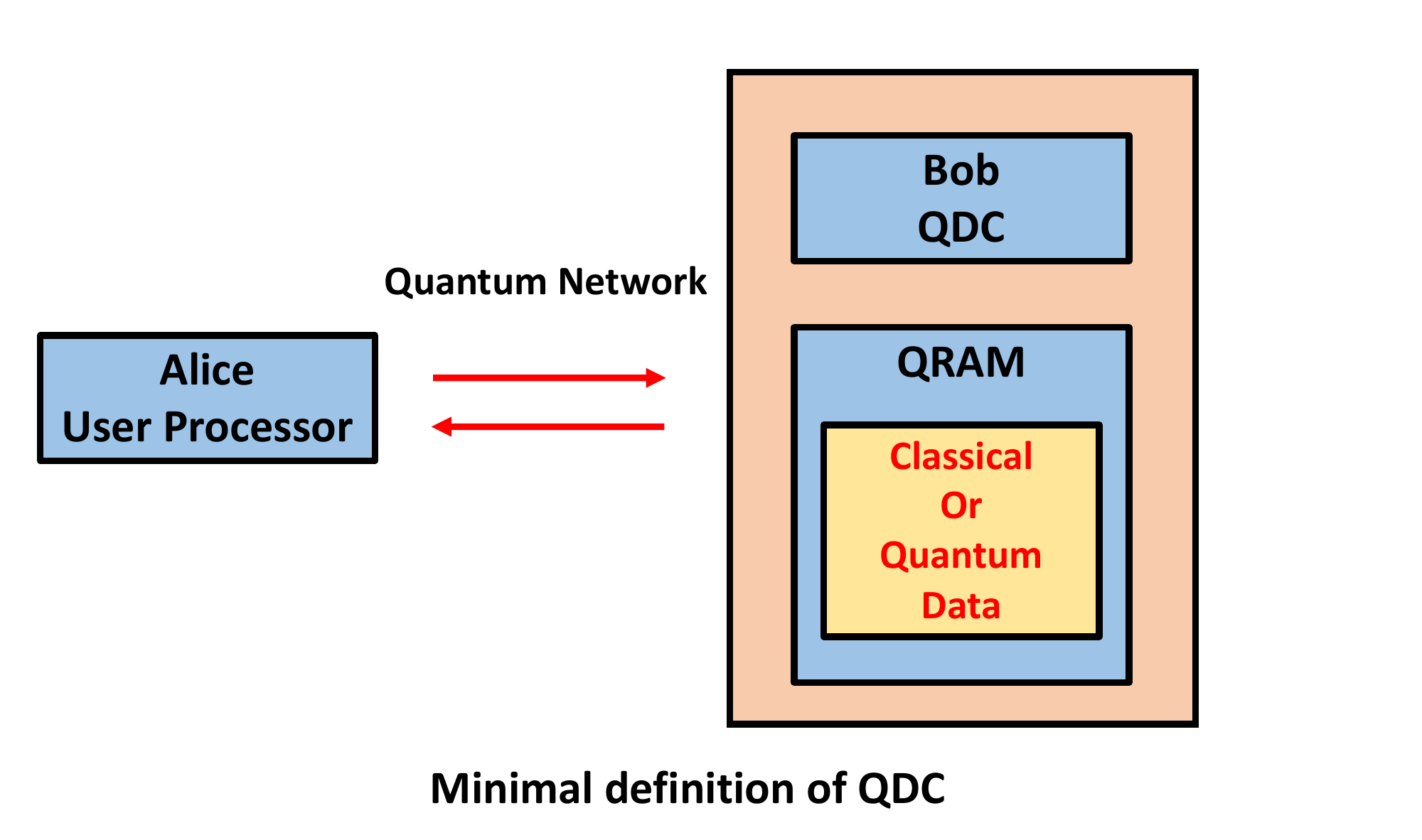}
\caption{The minimal definition of QDC contains the quantum network and QRAM. The data stored in QRAM can be either classical or quantum.}
\label{fig:qdc_minimal}
\end{center}  
\end{figure}

\textbf{\emph{General Theory}}. A minimal definition of a QDC is a quantum or classical database, equipped with QRAM, connected to a quantum (communication) network. The minimal function of a QDC is that Alice (the customer) is able to upload and download information (classical or quantum) by providing the address to the database (Bob), and Bob will provide the information to Alice through QRAM, sending it via the quantum network (see Figure \ref{fig:qdc_minimal}).

About the role of QRAM, there are numerous quantum algorithms that claim potential advantages against their classical counterparts, but those algorithms often implicitly require an interface between classical data and the quantum processor. The advantages of the computational complexity are estimated usually from the query complexity where the oracle provides this interface (see, for instance, \cite{wittek2014quantum}). Quantum Random Access Memory (QRAM, see \cite{nielsen2002quantum,giovannetti2008quantum,connorthesis}) is a general-purpose architecture that could serve as a realization of such oracles. 
    More specifically, QRAM allows a user to perform a superposition of queries to different elements of a data set stored in memory. The data itself can be either classical or quantum. In the case where the data is classical, the user provides an arbitrary superposition of addresses as input, and the QRAM returns an entangled state 
    where the addresses are correlated with the corresponding data:
    %More precisely, QRAM is defined by the following operation:
\begin{align}
\label{eq:QRAM_def_classical}
 %&\left| {{\psi _{{\rm{in }}}}} \right\rangle  =
 \sum\limits_{i = 0}^{N - 1} {{\alpha _i}} {\left| i \right\rangle ^{Q_1}}{\left| 0 \right\rangle ^{Q_2}} \to %\nonumber\\
 %& \left| {{\psi _{{\rm{out }}}}} \right\rangle  =
 \sum\limits_{i = 0}^{N - 1} {{\alpha _i}} {\left| i \right\rangle ^{Q_1}}{\left| {{x_i}} \right\rangle ^{Q_2}}~.
\end{align}  
Here, the superscripts $Q_1$ and $Q_2$ respectively denote the input and output qubit registers, $x_i$ denotes the $i$-th element of the classical data set, the $\alpha_i$ are generic coefficients, and $N$ is the size of the database. 
We emphasize the distinction between QRAM, defined via \Cref{eq:QRAM_def_classical}, and so-called random-access quantum memories~\cite{naik2017random,jiang2019experimental,langenfeld2020network}; the latter do not allow for accessing a superposition of multiple different data elements and hence are not sufficient for our purposes. Indeed, the ability to perform a superposition of queries as in \Cref{eq:QRAM_def_classical} is crucial to the applications we describe below. Moreover, in Appendix we will discuss in detail a quantum version of QRAM, and a formal definition of QDCs.

Note that our definition means that a quantum architecture could be qualified as QDC only if it has both QRAM and quantum networks implemented. Thus, if a device does not have either of them, it cannot be called a QDC according to our definition. For example, a standard quantum memory, as described in, e.g., Refs.~\cite{jiang2019experimental,langenfeld2020network}, coupled to a quantum network \emph{cannot} be called a QDC. This is because QRAM is more than just a quantum memory; QRAM requires that different elements of the memory can be queried in superposition as in \Cref{eq:QRAM_def_classical} (and the quantum version in Appendix). Moreover, there are extended parameters we could choose when we choose the circuit depth or the width of the QRAM implementation, see \cite{babbush2018encoding,low2018trading,berry2019qubitization,di2020fault,di2019methods,hann2021resilience,connorthesis,chen2021scalable}, but for latter applications, we assume our QRAM circuits to be shallow. Further, we assume that QRAM has been built in the fault-tolerant way and has been error-corrected. Building large-scale fault-tolerant QRAM is, in fact, a primary challenge in experiments \cite{arunachalam2015robustness}.

About the role of the quantum networks, they might be realizable in the future due to the fast development of quantum communication technology in recent years. Here, we are considering the service provided by QDC is centralized and has some physical distances from users. Thus, quantum states are supposed to be teleported through the quantum network from the user to the QDC or vice versa, where quantum teleportation technology includes the technologies of quantum satellites \cite{yin2017satellite,liao2017satellite,chen2021integrated}, quantum repeaters \cite{briegel1998quantum,duan2001long,kimble2008quantum,munro2015inside,muralidharan2014ultrafast,muralidharan2016optimal}, etc.

\textbf{\emph{QDC for quantum computing: fault-tolerant resource savings}}. As the first example about QDCs applied for quantum computing, we show how a QDC can provide resource savings in a fault-tolerant cost model to users running query-based quantum algorithms. There are two aspects of resource savings induced by QDC: hardware outsourcing and communication costs (see Appendix for an unification of space and time costs). The reason for the hardware outsourcing is simple. Imagine that we are doing fault-tolerant quantum computation with a significant amount of $T$-gates in the queries, which are considered to be expensive and require the magic state distillation~\cite{bravyi2005universal,bravyi2012magic}. 
Rather than preparing the requisite magic states themselves, the user could instead ask the QDC to prepare the magic states, thereby reducing the resources required of the user. 
This naive approach, however, has a high communication cost since each magic state would need to be sent over the quantum network from the QDC to the user.

In contrast to this naive approach, we propose that the user outsources entire oracle queries to the QDC. Outsourcing entire queries provides a particularly efficient way for users to offload large amounts of magic state distillation to the QDC with minimal communication cost.

More specifically, without the aid of a QDC, a user would be required to distill at least $\mathcal{O}(\sqrt{N})$~\cite{low2018trading} magic states in order to query a data set of size $N$ as in \Cref{eq:QRAM_def_classical}. In contrast, with a QDC, a user can outsource the query to the QDC: the QDC is responsible for implementing the query and distilling the associated magic states, while the user only incurs a $\mathcal{O}(\log N)$ communication cost. This communication cost is due to the fact that the input and outputs of the query must be sent between the user and the QDC. The user also benefits in that they are no longer responsible for the potentially large amount of ancillary qubits needed to implement a query~\cite{low2018trading,berry2019qubitization,di2020fault}; this hardware cost is paid by the QDC. Thus, this approach of outsourcing full queries to the QDC is exponentially more efficient than naive the approach described previously in terms of communication cost.

Both the savings from the hardware and the communication costs could be quantifies this benefit by the following example. Suppose, a user wishes to run a 100-qubit algorithm that requires $10^8$ $T$-gates when decomposed into Clifford $+T$ operations. Further, suppose that the user has a device with physical error rates of $p=10^{-3}$ and that the target failure probability for the entire computation is $<1\%$. To achieve this failure probability, we assume error correction is used, and that gates are implemented fault tolerantly. Non-Clifford gates are implemented fault-tolerantly with the aid of magic state distillation.

\begin{figure}[h!]
\begin{center}  
\includegraphics[width=0.5\textwidth]{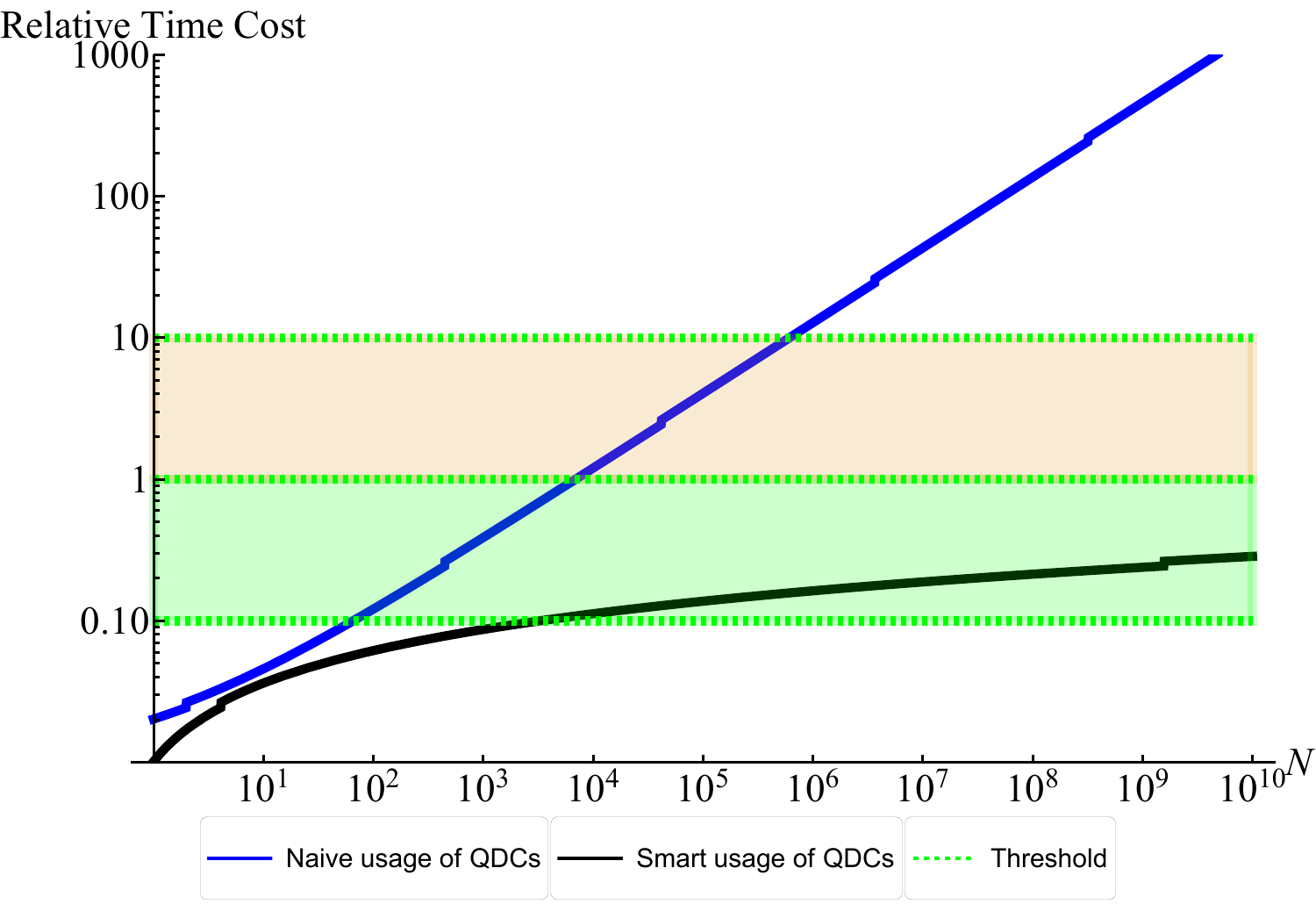}
\caption{The QDC-assisted relative time cost (the time cost from the user side with QDCs, divided by the one without QDCs) depending on the size of the data $N$. The dashed lines represent the threshold where QDC has the comparable performance as the situation without QDC, with the relative ratio $0.1$, $1$ and $10$. Different thresholds correspond to prefactors relating the delay factor and the data $N$, and the relative ratio choices might correspond to different physical hardware. We consider the situations where two different methods of usages of QDCs, and both of them prepare the magic state distillation in the QDC side. We replace the delay factor by the function $\sqrt{N}$ or $\log N$ directly where $N$ is the size of the data in those two methods. For the solid lines from up to down, we have naive (blue) and smart (black) usages of QDCs. Some sudden jumps in the plot are because of the even integer values of the code distance, and we defer more precise discussions in Appendix.}
\label{fig:QDCforcomputing}
\end{center}  
\end{figure}

A resource estimate for exactly this situation is performed in \cite{litinski2019game} for surface codes (see Appendix for a detailed review). The outsourcing can enable resource savings for the user (potentially both in the overall algorithm runtime and hardware cost).  To estimate these savings, we suppose that these queries are responsible for $99\%$ of the algorithm's $T$-state consumption (this is not an unreasonable supposition; see, for instance, \cite{berry2019qubitization}). According to \cite{litinski2019game} and Appendix, we observe that the user can now use a 15-to-1 magic state distillation scheme, as the user need only produce $\sim 10^6$ magic states, and the total probability that any such state is faulty is $10^6 \times 35p^3 \sim 0.01$, which is within the allowed error tolerance. As described in \cite{litinski2019game}, the number of surface code tiles required for computation and distillation with the 15-to-1 scheme is 164. With the distillation scheme selected, we can now estimate the required code distance $d$ and algorithm run time. The code distance must satisfy,
\begin{align}
    &(\text{\# tiles} = 164) \times (\text{\# code cycles}\times (1+\text{delay}))\nonumber\\
    & \times p_L(p,d) < 1\%~,
\end{align}
to guarantee that that the total error probability remains below $1\%$. Here $p_L$ is the logical error probability of a distance $d$ surface code with physical error probability $p$, which is approximately given by \cite{fowler2018low}. The parameter \emph{\# code cycles} is the number of surface code cycles required to distill $10^6$ magic states, i.e., the minimum number of cycles required to run the algorithm assuming instantaneous data center queries. In practice, however, the data center queries will not be instantaneous. Thus we add a \emph{delay} factor to the total number of cycles. This delay is related to the QDC's latency, $\tau$, and the exact amount of the delay depends not only on how the oracle is implemented by the QDC, but also on the communication time overhead. Moreover, with the $\sqrt{N}\to \log N$ arguments from the communication cost, we could use assumptions, $\text{delay factor}\sim \mathcal{O}({\sqrt{N}}) $ or $\text{delay factor}\sim \mathcal{O}({\log{N}})$, respectively, referring to the protocols where oracle queries are implemented by the user (with magic states sent one-by-one from the QDC) or where full queries are implemented by the QDC. In Figure \ref{fig:QDCforcomputing}, we plot the relative time costs depending on the data size $N$ of QDCs, where we show that QDCs provide significant time savings in some ranges of data sizes, where the communication cost savings could be exponential. Finally, although our calculation is query-based, there are proposals where a query-based approach could provide a unified framework for all quantum algorithms \cite{martyn2021grand}, despite the circuit depth for QRAMs should be shallow. \textcolor{black}{Finally, we emphasize that the advantage is only for outsourcing users instead of users combined with QDC. Moreover, the exponential savings of communication costs are from QRAM, instead of specific quantum or classical algorithms.} \textcolor{black}{As a summary, combinations of QRAM and quantum networks might lead significant benefits for outsourcing costs for $T$-gate preparations and magic state distillations from unique features of QRAM, leading to useful applications in quantum computing.} 

\textbf{\emph{QDC for quantum communication: multi-party private quantum communication}}. Quantum Private Query (QPQ) \cite{giovannetti2008queries}, a protocol combining QRAM and quantum networks, could already serve as an important application of QDC for quantum communication. Furthermore, the application of QDCs could be much broader to provide the users with fast and secure service. Based on QPQ and Quantum Secret Sharing from \cite{cleve1999share,hillery1999quantum}, we propose an original protocol, so-called multi-party private quantum communication, as an example of applications of QDCs. We provide discussions of QPQ and related concepts in Appendix as references \footnote{there are studies about limitations and insecurity concerns about concepts that are related to QPQs. In fact, there are no-go theorems \cite{mayers1997unconditionally,lo1997quantum,lo1997insecurity} about the imperfection of certain quantum computation and communication schemes. The QPQ protocol does not violate the no-go theorems \cite{lo1997insecurity} since the setup is different. In QPQ, we do not have the security requirement required for the no-theorem, where the user Alice cannot know the private key of QDCs, since QDCs do not have private keys. On the other hand, it will be interesting to understand better how those studies could potentially improve the capability of QDCs.}.

\begin{figure}
    \centering
    \includegraphics[width=0.4\textwidth]{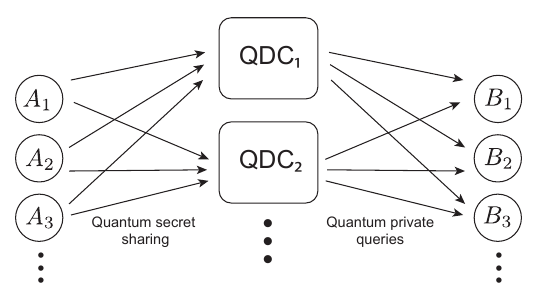}
    \caption{Multi-party private quantum communication protocol. A set of sending users $A_i$ communicates privately to a set of receiving users $B_i$ through untrusted, non-cooperating QDCs. The use of quantum secret sharing and quantum private queries guarantees that no QDC can learn what information was communicated nor where the information was sent. }
    \label{fig:multi-party}
\end{figure}

We present a novel protocol for multi-party private quantum communication using QDCs. We consider the situation in Figure \ref{fig:multi-party} where many sending users  (denoted $A_1$, $A_2$, etc) want to communicate privately to a set of receiving users (denoted $B_1$, $B_2$, etc; $A_i$ communicates with $B_j$, where $i\neq j$ in general). The communication occurs through two or more untrusted (but non-cooperating) QDCs. Importantly, it is assumed that the users do not share any initial secret keys or entangled qubits, and that the users do not possess any secure communication links between them (either classical or quantum); all communication takes place over the untrusted quantum network shared with the QDCs. 

The protocol is as follows. First, each sending user $A_i$ takes their quantum message, and decomposes it into several distinct parts using a quantum secret sharing protocol~\cite{cleve1999share,hillery1999quantum}. In isolation, each part of the secret message looks like a maximally mixed state, but when sufficiently many parts are assembled together, the original message can be perfectly recovered. Second, each sending user $A_i$ sends parts of their secrets to the QDCs, where they are stored in QRAM at a publicly announced address. No one QDC should receive enough parts of a secret to reconstruct the original message. Finally, the receiving users interrogate the QDCs using the quantum private queries protocol. Each user $B_j$ interrogates sufficiently many QDCs in order to retrieve enough parts of the secret to reconstruct $A_i$'s original message. Advantages of this protocol is explained in Appendix. \textcolor{black}{Moreover, a final note is that our protocol does not offer the communication between $A_k$ and $B_k$ protection from interception by $B_{j\ne k}$, rather it only provides privacy from untrusted QDCs}. \textcolor{black}{As a summary, combinations of QRAM and quantum networks could lead significant benefits for secure and private quantum communications, where quantum secret sharing and quantum private queries could serve as components.}

\begin{figure}
    \centering
    \includegraphics[width=0.5\textwidth]{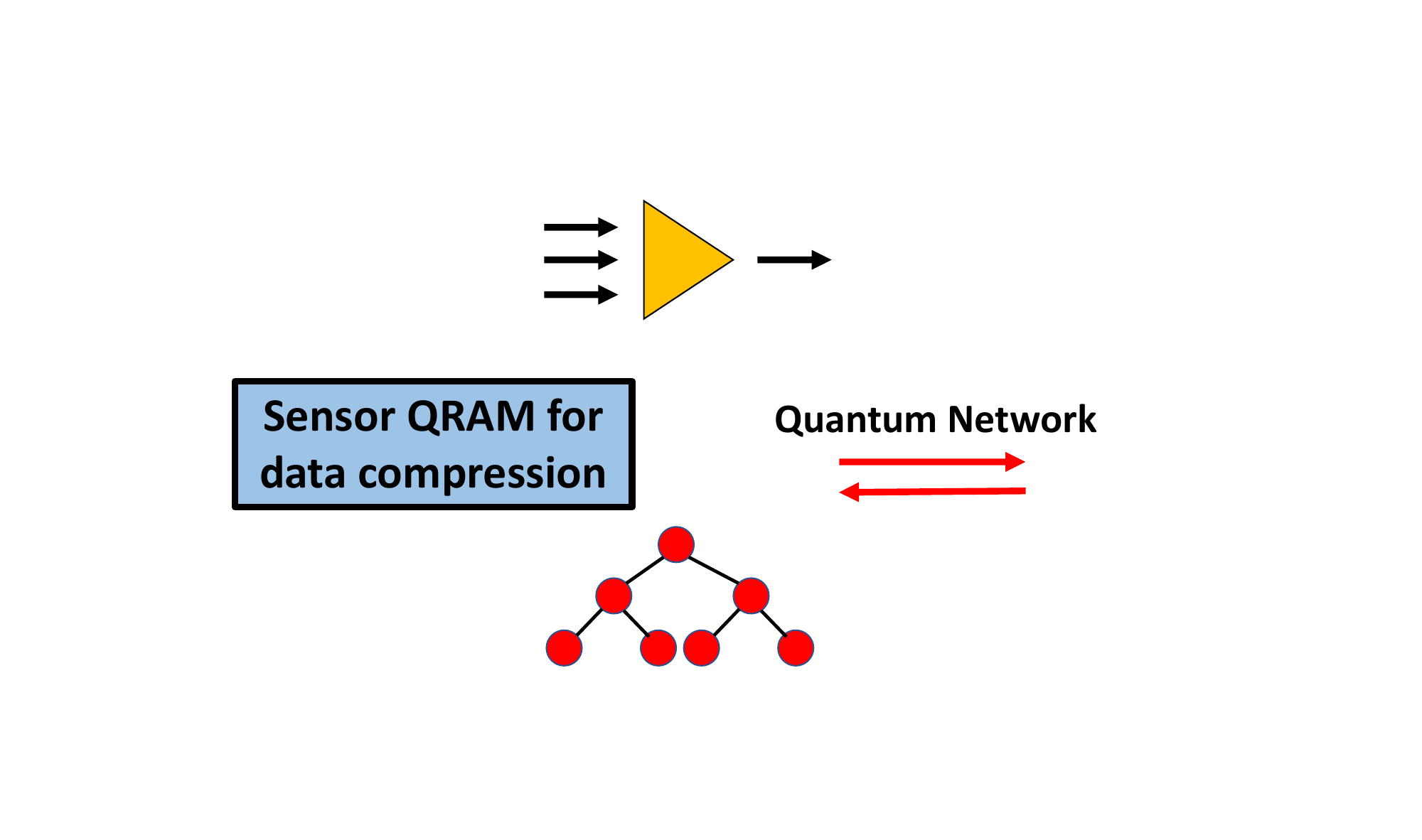}
    \caption{An illustration of QDC for quantum sensing: we use QDCs to perform quantum data compression that enables distributive sensing.}
    \label{fig:distribute}
\end{figure}

\textbf{\emph{QDC for quantum sensing: data compression and distributive sensing}}. In the context of quantum sensing, QDCs can be used to compress quantum data and signals, enabling more efficient communication in distributed sensing tasks (see Figure \ref{fig:distribute}).
%We discuss this application in further detail below, but we begin by first describing the underlying capability that enables these applications---the ability of a QDC to compress quantum data. In short, the reason why we need to compress the data during quantum sensing, is that the raw data has low entanglement entropy. Thus, it is resource-consuming to teleport such low entropy data directly. Hence we need quantum data compression to build more efficient distributed quantum sensors.
\begin{comment}
\begin{figure}
    \centering
    \includegraphics[width=0.5\textwidth]{figures/dissensing.pdf}
    \caption{An illustration of QDC for quantum sensing: we use QDCs to perform quantum data compression that enables distributive sensing.}
    \label{fig:distribute}
\end{figure}
\end{comment}

To start, we illustrate how QDCs can be used to compress quantum data through a simple example. Suppose that the quantum data held by the QDC is confined to the single-expectation subspace, spanned by states where only one of the $N$ qubits in the QDC's quantum memory is in the $\ket{1}$ state and all others are in $\ket{0}$. The state of the memory can then be written as 
\begin{align}
\ket{\psi_{\text{unary}}} = \sum_{i = 0}^{N-1} \alpha_i \bigotimes_{j = 1}^{N} \ket{\delta_{ij}}^{D_j} ,
\end{align}
where $D_j$ indicates the $j$-th qubit in the $N$-qubit quantum memory, and $\delta_{ij}$ is the Kronecker delta ($\delta_{ij} = 1$ for $i=j$ and $\delta_{ij} = 0$ otherwise). Though the entire Hilbert space of the $N$-qubit quantum memory has the dimension $2^N$, the single-excitation subspace has only dimension $N$. Thus, one could equivalently represent the above state using only $\log N$ qubits, as 
\begin{align}
\ket{\psi_{\text{binary}}}  = \sum_{i = 0}^{N-1} \alpha_i \ket{i}^{Q_1},
\end{align}
where $Q_1$ denotes a $\log N$-qubit register, and $\ket{i}^{Q_1}$ denotes the $i$-th basis state of this register. The two states $\ket{\psi_{\text{unary}}} $ and $\ket{\psi_{\text{both}}} $ contain the same quantum information (the $N$ complex coefficients $\alpha_i$) but encode this information in different ways.

%The mapping $\ket{\psi_{\text{unary}}} \rightarrow \ket{\psi_{\text{binary}}}$ constitutes compression of this quantum information because the information is mapped from an $N$-qubit encoding to a $\log N$-qubit encoding. The compressed state $\ket{\psi_{\text{binary}}} $ can be more efficiently stored or transmitted. 

A QDC can be used to realize the unary-to-binary compression described above, \textcolor{black}{where the precise form implemented using QRAM is originally constructed in our work}. The compression proceeds in two steps: first, the QDC performs an operation $U$ (defined below) that encodes the location of the single excitation into a $\log N$-qubit address register, then a single QRAM query is performed in order to extract the excitation from the memory.  In detail, the unitary $U$ enacts the operation,
\begin{align}
\label{eq:compression_U}
&U\left(\ket{0}^{Q_1}   \sum_{i = 0}^{N-1} \alpha_i \left[\bigotimes _{j = 1}^{N} \ket{\delta_{ij}}^{D_j} \right]\right) 
\nonumber \\
=& \sum_{i=0}^{N-1} \alpha_i  \ket{i}^{Q_1}  \left[\bigotimes_{j = 1}^{N} \ket{\delta_{ij}}^{D_j} \right].
\end{align} 
We note that the operation $U$ is not equivalent to a QRAM query,  so $U$ falls outside the scope of operations that a QDC can perform per the minimal definition. As we describe in Appendix, however, the operation $U$ can be straightforwardly implemented using only minor modifications to standard QRAM architectures. Next, a QRAM query extracts the single excitation from the quantum memory and stores it in an output register $Q_2$,
\begin{align}
    \label{eq:QRAM_compression}
    &\sum_{i=0}^{N-1} \alpha_i  \ket{i}^{Q_1} \ket{0}^{Q_2} \left[\bigotimes_{j = 1}^{N} \ket{\delta_{ij}}^{D_j} \right]
    \nonumber \\
    \rightarrow& \sum_{i=0}^{N-1} \alpha_i  \ket{i}^{Q_1}  \ket{1}^{Q_2} \left[\bigotimes_{j = 1}^{N} \ket{0}^{D_j} \right].
\end{align}
After this step, the $Q_2$ and $D_j$ registers are disentangled from the $Q_1$ register. The state of the $Q_1$ register is $\ket{\psi_{\text{binary}}}$, which constitutes the compressed representation of the quantum data that was originally stored in the QDC's memory. This compressed data may subsequently be stored, transmitted, or measured, depending on the application \footnote{Moreover, we remark that numerous extensions of this simple unary-to-binary compression are possible. For example, it is straightforward to generalize the above procedure to the case of multiple excitations, such that a QDC can be used to compress multi-excitation subspaces as well (see related discussions in Appendix about quantum data compression). QDCs could even be used for Schmacher coding~\cite{schumacher1995quantum}, universal source coding~\cite{jozsa1998universal, hayashi2002quantum,bennett2006universal}, or other quantum compression tasks. The ability to efficiently compress quantum data using QRAM and related architectures is studied more thoroughly in \cite{hann2022inprep}.}. Thus, QDCs can be used to reduce the entanglement cost for distributed sensing applications. See details in Appendix for explanations. \textcolor{black}{As a summary, combinations of QRAM and quantum networks could lead significant benefits for distributed quantum sensing, where QRAM designs could be helpful to perform quantum data compression and benefit entanglement cost reduction in distributive sensing applications.}

\textbf{\emph{Outlook and Conclusion}}. 
Our research on QDCs opens up a promising new direction in quantum information science (see details in Appendix for more applications). Recently, we have analyzed the favorable error scaling of QRAM that only scales poly-log with the size of the database \cite{hann2021resilience}, which implies that QDC might be an intermediate-term application without the requirement of full error correction. QDC provides an example of application-specific efficient architectural design, taking full advantage of shallow QRAM circuits and small overhead in quantum communication. Given the tree-like structure of QRAM, it will be interesting to explore the future possibility of distributed QDC, so that we may de-centralize the QRAM and perform the entire QRAM over the distributed quantum networks.
\\

\section*{Acknowledgement}
We thank Gideon Lee, John Preskill, Nicolas Sawaya, Xiaodi Wu, Xiao Yuan, Pei Zeng and Sisi Zhou for useful discussions. We specifically thank Isaac Chuang for inspiring discussions. JL is supported in part by International Business Machines (IBM) Quantum through the Chicago Quantum Exchange, and the Pritzker School of Molecular Engineering at the University of Chicago through AFOSR MURI (FA9550-21-1-0209). LJ acknowledges support from the ARO (W911NF-18-1-0020, W911NF-18-1-0212), ARO MURI (W911NF-16-1-0349, W911NF-21-1-0325), AFOSR MURI (FA9550-19-1-0399, FA9550-21-1-0209), AFRL (FA8649-21-P-0781), DoE Q-NEXT, NSF (OMA-1936118, EEC-1941583, OMA-2137642), NTT Research, and the Packard Foundation (2020-71479).

%\bibliography{ref}% Produces the bibliography via BibTeX.
\bibliographystyle{apsrev4-1}
\bibliography{citations.bib}

\begin{thebibliography}{87}%
\makeatletter
\providecommand \@ifxundefined [1]{%
 \@ifx{#1\undefined}
}%
\providecommand \@ifnum [1]{%
 \ifnum #1\expandafter \@firstoftwo
 \else \expandafter \@secondoftwo
 \fi
}%
\providecommand \@ifx [1]{%
 \ifx #1\expandafter \@firstoftwo
 \else \expandafter \@secondoftwo
 \fi
}%
\providecommand \natexlab [1]{#1}%
\providecommand \enquote  [1]{``#1''}%
\providecommand \bibnamefont  [1]{#1}%
\providecommand \bibfnamefont [1]{#1}%
\providecommand \citenamefont [1]{#1}%
\providecommand \href@noop [0]{\@secondoftwo}%
\providecommand \href [0]{\begingroup \@sanitize@url \@href}%
\providecommand \@href[1]{\@@startlink{#1}\@@href}%
\providecommand \@@href[1]{\endgroup#1\@@endlink}%
\providecommand \@sanitize@url [0]{\catcode `\\12\catcode `\$12\catcode
  `\&12\catcode `\#12\catcode `\^12\catcode `\_12\catcode `\%12\relax}%
\providecommand \@@startlink[1]{}%
\providecommand \@@endlink[0]{}%
\providecommand \url  [0]{\begingroup\@sanitize@url \@url }%
\providecommand \@url [1]{\endgroup\@href {#1}{\urlprefix }}%
\providecommand \urlprefix  [0]{URL }%
\providecommand \Eprint [0]{\href }%
\providecommand \doibase [0]{http://dx.doi.org/}%
\providecommand \selectlanguage [0]{\@gobble}%
\providecommand \bibinfo  [0]{\@secondoftwo}%
\providecommand \bibfield  [0]{\@secondoftwo}%
\providecommand \translation [1]{[#1]}%
\providecommand \BibitemOpen [0]{}%
\providecommand \bibitemStop [0]{}%
\providecommand \bibitemNoStop [0]{.\EOS\space}%
\providecommand \EOS [0]{\spacefactor3000\relax}%
\providecommand \BibitemShut  [1]{\csname bibitem#1\endcsname}%
\let\auto@bib@innerbib\@empty
%</preamble>
\bibitem [{\citenamefont {Feynman}(2018)}]{feynman2018simulating}%
  \BibitemOpen
  \bibfield  {author} {\bibinfo {author} {\bibfnamefont {R.~P.}\ \bibnamefont
  {Feynman}},\ }in\ \href@noop {} {\emph {\bibinfo {booktitle} {Feynman and
  computation}}}\ (\bibinfo  {publisher} {CRC Press},\ \bibinfo {year} {2018})\
  pp.\ \bibinfo {pages} {133--153}\BibitemShut {NoStop}%
\bibitem [{\citenamefont {Shor}(1999)}]{shor1999polynomial}%
  \BibitemOpen
  \bibfield  {author} {\bibinfo {author} {\bibfnamefont {P.~W.}\ \bibnamefont
  {Shor}},\ }\href@noop {} {\bibfield  {journal} {\bibinfo  {journal} {SIAM
  review}\ }\textbf {\bibinfo {volume} {41}},\ \bibinfo {pages} {303} (\bibinfo
  {year} {1999})}\BibitemShut {NoStop}%
\bibitem [{\citenamefont {Grover}(1997)}]{grover1997quantum}%
  \BibitemOpen
  \bibfield  {author} {\bibinfo {author} {\bibfnamefont {L.~K.}\ \bibnamefont
  {Grover}},\ }\href@noop {} {\bibfield  {journal} {\bibinfo  {journal}
  {Physical review letters}\ }\textbf {\bibinfo {volume} {79}},\ \bibinfo
  {pages} {325} (\bibinfo {year} {1997})}\BibitemShut {NoStop}%
\bibitem [{\citenamefont {Preskill}(2018)}]{preskill2018quantum}%
  \BibitemOpen
  \bibfield  {author} {\bibinfo {author} {\bibfnamefont {J.}~\bibnamefont
  {Preskill}},\ }\href@noop {} {\bibfield  {journal} {\bibinfo  {journal}
  {Quantum}\ }\textbf {\bibinfo {volume} {2}},\ \bibinfo {pages} {79} (\bibinfo
  {year} {2018})}\BibitemShut {NoStop}%
\bibitem [{\citenamefont {Wittek}(2014)}]{wittek2014quantum}%
  \BibitemOpen
  \bibfield  {author} {\bibinfo {author} {\bibfnamefont {P.}~\bibnamefont
  {Wittek}},\ }\href@noop {} {\emph {\bibinfo {title} {Quantum machine
  learning: what quantum computing means to data mining}}}\ (\bibinfo
  {publisher} {Academic Press},\ \bibinfo {year} {2014})\BibitemShut {NoStop}%
\bibitem [{\citenamefont {Biamonte}\ \emph {et~al.}(2017)\citenamefont
  {Biamonte}, \citenamefont {Wittek}, \citenamefont {Pancotti}, \citenamefont
  {Rebentrost}, \citenamefont {Wiebe},\ and\ \citenamefont
  {Lloyd}}]{biamonte2017quantum}%
  \BibitemOpen
  \bibfield  {author} {\bibinfo {author} {\bibfnamefont {J.}~\bibnamefont
  {Biamonte}}, \bibinfo {author} {\bibfnamefont {P.}~\bibnamefont {Wittek}},
  \bibinfo {author} {\bibfnamefont {N.}~\bibnamefont {Pancotti}}, \bibinfo
  {author} {\bibfnamefont {P.}~\bibnamefont {Rebentrost}}, \bibinfo {author}
  {\bibfnamefont {N.}~\bibnamefont {Wiebe}}, \ and\ \bibinfo {author}
  {\bibfnamefont {S.}~\bibnamefont {Lloyd}},\ }\href@noop {} {\bibfield
  {journal} {\bibinfo  {journal} {Nature}\ }\textbf {\bibinfo {volume} {549}},\
  \bibinfo {pages} {195} (\bibinfo {year} {2017})}\BibitemShut {NoStop}%
\bibitem [{\citenamefont {Gisin}\ \emph {et~al.}(2002)\citenamefont {Gisin},
  \citenamefont {Ribordy}, \citenamefont {Tittel},\ and\ \citenamefont
  {Zbinden}}]{gisin2002quantum}%
  \BibitemOpen
  \bibfield  {author} {\bibinfo {author} {\bibfnamefont {N.}~\bibnamefont
  {Gisin}}, \bibinfo {author} {\bibfnamefont {G.}~\bibnamefont {Ribordy}},
  \bibinfo {author} {\bibfnamefont {W.}~\bibnamefont {Tittel}}, \ and\ \bibinfo
  {author} {\bibfnamefont {H.}~\bibnamefont {Zbinden}},\ }\href@noop {}
  {\bibfield  {journal} {\bibinfo  {journal} {Reviews of modern physics}\
  }\textbf {\bibinfo {volume} {74}},\ \bibinfo {pages} {145} (\bibinfo {year}
  {2002})}\BibitemShut {NoStop}%
\bibitem [{\citenamefont {Cleve}\ \emph {et~al.}(1999)\citenamefont {Cleve},
  \citenamefont {Gottesman},\ and\ \citenamefont {Lo}}]{cleve1999share}%
  \BibitemOpen
  \bibfield  {author} {\bibinfo {author} {\bibfnamefont {R.}~\bibnamefont
  {Cleve}}, \bibinfo {author} {\bibfnamefont {D.}~\bibnamefont {Gottesman}}, \
  and\ \bibinfo {author} {\bibfnamefont {H.-K.}\ \bibnamefont {Lo}},\
  }\href@noop {} {\bibfield  {journal} {\bibinfo  {journal} {Physical Review
  Letters}\ }\textbf {\bibinfo {volume} {83}},\ \bibinfo {pages} {648}
  (\bibinfo {year} {1999})}\BibitemShut {NoStop}%
\bibitem [{\citenamefont {Hillery}\ \emph {et~al.}(1999)\citenamefont
  {Hillery}, \citenamefont {Bu{\v{z}}ek},\ and\ \citenamefont
  {Berthiaume}}]{hillery1999quantum}%
  \BibitemOpen
  \bibfield  {author} {\bibinfo {author} {\bibfnamefont {M.}~\bibnamefont
  {Hillery}}, \bibinfo {author} {\bibfnamefont {V.}~\bibnamefont
  {Bu{\v{z}}ek}}, \ and\ \bibinfo {author} {\bibfnamefont {A.}~\bibnamefont
  {Berthiaume}},\ }\href@noop {} {\bibfield  {journal} {\bibinfo  {journal}
  {Physical Review A}\ }\textbf {\bibinfo {volume} {59}},\ \bibinfo {pages}
  {1829} (\bibinfo {year} {1999})}\BibitemShut {NoStop}%
\bibitem [{\citenamefont {Kimble}(2008)}]{kimble2008quantum}%
  \BibitemOpen
  \bibfield  {author} {\bibinfo {author} {\bibfnamefont {H.~J.}\ \bibnamefont
  {Kimble}},\ }\href@noop {} {\bibfield  {journal} {\bibinfo  {journal}
  {Nature}\ }\textbf {\bibinfo {volume} {453}},\ \bibinfo {pages} {1023}
  (\bibinfo {year} {2008})}\BibitemShut {NoStop}%
\bibitem [{\citenamefont {Caleffi}\ \emph {et~al.}(2018)\citenamefont
  {Caleffi}, \citenamefont {Cacciapuoti},\ and\ \citenamefont
  {Bianchi}}]{caleffi2018quantum}%
  \BibitemOpen
  \bibfield  {author} {\bibinfo {author} {\bibfnamefont {M.}~\bibnamefont
  {Caleffi}}, \bibinfo {author} {\bibfnamefont {A.~S.}\ \bibnamefont
  {Cacciapuoti}}, \ and\ \bibinfo {author} {\bibfnamefont {G.}~\bibnamefont
  {Bianchi}},\ }in\ \href@noop {} {\emph {\bibinfo {booktitle} {Proceedings of
  the 5th ACM International Conference on Nanoscale Computing and
  Communication}}}\ (\bibinfo {year} {2018})\ pp.\ \bibinfo {pages}
  {1--4}\BibitemShut {NoStop}%
\bibitem [{\citenamefont {Muralidharan}\ \emph {et~al.}(2014)\citenamefont
  {Muralidharan}, \citenamefont {Kim}, \citenamefont {L{\"u}tkenhaus},
  \citenamefont {Lukin},\ and\ \citenamefont
  {Jiang}}]{muralidharan2014ultrafast}%
  \BibitemOpen
  \bibfield  {author} {\bibinfo {author} {\bibfnamefont {S.}~\bibnamefont
  {Muralidharan}}, \bibinfo {author} {\bibfnamefont {J.}~\bibnamefont {Kim}},
  \bibinfo {author} {\bibfnamefont {N.}~\bibnamefont {L{\"u}tkenhaus}},
  \bibinfo {author} {\bibfnamefont {M.~D.}\ \bibnamefont {Lukin}}, \ and\
  \bibinfo {author} {\bibfnamefont {L.}~\bibnamefont {Jiang}},\ }\href@noop {}
  {\bibfield  {journal} {\bibinfo  {journal} {Physical review letters}\
  }\textbf {\bibinfo {volume} {112}},\ \bibinfo {pages} {250501} (\bibinfo
  {year} {2014})}\BibitemShut {NoStop}%
\bibitem [{\citenamefont {Muralidharan}\ \emph {et~al.}(2016)\citenamefont
  {Muralidharan}, \citenamefont {Li}, \citenamefont {Kim}, \citenamefont
  {L{\"u}tkenhaus}, \citenamefont {Lukin},\ and\ \citenamefont
  {Jiang}}]{muralidharan2016optimal}%
  \BibitemOpen
  \bibfield  {author} {\bibinfo {author} {\bibfnamefont {S.}~\bibnamefont
  {Muralidharan}}, \bibinfo {author} {\bibfnamefont {L.}~\bibnamefont {Li}},
  \bibinfo {author} {\bibfnamefont {J.}~\bibnamefont {Kim}}, \bibinfo {author}
  {\bibfnamefont {N.}~\bibnamefont {L{\"u}tkenhaus}}, \bibinfo {author}
  {\bibfnamefont {M.~D.}\ \bibnamefont {Lukin}}, \ and\ \bibinfo {author}
  {\bibfnamefont {L.}~\bibnamefont {Jiang}},\ }\href@noop {} {\bibfield
  {journal} {\bibinfo  {journal} {Scientific reports}\ }\textbf {\bibinfo
  {volume} {6}},\ \bibinfo {pages} {1} (\bibinfo {year} {2016})}\BibitemShut
  {NoStop}%
\bibitem [{\citenamefont {Degen}\ \emph {et~al.}(2017)\citenamefont {Degen},
  \citenamefont {Reinhard},\ and\ \citenamefont
  {Cappellaro}}]{degen2017quantum}%
  \BibitemOpen
  \bibfield  {author} {\bibinfo {author} {\bibfnamefont {C.~L.}\ \bibnamefont
  {Degen}}, \bibinfo {author} {\bibfnamefont {F.}~\bibnamefont {Reinhard}}, \
  and\ \bibinfo {author} {\bibfnamefont {P.}~\bibnamefont {Cappellaro}},\
  }\href@noop {} {\bibfield  {journal} {\bibinfo  {journal} {Reviews of modern
  physics}\ }\textbf {\bibinfo {volume} {89}},\ \bibinfo {pages} {035002}
  (\bibinfo {year} {2017})}\BibitemShut {NoStop}%
\bibitem [{\citenamefont {Giovannetti}\ \emph {et~al.}(2006)\citenamefont
  {Giovannetti}, \citenamefont {Lloyd},\ and\ \citenamefont
  {Maccone}}]{giovannetti2006quantum}%
  \BibitemOpen
  \bibfield  {author} {\bibinfo {author} {\bibfnamefont {V.}~\bibnamefont
  {Giovannetti}}, \bibinfo {author} {\bibfnamefont {S.}~\bibnamefont {Lloyd}},
  \ and\ \bibinfo {author} {\bibfnamefont {L.}~\bibnamefont {Maccone}},\
  }\href@noop {} {\bibfield  {journal} {\bibinfo  {journal} {Physical review
  letters}\ }\textbf {\bibinfo {volume} {96}},\ \bibinfo {pages} {010401}
  (\bibinfo {year} {2006})}\BibitemShut {NoStop}%
\bibitem [{\citenamefont {Giovannetti}\ \emph {et~al.}(2011)\citenamefont
  {Giovannetti}, \citenamefont {Lloyd},\ and\ \citenamefont
  {Maccone}}]{giovannetti2011advances}%
  \BibitemOpen
  \bibfield  {author} {\bibinfo {author} {\bibfnamefont {V.}~\bibnamefont
  {Giovannetti}}, \bibinfo {author} {\bibfnamefont {S.}~\bibnamefont {Lloyd}},
  \ and\ \bibinfo {author} {\bibfnamefont {L.}~\bibnamefont {Maccone}},\
  }\href@noop {} {\bibfield  {journal} {\bibinfo  {journal} {Nature photonics}\
  }\textbf {\bibinfo {volume} {5}},\ \bibinfo {pages} {222} (\bibinfo {year}
  {2011})}\BibitemShut {NoStop}%
\bibitem [{\citenamefont {Giovannetti}\ \emph
  {et~al.}(2008{\natexlab{a}})\citenamefont {Giovannetti}, \citenamefont
  {Lloyd},\ and\ \citenamefont {Maccone}}]{giovannetti2008quantum}%
  \BibitemOpen
  \bibfield  {author} {\bibinfo {author} {\bibfnamefont {V.}~\bibnamefont
  {Giovannetti}}, \bibinfo {author} {\bibfnamefont {S.}~\bibnamefont {Lloyd}},
  \ and\ \bibinfo {author} {\bibfnamefont {L.}~\bibnamefont {Maccone}},\
  }\href@noop {} {\bibfield  {journal} {\bibinfo  {journal} {Physical review
  letters}\ }\textbf {\bibinfo {volume} {100}},\ \bibinfo {pages} {160501}
  (\bibinfo {year} {2008}{\natexlab{a}})}\BibitemShut {NoStop}%
\bibitem [{\citenamefont {Giovannetti}\ \emph
  {et~al.}(2008{\natexlab{b}})\citenamefont {Giovannetti}, \citenamefont
  {Lloyd},\ and\ \citenamefont {Maccone}}]{giovannetti2008architectures}%
  \BibitemOpen
  \bibfield  {author} {\bibinfo {author} {\bibfnamefont {V.}~\bibnamefont
  {Giovannetti}}, \bibinfo {author} {\bibfnamefont {S.}~\bibnamefont {Lloyd}},
  \ and\ \bibinfo {author} {\bibfnamefont {L.}~\bibnamefont {Maccone}},\
  }\href@noop {} {\bibfield  {journal} {\bibinfo  {journal} {Physical Review
  A}\ }\textbf {\bibinfo {volume} {78}},\ \bibinfo {pages} {052310} (\bibinfo
  {year} {2008}{\natexlab{b}})}\BibitemShut {NoStop}%
\bibitem [{\citenamefont {Hong}\ \emph {et~al.}(2012)\citenamefont {Hong},
  \citenamefont {Xiang}, \citenamefont {Zhu}, \citenamefont {Jiang},\ and\
  \citenamefont {Wu}}]{hong2012robust}%
  \BibitemOpen
  \bibfield  {author} {\bibinfo {author} {\bibfnamefont {F.-Y.}\ \bibnamefont
  {Hong}}, \bibinfo {author} {\bibfnamefont {Y.}~\bibnamefont {Xiang}},
  \bibinfo {author} {\bibfnamefont {Z.-Y.}\ \bibnamefont {Zhu}}, \bibinfo
  {author} {\bibfnamefont {L.-z.}\ \bibnamefont {Jiang}}, \ and\ \bibinfo
  {author} {\bibfnamefont {L.-n.}\ \bibnamefont {Wu}},\ }\href@noop {}
  {\bibfield  {journal} {\bibinfo  {journal} {Physical Review A}\ }\textbf
  {\bibinfo {volume} {86}},\ \bibinfo {pages} {010306} (\bibinfo {year}
  {2012})}\BibitemShut {NoStop}%
\bibitem [{\citenamefont {Arunachalam}\ \emph {et~al.}(2015)\citenamefont
  {Arunachalam}, \citenamefont {Gheorghiu}, \citenamefont {Jochym-O'Connor},
  \citenamefont {Mosca},\ and\ \citenamefont
  {Srinivasan}}]{arunachalam2015robustness}%
  \BibitemOpen
  \bibfield  {author} {\bibinfo {author} {\bibfnamefont {S.}~\bibnamefont
  {Arunachalam}}, \bibinfo {author} {\bibfnamefont {V.}~\bibnamefont
  {Gheorghiu}}, \bibinfo {author} {\bibfnamefont {T.}~\bibnamefont
  {Jochym-O'Connor}}, \bibinfo {author} {\bibfnamefont {M.}~\bibnamefont
  {Mosca}}, \ and\ \bibinfo {author} {\bibfnamefont {P.~V.}\ \bibnamefont
  {Srinivasan}},\ }\href@noop {} {\bibfield  {journal} {\bibinfo  {journal}
  {New Journal of Physics}\ }\textbf {\bibinfo {volume} {17}},\ \bibinfo
  {pages} {123010} (\bibinfo {year} {2015})}\BibitemShut {NoStop}%
\bibitem [{\citenamefont {Hann}\ \emph {et~al.}(2019)\citenamefont {Hann},
  \citenamefont {Zou}, \citenamefont {Zhang}, \citenamefont {Chu},
  \citenamefont {Schoelkopf}, \citenamefont {Girvin},\ and\ \citenamefont
  {Jiang}}]{hann2019hardware}%
  \BibitemOpen
  \bibfield  {author} {\bibinfo {author} {\bibfnamefont {C.~T.}\ \bibnamefont
  {Hann}}, \bibinfo {author} {\bibfnamefont {C.-L.}\ \bibnamefont {Zou}},
  \bibinfo {author} {\bibfnamefont {Y.}~\bibnamefont {Zhang}}, \bibinfo
  {author} {\bibfnamefont {Y.}~\bibnamefont {Chu}}, \bibinfo {author}
  {\bibfnamefont {R.~J.}\ \bibnamefont {Schoelkopf}}, \bibinfo {author}
  {\bibfnamefont {S.~M.}\ \bibnamefont {Girvin}}, \ and\ \bibinfo {author}
  {\bibfnamefont {L.}~\bibnamefont {Jiang}},\ }\href@noop {} {\bibfield
  {journal} {\bibinfo  {journal} {Physical review letters}\ }\textbf {\bibinfo
  {volume} {123}},\ \bibinfo {pages} {250501} (\bibinfo {year}
  {2019})}\BibitemShut {NoStop}%
\bibitem [{\citenamefont {Di~Matteo}\ \emph {et~al.}(2020)\citenamefont
  {Di~Matteo}, \citenamefont {Gheorghiu},\ and\ \citenamefont
  {Mosca}}]{di2020fault}%
  \BibitemOpen
  \bibfield  {author} {\bibinfo {author} {\bibfnamefont {O.}~\bibnamefont
  {Di~Matteo}}, \bibinfo {author} {\bibfnamefont {V.}~\bibnamefont
  {Gheorghiu}}, \ and\ \bibinfo {author} {\bibfnamefont {M.}~\bibnamefont
  {Mosca}},\ }\href@noop {} {\bibfield  {journal} {\bibinfo  {journal} {IEEE
  Transactions on Quantum Engineering}\ }\textbf {\bibinfo {volume} {1}},\
  \bibinfo {pages} {1} (\bibinfo {year} {2020})}\BibitemShut {NoStop}%
\bibitem [{\citenamefont {Paler}\ \emph {et~al.}(2020)\citenamefont {Paler},
  \citenamefont {Oumarou},\ and\ \citenamefont
  {Basmadjian}}]{paler2020parallelizing}%
  \BibitemOpen
  \bibfield  {author} {\bibinfo {author} {\bibfnamefont {A.}~\bibnamefont
  {Paler}}, \bibinfo {author} {\bibfnamefont {O.}~\bibnamefont {Oumarou}}, \
  and\ \bibinfo {author} {\bibfnamefont {R.}~\bibnamefont {Basmadjian}},\
  }\href@noop {} {\bibfield  {journal} {\bibinfo  {journal} {Physical Review
  A}\ }\textbf {\bibinfo {volume} {102}},\ \bibinfo {pages} {032608} (\bibinfo
  {year} {2020})}\BibitemShut {NoStop}%
\bibitem [{\citenamefont {Hann}\ \emph {et~al.}(2021)\citenamefont {Hann},
  \citenamefont {Lee}, \citenamefont {Girvin},\ and\ \citenamefont
  {Jiang}}]{hann2021resilience}%
  \BibitemOpen
  \bibfield  {author} {\bibinfo {author} {\bibfnamefont {C.~T.}\ \bibnamefont
  {Hann}}, \bibinfo {author} {\bibfnamefont {G.}~\bibnamefont {Lee}}, \bibinfo
  {author} {\bibfnamefont {S.}~\bibnamefont {Girvin}}, \ and\ \bibinfo {author}
  {\bibfnamefont {L.}~\bibnamefont {Jiang}},\ }\href@noop {} {\bibfield
  {journal} {\bibinfo  {journal} {PRX Quantum}\ }\textbf {\bibinfo {volume}
  {2}},\ \bibinfo {pages} {020311} (\bibinfo {year} {2021})}\BibitemShut
  {NoStop}%
\bibitem [{\citenamefont {Hann}(2021)}]{connorthesis}%
  \BibitemOpen
  \bibfield  {author} {\bibinfo {author} {\bibfnamefont {C.}~\bibnamefont
  {Hann}},\ }\emph {\bibinfo {title} {Practicality of Quantum Random Access
  Memory}},\ \href@noop {} {Ph.D. thesis},\ \bibinfo  {school} {Yale
  University} (\bibinfo {year} {2021})\BibitemShut {NoStop}%
\bibitem [{\citenamefont {Giovannetti}\ \emph
  {et~al.}(2008{\natexlab{c}})\citenamefont {Giovannetti}, \citenamefont
  {Lloyd},\ and\ \citenamefont {Maccone}}]{giovannetti2008queries}%
  \BibitemOpen
  \bibfield  {author} {\bibinfo {author} {\bibfnamefont {V.}~\bibnamefont
  {Giovannetti}}, \bibinfo {author} {\bibfnamefont {S.}~\bibnamefont {Lloyd}},
  \ and\ \bibinfo {author} {\bibfnamefont {L.}~\bibnamefont {Maccone}},\
  }\href@noop {} {\bibfield  {journal} {\bibinfo  {journal} {Physical review
  letters}\ }\textbf {\bibinfo {volume} {100}},\ \bibinfo {pages} {230502}
  (\bibinfo {year} {2008}{\natexlab{c}})}\BibitemShut {NoStop}%
\bibitem [{\citenamefont {Nielsen}\ and\ \citenamefont
  {Chuang}(2002)}]{nielsen2002quantum}%
  \BibitemOpen
  \bibfield  {author} {\bibinfo {author} {\bibfnamefont {M.~A.}\ \bibnamefont
  {Nielsen}}\ and\ \bibinfo {author} {\bibfnamefont {I.}~\bibnamefont
  {Chuang}},\ }\href@noop {} {\enquote {\bibinfo {title} {Quantum computation
  and quantum information},}\ } (\bibinfo {year} {2002})\BibitemShut {NoStop}%
\bibitem [{\citenamefont {Naik}\ \emph {et~al.}(2017)\citenamefont {Naik},
  \citenamefont {Leung}, \citenamefont {Chakram}, \citenamefont {Groszkowski},
  \citenamefont {Lu}, \citenamefont {Earnest}, \citenamefont {McKay},
  \citenamefont {Koch},\ and\ \citenamefont {Schuster}}]{naik2017random}%
  \BibitemOpen
  \bibfield  {author} {\bibinfo {author} {\bibfnamefont {R.}~\bibnamefont
  {Naik}}, \bibinfo {author} {\bibfnamefont {N.}~\bibnamefont {Leung}},
  \bibinfo {author} {\bibfnamefont {S.}~\bibnamefont {Chakram}}, \bibinfo
  {author} {\bibfnamefont {P.}~\bibnamefont {Groszkowski}}, \bibinfo {author}
  {\bibfnamefont {Y.}~\bibnamefont {Lu}}, \bibinfo {author} {\bibfnamefont
  {N.}~\bibnamefont {Earnest}}, \bibinfo {author} {\bibfnamefont
  {D.}~\bibnamefont {McKay}}, \bibinfo {author} {\bibfnamefont
  {J.}~\bibnamefont {Koch}}, \ and\ \bibinfo {author} {\bibfnamefont
  {D.}~\bibnamefont {Schuster}},\ }\href@noop {} {\bibfield  {journal}
  {\bibinfo  {journal} {Nature communications}\ }\textbf {\bibinfo {volume}
  {8}},\ \bibinfo {pages} {1} (\bibinfo {year} {2017})}\BibitemShut {NoStop}%
\bibitem [{\citenamefont {Jiang}\ \emph {et~al.}(2019)\citenamefont {Jiang},
  \citenamefont {Pu}, \citenamefont {Chang}, \citenamefont {Li}, \citenamefont
  {Zhang},\ and\ \citenamefont {Duan}}]{jiang2019experimental}%
  \BibitemOpen
  \bibfield  {author} {\bibinfo {author} {\bibfnamefont {N.}~\bibnamefont
  {Jiang}}, \bibinfo {author} {\bibfnamefont {Y.-F.}\ \bibnamefont {Pu}},
  \bibinfo {author} {\bibfnamefont {W.}~\bibnamefont {Chang}}, \bibinfo
  {author} {\bibfnamefont {C.}~\bibnamefont {Li}}, \bibinfo {author}
  {\bibfnamefont {S.}~\bibnamefont {Zhang}}, \ and\ \bibinfo {author}
  {\bibfnamefont {L.-M.}\ \bibnamefont {Duan}},\ }\href@noop {} {\bibfield
  {journal} {\bibinfo  {journal} {npj Quantum Information}\ }\textbf {\bibinfo
  {volume} {5}},\ \bibinfo {pages} {1} (\bibinfo {year} {2019})}\BibitemShut
  {NoStop}%
\bibitem [{\citenamefont {Langenfeld}\ \emph {et~al.}(2020)\citenamefont
  {Langenfeld}, \citenamefont {Morin}, \citenamefont {K{\"o}rber},\ and\
  \citenamefont {Rempe}}]{langenfeld2020network}%
  \BibitemOpen
  \bibfield  {author} {\bibinfo {author} {\bibfnamefont {S.}~\bibnamefont
  {Langenfeld}}, \bibinfo {author} {\bibfnamefont {O.}~\bibnamefont {Morin}},
  \bibinfo {author} {\bibfnamefont {M.}~\bibnamefont {K{\"o}rber}}, \ and\
  \bibinfo {author} {\bibfnamefont {G.}~\bibnamefont {Rempe}},\ }\href@noop {}
  {\bibfield  {journal} {\bibinfo  {journal} {npj Quantum Information}\
  }\textbf {\bibinfo {volume} {6}},\ \bibinfo {pages} {1} (\bibinfo {year}
  {2020})}\BibitemShut {NoStop}%
\bibitem [{\citenamefont {Babbush}\ \emph {et~al.}(2018)\citenamefont
  {Babbush}, \citenamefont {Gidney}, \citenamefont {Berry}, \citenamefont
  {Wiebe}, \citenamefont {McClean}, \citenamefont {Paler}, \citenamefont
  {Fowler},\ and\ \citenamefont {Neven}}]{babbush2018encoding}%
  \BibitemOpen
  \bibfield  {author} {\bibinfo {author} {\bibfnamefont {R.}~\bibnamefont
  {Babbush}}, \bibinfo {author} {\bibfnamefont {C.}~\bibnamefont {Gidney}},
  \bibinfo {author} {\bibfnamefont {D.~W.}\ \bibnamefont {Berry}}, \bibinfo
  {author} {\bibfnamefont {N.}~\bibnamefont {Wiebe}}, \bibinfo {author}
  {\bibfnamefont {J.}~\bibnamefont {McClean}}, \bibinfo {author} {\bibfnamefont
  {A.}~\bibnamefont {Paler}}, \bibinfo {author} {\bibfnamefont
  {A.}~\bibnamefont {Fowler}}, \ and\ \bibinfo {author} {\bibfnamefont
  {H.}~\bibnamefont {Neven}},\ }\href@noop {} {\bibfield  {journal} {\bibinfo
  {journal} {Physical Review X}\ }\textbf {\bibinfo {volume} {8}},\ \bibinfo
  {pages} {041015} (\bibinfo {year} {2018})}\BibitemShut {NoStop}%
\bibitem [{\citenamefont {Low}\ \emph {et~al.}(2018)\citenamefont {Low},
  \citenamefont {Kliuchnikov},\ and\ \citenamefont
  {Schaeffer}}]{low2018trading}%
  \BibitemOpen
  \bibfield  {author} {\bibinfo {author} {\bibfnamefont {G.~H.}\ \bibnamefont
  {Low}}, \bibinfo {author} {\bibfnamefont {V.}~\bibnamefont {Kliuchnikov}}, \
  and\ \bibinfo {author} {\bibfnamefont {L.}~\bibnamefont {Schaeffer}},\
  }\href@noop {} {\bibfield  {journal} {\bibinfo  {journal} {arXiv preprint
  arXiv:1812.00954}\ } (\bibinfo {year} {2018})}\BibitemShut {NoStop}%
\bibitem [{\citenamefont {Berry}\ \emph {et~al.}(2019)\citenamefont {Berry},
  \citenamefont {Gidney}, \citenamefont {Motta}, \citenamefont {McClean},\ and\
  \citenamefont {Babbush}}]{berry2019qubitization}%
  \BibitemOpen
  \bibfield  {author} {\bibinfo {author} {\bibfnamefont {D.~W.}\ \bibnamefont
  {Berry}}, \bibinfo {author} {\bibfnamefont {C.}~\bibnamefont {Gidney}},
  \bibinfo {author} {\bibfnamefont {M.}~\bibnamefont {Motta}}, \bibinfo
  {author} {\bibfnamefont {J.~R.}\ \bibnamefont {McClean}}, \ and\ \bibinfo
  {author} {\bibfnamefont {R.}~\bibnamefont {Babbush}},\ }\href@noop {}
  {\bibfield  {journal} {\bibinfo  {journal} {Quantum}\ }\textbf {\bibinfo
  {volume} {3}},\ \bibinfo {pages} {208} (\bibinfo {year} {2019})}\BibitemShut
  {NoStop}%
\bibitem [{\citenamefont {Di~Matteo}(2019)}]{di2019methods}%
  \BibitemOpen
  \bibfield  {author} {\bibinfo {author} {\bibfnamefont {O.}~\bibnamefont
  {Di~Matteo}},\ }\href@noop {} {\emph {\bibinfo {title} {Methods for parallel
  quantum circuit synthesis, fault-tolerant quantum RAM, and quantum state
  tomography}}}\ (\bibinfo  {publisher} {University of Waterloo},\ \bibinfo
  {year} {2019})\BibitemShut {NoStop}%
\bibitem [{\citenamefont {Chen}\ \emph
  {et~al.}(2021{\natexlab{a}})\citenamefont {Chen}, \citenamefont {Dai},
  \citenamefont {Errando-Herranz}, \citenamefont {Lloyd},\ and\ \citenamefont
  {Englund}}]{chen2021scalable}%
  \BibitemOpen
  \bibfield  {author} {\bibinfo {author} {\bibfnamefont {K.~C.}\ \bibnamefont
  {Chen}}, \bibinfo {author} {\bibfnamefont {W.}~\bibnamefont {Dai}}, \bibinfo
  {author} {\bibfnamefont {C.}~\bibnamefont {Errando-Herranz}}, \bibinfo
  {author} {\bibfnamefont {S.}~\bibnamefont {Lloyd}}, \ and\ \bibinfo {author}
  {\bibfnamefont {D.}~\bibnamefont {Englund}},\ }\href@noop {} {\bibfield
  {journal} {\bibinfo  {journal} {PRX Quantum}\ }\textbf {\bibinfo {volume}
  {2}},\ \bibinfo {pages} {030319} (\bibinfo {year}
  {2021}{\natexlab{a}})}\BibitemShut {NoStop}%
\bibitem [{\citenamefont {Yin}\ \emph {et~al.}(2017)\citenamefont {Yin},
  \citenamefont {Cao}, \citenamefont {Li}, \citenamefont {Liao}, \citenamefont
  {Zhang}, \citenamefont {Ren}, \citenamefont {Cai}, \citenamefont {Liu},
  \citenamefont {Li}, \citenamefont {Dai} \emph {et~al.}}]{yin2017satellite}%
  \BibitemOpen
  \bibfield  {author} {\bibinfo {author} {\bibfnamefont {J.}~\bibnamefont
  {Yin}}, \bibinfo {author} {\bibfnamefont {Y.}~\bibnamefont {Cao}}, \bibinfo
  {author} {\bibfnamefont {Y.-H.}\ \bibnamefont {Li}}, \bibinfo {author}
  {\bibfnamefont {S.-K.}\ \bibnamefont {Liao}}, \bibinfo {author}
  {\bibfnamefont {L.}~\bibnamefont {Zhang}}, \bibinfo {author} {\bibfnamefont
  {J.-G.}\ \bibnamefont {Ren}}, \bibinfo {author} {\bibfnamefont {W.-Q.}\
  \bibnamefont {Cai}}, \bibinfo {author} {\bibfnamefont {W.-Y.}\ \bibnamefont
  {Liu}}, \bibinfo {author} {\bibfnamefont {B.}~\bibnamefont {Li}}, \bibinfo
  {author} {\bibfnamefont {H.}~\bibnamefont {Dai}},  \emph {et~al.},\
  }\href@noop {} {\bibfield  {journal} {\bibinfo  {journal} {Science}\ }\textbf
  {\bibinfo {volume} {356}},\ \bibinfo {pages} {1140} (\bibinfo {year}
  {2017})}\BibitemShut {NoStop}%
\bibitem [{\citenamefont {Liao}\ \emph {et~al.}(2017)\citenamefont {Liao},
  \citenamefont {Cai}, \citenamefont {Liu}, \citenamefont {Zhang},
  \citenamefont {Li}, \citenamefont {Ren}, \citenamefont {Yin}, \citenamefont
  {Shen}, \citenamefont {Cao}, \citenamefont {Li} \emph
  {et~al.}}]{liao2017satellite}%
  \BibitemOpen
  \bibfield  {author} {\bibinfo {author} {\bibfnamefont {S.-K.}\ \bibnamefont
  {Liao}}, \bibinfo {author} {\bibfnamefont {W.-Q.}\ \bibnamefont {Cai}},
  \bibinfo {author} {\bibfnamefont {W.-Y.}\ \bibnamefont {Liu}}, \bibinfo
  {author} {\bibfnamefont {L.}~\bibnamefont {Zhang}}, \bibinfo {author}
  {\bibfnamefont {Y.}~\bibnamefont {Li}}, \bibinfo {author} {\bibfnamefont
  {J.-G.}\ \bibnamefont {Ren}}, \bibinfo {author} {\bibfnamefont
  {J.}~\bibnamefont {Yin}}, \bibinfo {author} {\bibfnamefont {Q.}~\bibnamefont
  {Shen}}, \bibinfo {author} {\bibfnamefont {Y.}~\bibnamefont {Cao}}, \bibinfo
  {author} {\bibfnamefont {Z.-P.}\ \bibnamefont {Li}},  \emph {et~al.},\
  }\href@noop {} {\bibfield  {journal} {\bibinfo  {journal} {Nature}\ }\textbf
  {\bibinfo {volume} {549}},\ \bibinfo {pages} {43} (\bibinfo {year}
  {2017})}\BibitemShut {NoStop}%
\bibitem [{\citenamefont {Chen}\ \emph
  {et~al.}(2021{\natexlab{b}})\citenamefont {Chen}, \citenamefont {Zhang},
  \citenamefont {Chen}, \citenamefont {Cai}, \citenamefont {Liao},
  \citenamefont {Zhang}, \citenamefont {Chen}, \citenamefont {Yin},
  \citenamefont {Ren}, \citenamefont {Chen} \emph
  {et~al.}}]{chen2021integrated}%
  \BibitemOpen
  \bibfield  {author} {\bibinfo {author} {\bibfnamefont {Y.-A.}\ \bibnamefont
  {Chen}}, \bibinfo {author} {\bibfnamefont {Q.}~\bibnamefont {Zhang}},
  \bibinfo {author} {\bibfnamefont {T.-Y.}\ \bibnamefont {Chen}}, \bibinfo
  {author} {\bibfnamefont {W.-Q.}\ \bibnamefont {Cai}}, \bibinfo {author}
  {\bibfnamefont {S.-K.}\ \bibnamefont {Liao}}, \bibinfo {author}
  {\bibfnamefont {J.}~\bibnamefont {Zhang}}, \bibinfo {author} {\bibfnamefont
  {K.}~\bibnamefont {Chen}}, \bibinfo {author} {\bibfnamefont {J.}~\bibnamefont
  {Yin}}, \bibinfo {author} {\bibfnamefont {J.-G.}\ \bibnamefont {Ren}},
  \bibinfo {author} {\bibfnamefont {Z.}~\bibnamefont {Chen}},  \emph {et~al.},\
  }\href@noop {} {\bibfield  {journal} {\bibinfo  {journal} {Nature}\ }\textbf
  {\bibinfo {volume} {589}},\ \bibinfo {pages} {214} (\bibinfo {year}
  {2021}{\natexlab{b}})}\BibitemShut {NoStop}%
\bibitem [{\citenamefont {Briegel}\ \emph {et~al.}(1998)\citenamefont
  {Briegel}, \citenamefont {Dur}, \citenamefont {Cirac},\ and\ \citenamefont
  {Zoller}}]{briegel1998quantum}%
  \BibitemOpen
  \bibfield  {author} {\bibinfo {author} {\bibfnamefont {H.-J.}\ \bibnamefont
  {Briegel}}, \bibinfo {author} {\bibfnamefont {W.}~\bibnamefont {Dur}},
  \bibinfo {author} {\bibfnamefont {J.~I.}\ \bibnamefont {Cirac}}, \ and\
  \bibinfo {author} {\bibfnamefont {P.}~\bibnamefont {Zoller}},\ }\href@noop {}
  {\bibfield  {journal} {\bibinfo  {journal} {Physical Review Letters}\
  }\textbf {\bibinfo {volume} {81}},\ \bibinfo {pages} {5932} (\bibinfo {year}
  {1998})}\BibitemShut {NoStop}%
\bibitem [{\citenamefont {Duan}\ \emph {et~al.}(2001)\citenamefont {Duan},
  \citenamefont {Lukin}, \citenamefont {Cirac},\ and\ \citenamefont
  {Zoller}}]{duan2001long}%
  \BibitemOpen
  \bibfield  {author} {\bibinfo {author} {\bibfnamefont {L.-M.}\ \bibnamefont
  {Duan}}, \bibinfo {author} {\bibfnamefont {M.~D.}\ \bibnamefont {Lukin}},
  \bibinfo {author} {\bibfnamefont {J.~I.}\ \bibnamefont {Cirac}}, \ and\
  \bibinfo {author} {\bibfnamefont {P.}~\bibnamefont {Zoller}},\ }\href@noop {}
  {\bibfield  {journal} {\bibinfo  {journal} {Nature}\ }\textbf {\bibinfo
  {volume} {414}},\ \bibinfo {pages} {413} (\bibinfo {year}
  {2001})}\BibitemShut {NoStop}%
\bibitem [{\citenamefont {Munro}\ \emph {et~al.}(2015)\citenamefont {Munro},
  \citenamefont {Azuma}, \citenamefont {Tamaki},\ and\ \citenamefont
  {Nemoto}}]{munro2015inside}%
  \BibitemOpen
  \bibfield  {author} {\bibinfo {author} {\bibfnamefont {W.~J.}\ \bibnamefont
  {Munro}}, \bibinfo {author} {\bibfnamefont {K.}~\bibnamefont {Azuma}},
  \bibinfo {author} {\bibfnamefont {K.}~\bibnamefont {Tamaki}}, \ and\ \bibinfo
  {author} {\bibfnamefont {K.}~\bibnamefont {Nemoto}},\ }\href@noop {}
  {\bibfield  {journal} {\bibinfo  {journal} {IEEE Journal of Selected Topics
  in Quantum Electronics}\ }\textbf {\bibinfo {volume} {21}},\ \bibinfo {pages}
  {78} (\bibinfo {year} {2015})}\BibitemShut {NoStop}%
\bibitem [{\citenamefont {Bravyi}\ and\ \citenamefont
  {Kitaev}(2005)}]{bravyi2005universal}%
  \BibitemOpen
  \bibfield  {author} {\bibinfo {author} {\bibfnamefont {S.}~\bibnamefont
  {Bravyi}}\ and\ \bibinfo {author} {\bibfnamefont {A.}~\bibnamefont
  {Kitaev}},\ }\href@noop {} {\bibfield  {journal} {\bibinfo  {journal}
  {Physical Review A}\ }\textbf {\bibinfo {volume} {71}},\ \bibinfo {pages}
  {022316} (\bibinfo {year} {2005})}\BibitemShut {NoStop}%
\bibitem [{\citenamefont {Bravyi}\ and\ \citenamefont
  {Haah}(2012)}]{bravyi2012magic}%
  \BibitemOpen
  \bibfield  {author} {\bibinfo {author} {\bibfnamefont {S.}~\bibnamefont
  {Bravyi}}\ and\ \bibinfo {author} {\bibfnamefont {J.}~\bibnamefont {Haah}},\
  }\href@noop {} {\bibfield  {journal} {\bibinfo  {journal} {Physical Review
  A}\ }\textbf {\bibinfo {volume} {86}},\ \bibinfo {pages} {052329} (\bibinfo
  {year} {2012})}\BibitemShut {NoStop}%
\bibitem [{\citenamefont {Litinski}(2019)}]{litinski2019game}%
  \BibitemOpen
  \bibfield  {author} {\bibinfo {author} {\bibfnamefont {D.}~\bibnamefont
  {Litinski}},\ }\href@noop {} {\bibfield  {journal} {\bibinfo  {journal}
  {Quantum}\ }\textbf {\bibinfo {volume} {3}},\ \bibinfo {pages} {128}
  (\bibinfo {year} {2019})}\BibitemShut {NoStop}%
\bibitem [{\citenamefont {Fowler}\ and\ \citenamefont
  {Gidney}(2018)}]{fowler2018low}%
  \BibitemOpen
  \bibfield  {author} {\bibinfo {author} {\bibfnamefont {A.~G.}\ \bibnamefont
  {Fowler}}\ and\ \bibinfo {author} {\bibfnamefont {C.}~\bibnamefont
  {Gidney}},\ }\href@noop {} {\bibfield  {journal} {\bibinfo  {journal} {arXiv
  preprint arXiv:1808.06709}\ } (\bibinfo {year} {2018})}\BibitemShut {NoStop}%
\bibitem [{\citenamefont {Martyn}\ \emph {et~al.}(2021)\citenamefont {Martyn},
  \citenamefont {Rossi}, \citenamefont {Tan},\ and\ \citenamefont
  {Chuang}}]{martyn2021grand}%
  \BibitemOpen
  \bibfield  {author} {\bibinfo {author} {\bibfnamefont {J.~M.}\ \bibnamefont
  {Martyn}}, \bibinfo {author} {\bibfnamefont {Z.~M.}\ \bibnamefont {Rossi}},
  \bibinfo {author} {\bibfnamefont {A.~K.}\ \bibnamefont {Tan}}, \ and\
  \bibinfo {author} {\bibfnamefont {I.~L.}\ \bibnamefont {Chuang}},\
  }\href@noop {} {\bibfield  {journal} {\bibinfo  {journal} {PRX Quantum}\
  }\textbf {\bibinfo {volume} {2}},\ \bibinfo {pages} {040203} (\bibinfo {year}
  {2021})}\BibitemShut {NoStop}%
\bibitem [{Note1()}]{Note1}%
  \BibitemOpen
  \bibinfo {note} {There are studies about limitations and insecurity concerns
  about concepts that are related to QPQs. In fact, there are no-go theorems
  \cite {mayers1997unconditionally,lo1997quantum,lo1997insecurity} about the
  imperfection of certain quantum computation and communication schemes. The
  QPQ protocol does not violate the no-go theorems \cite {lo1997insecurity}
  since the setup is different. In QPQ, we do not have the security requirement
  required for the no-theorem, where the user Alice cannot know the private key
  of QDCs, since QDCs do not have private keys. On the other hand, it will be
  interesting to understand better how those studies could potentially improve
  the capability of QDCs.}\BibitemShut {Stop}%
\bibitem [{Note2()}]{Note2}%
  \BibitemOpen
  \bibinfo {note} {Moreover, we remark that numerous extensions of this simple
  unary-to-binary compression are possible. For example, it is straightforward
  to generalize the above procedure to the case of multiple excitations, such
  that a QDC can be used to compress multi-excitation subspaces as well (see
  related discussions in Appendix about quantum data compression). QDCs could
  even be used for Schmacher coding~\cite {schumacher1995quantum}, universal
  source coding~\cite {jozsa1998universal,
  hayashi2002quantum,bennett2006universal}, or other quantum compression tasks.
  The ability to efficiently compress quantum data using QRAM and related
  architectures is studied more thoroughly in \cite
  {hann2022inprep}.}\BibitemShut {Stop}%
\bibitem [{\citenamefont {Mayers}(1997)}]{mayers1997unconditionally}%
  \BibitemOpen
  \bibfield  {author} {\bibinfo {author} {\bibfnamefont {D.}~\bibnamefont
  {Mayers}},\ }\href@noop {} {\bibfield  {journal} {\bibinfo  {journal}
  {Physical review letters}\ }\textbf {\bibinfo {volume} {78}},\ \bibinfo
  {pages} {3414} (\bibinfo {year} {1997})}\BibitemShut {NoStop}%
\bibitem [{\citenamefont {Lo}\ and\ \citenamefont
  {Chau}(1997)}]{lo1997quantum}%
  \BibitemOpen
  \bibfield  {author} {\bibinfo {author} {\bibfnamefont {H.-K.}\ \bibnamefont
  {Lo}}\ and\ \bibinfo {author} {\bibfnamefont {H.~F.}\ \bibnamefont {Chau}},\
  }\href@noop {} {\bibfield  {journal} {\bibinfo  {journal} {Physical Review
  Letters}\ }\textbf {\bibinfo {volume} {78}},\ \bibinfo {pages} {3410}
  (\bibinfo {year} {1997})}\BibitemShut {NoStop}%
\bibitem [{\citenamefont {Lo}(1997)}]{lo1997insecurity}%
  \BibitemOpen
  \bibfield  {author} {\bibinfo {author} {\bibfnamefont {H.-K.}\ \bibnamefont
  {Lo}},\ }\href@noop {} {\bibfield  {journal} {\bibinfo  {journal} {Physical
  Review A}\ }\textbf {\bibinfo {volume} {56}},\ \bibinfo {pages} {1154}
  (\bibinfo {year} {1997})}\BibitemShut {NoStop}%
\bibitem [{\citenamefont {Schumacher}(1995)}]{schumacher1995quantum}%
  \BibitemOpen
  \bibfield  {author} {\bibinfo {author} {\bibfnamefont {B.}~\bibnamefont
  {Schumacher}},\ }\href@noop {} {\bibfield  {journal} {\bibinfo  {journal}
  {Physical Review A}\ }\textbf {\bibinfo {volume} {51}},\ \bibinfo {pages}
  {2738} (\bibinfo {year} {1995})}\BibitemShut {NoStop}%
\bibitem [{\citenamefont {Jozsa}\ \emph {et~al.}(1998)\citenamefont {Jozsa},
  \citenamefont {Horodecki}, \citenamefont {Horodecki},\ and\ \citenamefont
  {Horodecki}}]{jozsa1998universal}%
  \BibitemOpen
  \bibfield  {author} {\bibinfo {author} {\bibfnamefont {R.}~\bibnamefont
  {Jozsa}}, \bibinfo {author} {\bibfnamefont {M.}~\bibnamefont {Horodecki}},
  \bibinfo {author} {\bibfnamefont {P.}~\bibnamefont {Horodecki}}, \ and\
  \bibinfo {author} {\bibfnamefont {R.}~\bibnamefont {Horodecki}},\ }\href@noop
  {} {\bibfield  {journal} {\bibinfo  {journal} {Physical review letters}\
  }\textbf {\bibinfo {volume} {81}},\ \bibinfo {pages} {1714} (\bibinfo {year}
  {1998})}\BibitemShut {NoStop}%
\bibitem [{\citenamefont {Hayashi}\ and\ \citenamefont
  {Matsumoto}(2002)}]{hayashi2002quantum}%
  \BibitemOpen
  \bibfield  {author} {\bibinfo {author} {\bibfnamefont {M.}~\bibnamefont
  {Hayashi}}\ and\ \bibinfo {author} {\bibfnamefont {K.}~\bibnamefont
  {Matsumoto}},\ }\href@noop {} {\bibfield  {journal} {\bibinfo  {journal}
  {Physical Review A}\ }\textbf {\bibinfo {volume} {66}},\ \bibinfo {pages}
  {022311} (\bibinfo {year} {2002})}\BibitemShut {NoStop}%
\bibitem [{\citenamefont {Bennett}\ \emph {et~al.}(2006)\citenamefont
  {Bennett}, \citenamefont {Harrow},\ and\ \citenamefont
  {Lloyd}}]{bennett2006universal}%
  \BibitemOpen
  \bibfield  {author} {\bibinfo {author} {\bibfnamefont {C.~H.}\ \bibnamefont
  {Bennett}}, \bibinfo {author} {\bibfnamefont {A.~W.}\ \bibnamefont {Harrow}},
  \ and\ \bibinfo {author} {\bibfnamefont {S.}~\bibnamefont {Lloyd}},\
  }\href@noop {} {\bibfield  {journal} {\bibinfo  {journal} {Physical Review
  A}\ }\textbf {\bibinfo {volume} {73}},\ \bibinfo {pages} {032336} (\bibinfo
  {year} {2006})}\BibitemShut {NoStop}%
\bibitem [{\citenamefont {Hann}\ \emph {et~al.}()\citenamefont {Hann} \emph
  {et~al.}}]{hann2022inprep}%
  \BibitemOpen
  \bibfield  {author} {\bibinfo {author} {\bibfnamefont {C.~T.}\ \bibnamefont
  {Hann}} \emph {et~al.},\ }\href@noop {} {}\bibinfo {note} {In preparation
  (2022).}\BibitemShut {Stop}%
\bibitem [{\citenamefont {Low}\ and\ \citenamefont
  {Chuang}(2019)}]{low2019hamiltonian}%
  \BibitemOpen
  \bibfield  {author} {\bibinfo {author} {\bibfnamefont {G.~H.}\ \bibnamefont
  {Low}}\ and\ \bibinfo {author} {\bibfnamefont {I.~L.}\ \bibnamefont
  {Chuang}},\ }\href@noop {} {\bibfield  {journal} {\bibinfo  {journal}
  {Quantum}\ }\textbf {\bibinfo {volume} {3}},\ \bibinfo {pages} {163}
  (\bibinfo {year} {2019})}\BibitemShut {NoStop}%
\bibitem [{\citenamefont {Low}\ \emph {et~al.}(2016)\citenamefont {Low},
  \citenamefont {Yoder},\ and\ \citenamefont {Chuang}}]{LowQSPprx}%
  \BibitemOpen
  \bibfield  {author} {\bibinfo {author} {\bibfnamefont {G.~H.}\ \bibnamefont
  {Low}}, \bibinfo {author} {\bibfnamefont {T.~J.}\ \bibnamefont {Yoder}}, \
  and\ \bibinfo {author} {\bibfnamefont {I.~L.}\ \bibnamefont {Chuang}},\
  }\href {\doibase 10.1103/PhysRevX.6.041067} {\bibfield  {journal} {\bibinfo
  {journal} {Phys. Rev. X}\ }\textbf {\bibinfo {volume} {6}},\ \bibinfo {pages}
  {041067} (\bibinfo {year} {2016})}\BibitemShut {NoStop}%
\bibitem [{\citenamefont {Low}\ and\ \citenamefont
  {Chuang}(2017)}]{low2017optimal}%
  \BibitemOpen
  \bibfield  {author} {\bibinfo {author} {\bibfnamefont {G.~H.}\ \bibnamefont
  {Low}}\ and\ \bibinfo {author} {\bibfnamefont {I.~L.}\ \bibnamefont
  {Chuang}},\ }\href@noop {} {\bibfield  {journal} {\bibinfo  {journal}
  {Physical review letters}\ }\textbf {\bibinfo {volume} {118}},\ \bibinfo
  {pages} {010501} (\bibinfo {year} {2017})}\BibitemShut {NoStop}%
\bibitem [{\citenamefont {Giovannetti}\ \emph {et~al.}(2013)\citenamefont
  {Giovannetti}, \citenamefont {Maccone}, \citenamefont {Morimae},\ and\
  \citenamefont {Rudolph}}]{giovannetti2013efficient}%
  \BibitemOpen
  \bibfield  {author} {\bibinfo {author} {\bibfnamefont {V.}~\bibnamefont
  {Giovannetti}}, \bibinfo {author} {\bibfnamefont {L.}~\bibnamefont
  {Maccone}}, \bibinfo {author} {\bibfnamefont {T.}~\bibnamefont {Morimae}}, \
  and\ \bibinfo {author} {\bibfnamefont {T.~G.}\ \bibnamefont {Rudolph}},\
  }\href@noop {} {\bibfield  {journal} {\bibinfo  {journal} {Physical review
  letters}\ }\textbf {\bibinfo {volume} {111}},\ \bibinfo {pages} {230501}
  (\bibinfo {year} {2013})}\BibitemShut {NoStop}%
\bibitem [{\citenamefont {Fitzsimons}(2017)}]{fitzsimons2017private}%
  \BibitemOpen
  \bibfield  {author} {\bibinfo {author} {\bibfnamefont {J.~F.}\ \bibnamefont
  {Fitzsimons}},\ }\href@noop {} {\bibfield  {journal} {\bibinfo  {journal}
  {npj Quantum Information}\ }\textbf {\bibinfo {volume} {3}},\ \bibinfo
  {pages} {1} (\bibinfo {year} {2017})}\BibitemShut {NoStop}%
\bibitem [{\citenamefont {Preskill}(1999)}]{preskill1999plug}%
  \BibitemOpen
  \bibfield  {author} {\bibinfo {author} {\bibfnamefont {J.}~\bibnamefont
  {Preskill}},\ }\href@noop {} {\bibfield  {journal} {\bibinfo  {journal}
  {Nature}\ }\textbf {\bibinfo {volume} {402}},\ \bibinfo {pages} {357}
  (\bibinfo {year} {1999})}\BibitemShut {NoStop}%
\bibitem [{\citenamefont {Gottesman}\ and\ \citenamefont
  {Chuang}(1999)}]{gottesman1999demonstrating}%
  \BibitemOpen
  \bibfield  {author} {\bibinfo {author} {\bibfnamefont {D.}~\bibnamefont
  {Gottesman}}\ and\ \bibinfo {author} {\bibfnamefont {I.~L.}\ \bibnamefont
  {Chuang}},\ }\href@noop {} {\bibfield  {journal} {\bibinfo  {journal}
  {Nature}\ }\textbf {\bibinfo {volume} {402}},\ \bibinfo {pages} {390}
  (\bibinfo {year} {1999})}\BibitemShut {NoStop}%
\bibitem [{\citenamefont {Vadhan}(2017)}]{vadhan2017complexity}%
  \BibitemOpen
  \bibfield  {author} {\bibinfo {author} {\bibfnamefont {S.}~\bibnamefont
  {Vadhan}},\ }in\ \href@noop {} {\emph {\bibinfo {booktitle} {Tutorials on the
  Foundations of Cryptography}}}\ (\bibinfo  {publisher} {Springer},\ \bibinfo
  {year} {2017})\ pp.\ \bibinfo {pages} {347--450}\BibitemShut {NoStop}%
\bibitem [{\citenamefont {Lloyd}\ \emph {et~al.}(2014)\citenamefont {Lloyd},
  \citenamefont {Mohseni},\ and\ \citenamefont
  {Rebentrost}}]{lloyd2014quantum}%
  \BibitemOpen
  \bibfield  {author} {\bibinfo {author} {\bibfnamefont {S.}~\bibnamefont
  {Lloyd}}, \bibinfo {author} {\bibfnamefont {M.}~\bibnamefont {Mohseni}}, \
  and\ \bibinfo {author} {\bibfnamefont {P.}~\bibnamefont {Rebentrost}},\
  }\href@noop {} {\bibfield  {journal} {\bibinfo  {journal} {Nature Physics}\
  }\textbf {\bibinfo {volume} {10}},\ \bibinfo {pages} {631} (\bibinfo {year}
  {2014})}\BibitemShut {NoStop}%
\bibitem [{\citenamefont {Lloyd}(1996)}]{uni}%
  \BibitemOpen
  \bibfield  {author} {\bibinfo {author} {\bibfnamefont {S.}~\bibnamefont
  {Lloyd}},\ }\href@noop {} {\bibfield  {journal} {\bibinfo  {journal}
  {Science}\ ,\ \bibinfo {pages} {1073}} (\bibinfo {year} {1996})}\BibitemShut
  {NoStop}%
\bibitem [{\citenamefont {Suzuki}(1976)}]{suzuki1976trotter}%
  \BibitemOpen
  \bibfield  {author} {\bibinfo {author} {\bibfnamefont {M.}~\bibnamefont
  {Suzuki}},\ }\href {\doibase 10.1007/BF01609348} {\bibfield  {journal}
  {\bibinfo  {journal} {Communications in Mathematical Physics}\ }\textbf
  {\bibinfo {volume} {51}},\ \bibinfo {pages} {183} (\bibinfo {year}
  {1976})}\BibitemShut {NoStop}%
\bibitem [{\citenamefont {Childs}\ \emph {et~al.}(2019)\citenamefont {Childs},
  \citenamefont {Su}, \citenamefont {Tran}, \citenamefont {Wiebe},\ and\
  \citenamefont {Zhu}}]{tro}%
  \BibitemOpen
  \bibfield  {author} {\bibinfo {author} {\bibfnamefont {A.~M.}\ \bibnamefont
  {Childs}}, \bibinfo {author} {\bibfnamefont {Y.}~\bibnamefont {Su}}, \bibinfo
  {author} {\bibfnamefont {M.~C.}\ \bibnamefont {Tran}}, \bibinfo {author}
  {\bibfnamefont {N.}~\bibnamefont {Wiebe}}, \ and\ \bibinfo {author}
  {\bibfnamefont {S.}~\bibnamefont {Zhu}},\ }\href@noop {} {\bibfield
  {journal} {\bibinfo  {journal} {arXiv preprint arXiv:1912.08854}\ } (\bibinfo
  {year} {2019})}\BibitemShut {NoStop}%
\bibitem [{\citenamefont {Childs}\ and\ \citenamefont
  {Kothari}(2011)}]{childs2011walks}%
  \BibitemOpen
  \bibfield  {author} {\bibinfo {author} {\bibfnamefont {A.~M.}\ \bibnamefont
  {Childs}}\ and\ \bibinfo {author} {\bibfnamefont {R.}~\bibnamefont
  {Kothari}},\ }in\ \href@noop {} {\emph {\bibinfo {booktitle} {Theory of
  Quantum Computation, Communication, and Cryptography}}},\ \bibinfo {editor}
  {edited by\ \bibinfo {editor} {\bibfnamefont {W.}~\bibnamefont {van Dam}},
  \bibinfo {editor} {\bibfnamefont {V.~M.}\ \bibnamefont {Kendon}}, \ and\
  \bibinfo {editor} {\bibfnamefont {S.}~\bibnamefont {Severini}}}\ (\bibinfo
  {publisher} {Springer Berlin Heidelberg},\ \bibinfo {address} {Berlin,
  Heidelberg},\ \bibinfo {year} {2011})\ pp.\ \bibinfo {pages}
  {94--103}\BibitemShut {NoStop}%
\bibitem [{\citenamefont {Berry}\ \emph
  {et~al.}(2015{\natexlab{a}})\citenamefont {Berry}, \citenamefont {Childs},\
  and\ \citenamefont {Kothari}}]{berry2015hamiltonian}%
  \BibitemOpen
  \bibfield  {author} {\bibinfo {author} {\bibfnamefont {D.~W.}\ \bibnamefont
  {Berry}}, \bibinfo {author} {\bibfnamefont {A.~M.}\ \bibnamefont {Childs}}, \
  and\ \bibinfo {author} {\bibfnamefont {R.}~\bibnamefont {Kothari}},\ }in\
  \href@noop {} {\emph {\bibinfo {booktitle} {Foundations of Computer Science
  (FOCS), 2015 IEEE 56th Annual Symposium on}}}\ (\bibinfo {organization}
  {IEEE},\ \bibinfo {year} {2015})\ pp.\ \bibinfo {pages}
  {792--809}\BibitemShut {NoStop}%
\bibitem [{\citenamefont {Low}\ \emph {et~al.}(2019)\citenamefont {Low},
  \citenamefont {Kliuchnikov},\ and\ \citenamefont
  {Wiebe}}]{low2019multiproduct}%
  \BibitemOpen
  \bibfield  {author} {\bibinfo {author} {\bibfnamefont {G.~H.}\ \bibnamefont
  {Low}}, \bibinfo {author} {\bibfnamefont {V.}~\bibnamefont {Kliuchnikov}}, \
  and\ \bibinfo {author} {\bibfnamefont {N.}~\bibnamefont {Wiebe}},\
  }\href@noop {} {\  (\bibinfo {year} {2019})},\ \Eprint
  {http://arxiv.org/abs/arXiv:1907.11679} {arXiv:1907.11679} \BibitemShut
  {NoStop}%
\bibitem [{\citenamefont {Childs}\ and\ \citenamefont
  {Wiebe}(2012)}]{childs2012multiproduct}%
  \BibitemOpen
  \bibfield  {author} {\bibinfo {author} {\bibfnamefont {A.~M.}\ \bibnamefont
  {Childs}}\ and\ \bibinfo {author} {\bibfnamefont {N.}~\bibnamefont {Wiebe}},\
  }\href {http://dl.acm.org/citation.cfm?id=2481569.2481570} {\bibfield
  {journal} {\bibinfo  {journal} {Quantum Info. Comput.}\ }\textbf {\bibinfo
  {volume} {12}},\ \bibinfo {pages} {901} (\bibinfo {year} {2012})}\BibitemShut
  {NoStop}%
\bibitem [{\citenamefont {Berry}\ and\ \citenamefont
  {Childs}(2012)}]{berry2012black}%
  \BibitemOpen
  \bibfield  {author} {\bibinfo {author} {\bibfnamefont {D.~W.}\ \bibnamefont
  {Berry}}\ and\ \bibinfo {author} {\bibfnamefont {A.~M.}\ \bibnamefont
  {Childs}},\ }\href {http://dl.acm.org/citation.cfm?id=2231036.2231040}
  {\bibfield  {journal} {\bibinfo  {journal} {Quantum Info. Comput.}\ }\textbf
  {\bibinfo {volume} {12}},\ \bibinfo {pages} {29} (\bibinfo {year}
  {2012})}\BibitemShut {NoStop}%
\bibitem [{\citenamefont {Berry}\ \emph
  {et~al.}(2015{\natexlab{b}})\citenamefont {Berry}, \citenamefont {Childs},
  \citenamefont {Cleve}, \citenamefont {Kothari},\ and\ \citenamefont
  {Somma}}]{berry2015simulating}%
  \BibitemOpen
  \bibfield  {author} {\bibinfo {author} {\bibfnamefont {D.~W.}\ \bibnamefont
  {Berry}}, \bibinfo {author} {\bibfnamefont {A.~M.}\ \bibnamefont {Childs}},
  \bibinfo {author} {\bibfnamefont {R.}~\bibnamefont {Cleve}}, \bibinfo
  {author} {\bibfnamefont {R.}~\bibnamefont {Kothari}}, \ and\ \bibinfo
  {author} {\bibfnamefont {R.~D.}\ \bibnamefont {Somma}},\ }\href@noop {}
  {\bibfield  {journal} {\bibinfo  {journal} {Physical review letters}\
  }\textbf {\bibinfo {volume} {114}},\ \bibinfo {pages} {090502} (\bibinfo
  {year} {2015}{\natexlab{b}})}\BibitemShut {NoStop}%
\bibitem [{\citenamefont {Berry}\ \emph {et~al.}(2017)\citenamefont {Berry},
  \citenamefont {Childs}, \citenamefont {Cleve}, \citenamefont {Kothari},\ and\
  \citenamefont {Somma}}]{spar}%
  \BibitemOpen
  \bibfield  {author} {\bibinfo {author} {\bibfnamefont {D.~W.}\ \bibnamefont
  {Berry}}, \bibinfo {author} {\bibfnamefont {A.~M.}\ \bibnamefont {Childs}},
  \bibinfo {author} {\bibfnamefont {R.}~\bibnamefont {Cleve}}, \bibinfo
  {author} {\bibfnamefont {R.}~\bibnamefont {Kothari}}, \ and\ \bibinfo
  {author} {\bibfnamefont {R.~D.}\ \bibnamefont {Somma}},\ }in\ \href@noop {}
  {\emph {\bibinfo {booktitle} {Forum of Mathematics, Sigma}}},\ Vol.~\bibinfo
  {volume} {5}\ (\bibinfo {organization} {Cambridge University Press},\
  \bibinfo {year} {2017})\BibitemShut {NoStop}%
\bibitem [{\citenamefont {Zhang}\ \emph {et~al.}(2018)\citenamefont {Zhang},
  \citenamefont {Zou},\ and\ \citenamefont {Jiang}}]{zhang2018quantum}%
  \BibitemOpen
  \bibfield  {author} {\bibinfo {author} {\bibfnamefont {M.}~\bibnamefont
  {Zhang}}, \bibinfo {author} {\bibfnamefont {C.-L.}\ \bibnamefont {Zou}}, \
  and\ \bibinfo {author} {\bibfnamefont {L.}~\bibnamefont {Jiang}},\
  }\href@noop {} {\bibfield  {journal} {\bibinfo  {journal} {Physical review
  letters}\ }\textbf {\bibinfo {volume} {120}},\ \bibinfo {pages} {020502}
  (\bibinfo {year} {2018})}\BibitemShut {NoStop}%
\bibitem [{\citenamefont {Herr}\ \emph {et~al.}(2017)\citenamefont {Herr},
  \citenamefont {Nori},\ and\ \citenamefont {Devitt}}]{herr2017lattice}%
  \BibitemOpen
  \bibfield  {author} {\bibinfo {author} {\bibfnamefont {D.}~\bibnamefont
  {Herr}}, \bibinfo {author} {\bibfnamefont {F.}~\bibnamefont {Nori}}, \ and\
  \bibinfo {author} {\bibfnamefont {S.~J.}\ \bibnamefont {Devitt}},\
  }\href@noop {} {\bibfield  {journal} {\bibinfo  {journal} {New Journal of
  Physics}\ }\textbf {\bibinfo {volume} {19}},\ \bibinfo {pages} {013034}
  (\bibinfo {year} {2017})}\BibitemShut {NoStop}%
\bibitem [{\citenamefont {Giovannetti}\ \emph {et~al.}(2010)\citenamefont
  {Giovannetti}, \citenamefont {Lloyd},\ and\ \citenamefont
  {Maccone}}]{giovannetti2010quantum}%
  \BibitemOpen
  \bibfield  {author} {\bibinfo {author} {\bibfnamefont {V.}~\bibnamefont
  {Giovannetti}}, \bibinfo {author} {\bibfnamefont {S.}~\bibnamefont {Lloyd}},
  \ and\ \bibinfo {author} {\bibfnamefont {L.}~\bibnamefont {Maccone}},\
  }\href@noop {} {\bibfield  {journal} {\bibinfo  {journal} {IEEE Transactions
  on Information Theory}\ }\textbf {\bibinfo {volume} {56}},\ \bibinfo {pages}
  {3465} (\bibinfo {year} {2010})}\BibitemShut {NoStop}%
\bibitem [{\citenamefont {Arrazola}\ and\ \citenamefont
  {Scarani}(2016)}]{arrazola2016covert}%
  \BibitemOpen
  \bibfield  {author} {\bibinfo {author} {\bibfnamefont {J.~M.}\ \bibnamefont
  {Arrazola}}\ and\ \bibinfo {author} {\bibfnamefont {V.}~\bibnamefont
  {Scarani}},\ }\href@noop {} {\bibfield  {journal} {\bibinfo  {journal}
  {Physical review letters}\ }\textbf {\bibinfo {volume} {117}},\ \bibinfo
  {pages} {250503} (\bibinfo {year} {2016})}\BibitemShut {NoStop}%
\bibitem [{\citenamefont {Gottesman}\ \emph {et~al.}(2012)\citenamefont
  {Gottesman}, \citenamefont {Jennewein},\ and\ \citenamefont
  {Croke}}]{gottesman2012longer}%
  \BibitemOpen
  \bibfield  {author} {\bibinfo {author} {\bibfnamefont {D.}~\bibnamefont
  {Gottesman}}, \bibinfo {author} {\bibfnamefont {T.}~\bibnamefont
  {Jennewein}}, \ and\ \bibinfo {author} {\bibfnamefont {S.}~\bibnamefont
  {Croke}},\ }\href@noop {} {\bibfield  {journal} {\bibinfo  {journal}
  {Physical review letters}\ }\textbf {\bibinfo {volume} {109}},\ \bibinfo
  {pages} {070503} (\bibinfo {year} {2012})}\BibitemShut {NoStop}%
\bibitem [{\citenamefont {Khabiboulline}\ \emph
  {et~al.}(2019{\natexlab{a}})\citenamefont {Khabiboulline}, \citenamefont
  {Borregaard}, \citenamefont {De~Greve},\ and\ \citenamefont
  {Lukin}}]{khabiboulline2019optical}%
  \BibitemOpen
  \bibfield  {author} {\bibinfo {author} {\bibfnamefont {E.~T.}\ \bibnamefont
  {Khabiboulline}}, \bibinfo {author} {\bibfnamefont {J.}~\bibnamefont
  {Borregaard}}, \bibinfo {author} {\bibfnamefont {K.}~\bibnamefont
  {De~Greve}}, \ and\ \bibinfo {author} {\bibfnamefont {M.~D.}\ \bibnamefont
  {Lukin}},\ }\href@noop {} {\bibfield  {journal} {\bibinfo  {journal}
  {Physical review letters}\ }\textbf {\bibinfo {volume} {123}},\ \bibinfo
  {pages} {070504} (\bibinfo {year} {2019}{\natexlab{a}})}\BibitemShut
  {NoStop}%
\bibitem [{\citenamefont {Khabiboulline}\ \emph
  {et~al.}(2019{\natexlab{b}})\citenamefont {Khabiboulline}, \citenamefont
  {Borregaard}, \citenamefont {De~Greve},\ and\ \citenamefont
  {Lukin}}]{khabiboulline2019quantum}%
  \BibitemOpen
  \bibfield  {author} {\bibinfo {author} {\bibfnamefont {E.~T.}\ \bibnamefont
  {Khabiboulline}}, \bibinfo {author} {\bibfnamefont {J.}~\bibnamefont
  {Borregaard}}, \bibinfo {author} {\bibfnamefont {K.}~\bibnamefont
  {De~Greve}}, \ and\ \bibinfo {author} {\bibfnamefont {M.~D.}\ \bibnamefont
  {Lukin}},\ }\href@noop {} {\bibfield  {journal} {\bibinfo  {journal}
  {Physical Review A}\ }\textbf {\bibinfo {volume} {100}},\ \bibinfo {pages}
  {022316} (\bibinfo {year} {2019}{\natexlab{b}})}\BibitemShut {NoStop}%
\bibitem [{\citenamefont {Acin}(2001)}]{acin2001statistical}%
  \BibitemOpen
  \bibfield  {author} {\bibinfo {author} {\bibfnamefont {A.}~\bibnamefont
  {Acin}},\ }\href@noop {} {\bibfield  {journal} {\bibinfo  {journal} {Physical
  review letters}\ }\textbf {\bibinfo {volume} {87}},\ \bibinfo {pages}
  {177901} (\bibinfo {year} {2001})}\BibitemShut {NoStop}%
\bibitem [{\citenamefont {Duan}\ \emph {et~al.}(2007)\citenamefont {Duan},
  \citenamefont {Feng},\ and\ \citenamefont {Ying}}]{duan2007entanglement}%
  \BibitemOpen
  \bibfield  {author} {\bibinfo {author} {\bibfnamefont {R.}~\bibnamefont
  {Duan}}, \bibinfo {author} {\bibfnamefont {Y.}~\bibnamefont {Feng}}, \ and\
  \bibinfo {author} {\bibfnamefont {M.}~\bibnamefont {Ying}},\ }\href@noop {}
  {\bibfield  {journal} {\bibinfo  {journal} {Physical review letters}\
  }\textbf {\bibinfo {volume} {98}},\ \bibinfo {pages} {100503} (\bibinfo
  {year} {2007})}\BibitemShut {NoStop}%
\bibitem [{\citenamefont {Duan}\ \emph {et~al.}(2009)\citenamefont {Duan},
  \citenamefont {Feng},\ and\ \citenamefont {Ying}}]{duan2009perfect}%
  \BibitemOpen
  \bibfield  {author} {\bibinfo {author} {\bibfnamefont {R.}~\bibnamefont
  {Duan}}, \bibinfo {author} {\bibfnamefont {Y.}~\bibnamefont {Feng}}, \ and\
  \bibinfo {author} {\bibfnamefont {M.}~\bibnamefont {Ying}},\ }\href@noop {}
  {\bibfield  {journal} {\bibinfo  {journal} {Physical Review Letters}\
  }\textbf {\bibinfo {volume} {103}},\ \bibinfo {pages} {210501} (\bibinfo
  {year} {2009})}\BibitemShut {NoStop}%
\bibitem [{\citenamefont {Zhou}\ and\ \citenamefont
  {Jiang}(2021)}]{zhou2021asymptotic}%
  \BibitemOpen
  \bibfield  {author} {\bibinfo {author} {\bibfnamefont {S.}~\bibnamefont
  {Zhou}}\ and\ \bibinfo {author} {\bibfnamefont {L.}~\bibnamefont {Jiang}},\
  }\href@noop {} {\bibfield  {journal} {\bibinfo  {journal} {PRX Quantum}\
  }\textbf {\bibinfo {volume} {2}},\ \bibinfo {pages} {010343} (\bibinfo {year}
  {2021})}\BibitemShut {NoStop}%
\bibitem [{\citenamefont {Rossi}\ \emph {et~al.}(2021)\citenamefont {Rossi},
  \citenamefont {Yu}, \citenamefont {Chuang},\ and\ \citenamefont
  {Sugiura}}]{rossi2021quantum}%
  \BibitemOpen
  \bibfield  {author} {\bibinfo {author} {\bibfnamefont {Z.~M.}\ \bibnamefont
  {Rossi}}, \bibinfo {author} {\bibfnamefont {J.}~\bibnamefont {Yu}}, \bibinfo
  {author} {\bibfnamefont {I.~L.}\ \bibnamefont {Chuang}}, \ and\ \bibinfo
  {author} {\bibfnamefont {S.}~\bibnamefont {Sugiura}},\ }\href@noop {}
  {\bibfield  {journal} {\bibinfo  {journal} {arXiv preprint arXiv:2105.08707}\
  } (\bibinfo {year} {2021})}\BibitemShut {NoStop}%
\end{thebibliography}%

%\pagebreak
%\clearpage
%\foreach \x in {1,...,\the\pdflastximagepages}
%{
%	\clearpage
%	\includepdf[pages={\x,{}}]{QDC_technical_SM_PRL.pdf}
%}
\clearpage

\pagebreak

\onecolumngrid
\appendix

\vspace{0.5in}

\begin{center}
	{\Large \bf Appendix}
\end{center}

In Appendix, we provide some necessary information and related results about QDCs. The structure of Appendix is the following. In Section \ref{qdcmainsec} we give an introduction on several basic aspects of QDCs. Specifically, In Section \ref{writing} we give some perspectives about the writing function in QDCs. In Section \ref{comp}, \ref{comm} and \ref{sensing}, we discuss several explicit examples of QDCs applied in quantum computing, quantum communication, and quantum sensing, including Section \ref{oraclegeneral} and \ref{oracleappend} about a review of quantum simulation and qubitization algorithms \cite{low2019hamiltonian,LowQSPprx,low2017optimal} where QDCs are used to provide the oracle, Section \ref{multipleusers} about QDCs for computing with multiple users, Section \ref{Tgatesec} about a short introduction of the surface code and $T$-gate counting formalism developed in \cite{litinski2019game} and used in the $T$-gate example in the main text, Section \ref{qpq}, \ref{blindQC} about details of Quantum Private Query and blind quantum computing, Section \ref{datacomp} about some details of quantum data compression, and Section \ref{channelappend} about channel discrimination in quantum sensing and QDCs.

\section{More about QDCs}\label{qdcmainsec}
\subsection{Comments}\label{sec:comments}
Here we comment on some general perspectives about QDCs. We note also that some architectures without either QRAM or quantum networks could still be defined as QDCs. In Section \ref{channelappend}, we describe an application in quantum sensing where QRAMs are not necessarily used, but one could still use QDC architectures to realize it. 
    
One might be curious about how QDCs are different from a generalized version of quantum computers. Here, we should clarify that, of course, we could develop a Universal Quantum Computing (UQC) device that is associated with QDC. However, it is not necessary, and our QDC construction could directly serve remote users with their own quantum computation architectures. In fact, some of our examples mentioned later do not require UQC power for QDC, for instance, QDCs for the $T$-gate counting that have been discussed in the main text.

\begin{figure}[ht]
\centering
%\rule{\linewidth}{3cm}
\includegraphics[width=0.6\textwidth]{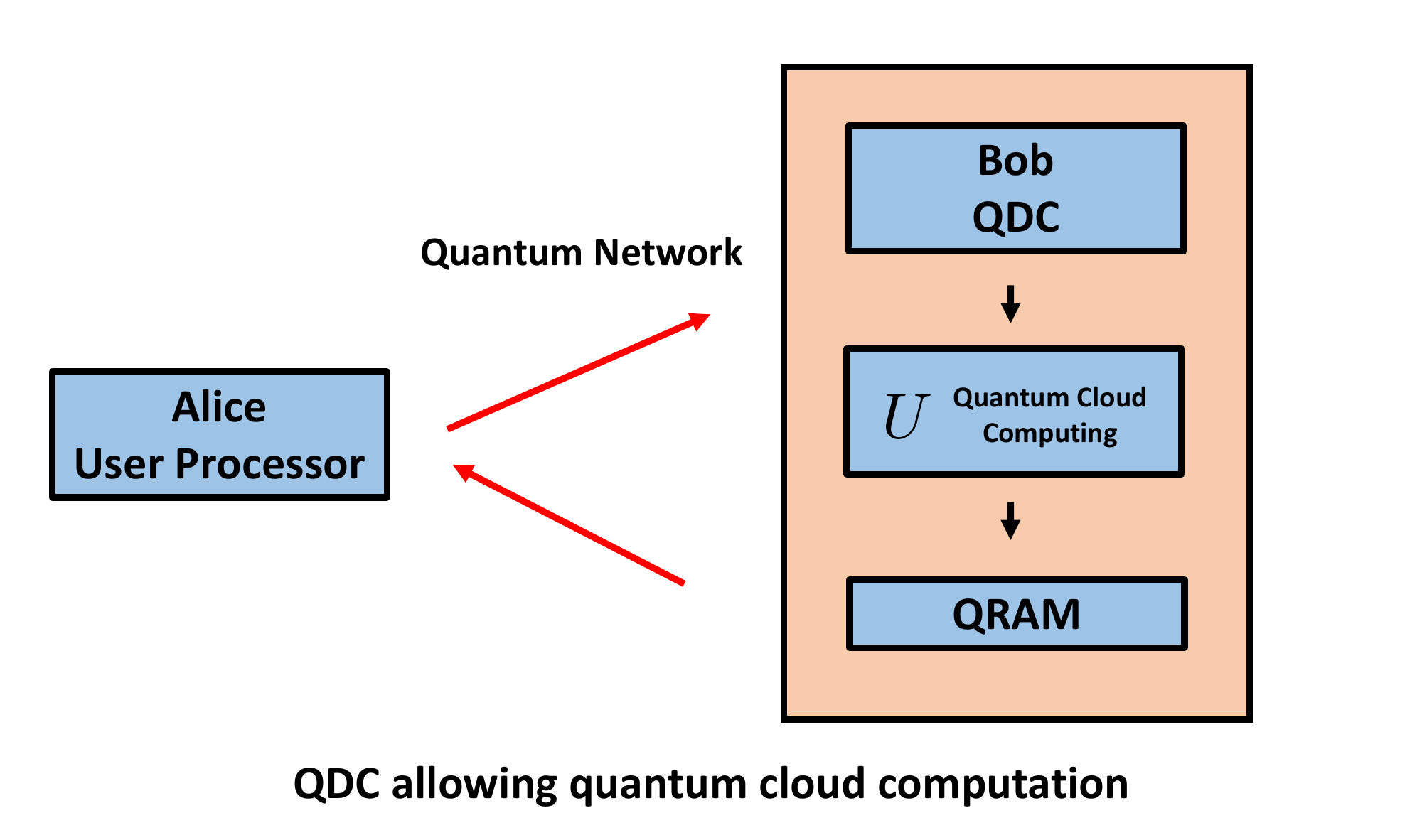}
\caption{QDCs could allow quantum cloud computing on the server.}
\label{fig:qdc_features}
\end{figure}

Moreover, QDCs could not only take just the above minimal definition, but also more general forms. For example, QDCs equipped with UQC could also perform quantum cloud computation (see Figure \ref{fig:qdc_features}). Quantum computers are hard to realize, and it is natural to consider remote cloud services running in QDCs and provide the results of computations to remote users. Moreover, if users wish to keep the privacy, quantum blind computation \cite{giovannetti2013efficient,fitzsimons2017private} could be performed in QDCs with the help of quantum networks \cite{fitzsimons2017private}, see further discussion in \Cref{comm}.

Finally, we also note that our proposal is also closely related to the idea of \emph{disposable quantum software} proposed in \cite{preskill1999plug} by Preskill. The quantum software, defined in \cite{preskill1999plug}, is a fragile quantum state that is hard to maintain by users, so they prefer to buy such a state through the quantum network. The early quantum teleportation scheme based on \cite{gottesman1999demonstrating} provides significant power for quantum devices by combining UQC and quantum communication, inspiring the observation in \cite{preskill1999plug} for quantum software. Our definition of QDC, including the functioning of UQC, could be a particular realization of disposable quantum software systems. However, we are more emphasizing the ingredient from QRAM, enabling extra capabilities for computation, communication, and sensing.  

Finally, we give an explicit definition of QRAM for quantum data, and provide a formal definition of QDCs. In this case, the user provides an arbitrary superposition of addresses as input, but the QRAM now returns the quantum state that was stored in the memory location specified by the address. More precisely, if the QRAM holds an arbitrary product state, $\bigotimes_{j=1}^N\ket{\psi_j}^{D_j}$, where $D_j$ denotes the $j$-th cell of the memory, then a QRAM query enacts the operation
\begin{align}
\label{eq:QRAM_def_quantum}
&\sum\limits_{i = 0}^{N - 1} {{\alpha _i}} {\left| i \right\rangle ^{Q_1}}{\left| 0 \right\rangle ^{Q_2}} \left[ \bigotimes_{j=1}^{N}\ket{\psi_j}^{D_j} \right]
\nonumber \\ 
\to & \sum\limits_{i = 0}^{N - 1} {{\alpha _i}} {\left| i \right\rangle ^{Q_1}}{\left| \psi_i \right\rangle ^{Q_2}} \left[ \bigotimes_{j=1}^{N}\ket{\overline{\psi}^{(i)}_j}^{D_j} \right],
\end{align}
which, conditioned on register $Q_1$ being in state $\ket{i}$, swaps the state of register $Q_2$ and the $i$-th cell of the quantum memory. Here,  $\ket{\overline{\psi}^{(i)}_j} = \ket{0}$ for $i=j$ and  $\ket{\overline{\psi}^{(i)}_j} = \ket{\psi_j}$ otherwise. By linearity, this definition also defines the query operation when the QRAM holds an entangled state. Note that, when the data is quantum, a QRAM query generally leaves the $Q_1$ and $Q_2$ registers entangled with the data. The difference between the classical and quantum operations will be manifest when we try to ``write" the data in QDCs (see Appendix for a detailed discussion).  

We now give a formal definition of a QDC, and we expand on aspects of this definition below.
\begin{definition}
A QDC, $\mathcal{D} = \{\mathcal{R}, \mathcal{I}\}$, consists of a QRAM, $\mathcal R$, coupled to a quantum communication network, $\mathcal I$,  and queries to the QDC can be performed in three steps: (1) a remote user uses $\mathcal I$ to send a quantum query to the QDC; (2) the QDC executes query using $\mathcal R$, as in either classical (in the main text) or \Cref{eq:QRAM_def_quantum}; (3) the QDC uses $\mathcal I$ to send input and output qubit registers, $Q_1$ and $Q_2$, back to the user. $\mathcal D$ is characterized by four key parameters: the size of the database $N$, the error in the query $\epsilon$, the latency $\tau$ (time cost of a single query), and the throughput $T$ (number of queries performed per unit time).  
%  ${D_Q} = \left( {{U_{{\rm{QRAM}}}},I} \right)$, where $U_{\rm{QRAM}}$ is the QRAM unitary that carries either the classical or quantum data, and $I$ is a quantum network. For the user, the throughput of $D_Q$ is given by the total running time $T_{\rm{total}}$, the precision (error) $\epsilon$, and the size of the data $N$. Moreover, $D_Q$ has a collection of latent parameters called $\theta$. 
\end{definition}

\subsection{Cost estimation for QDCs}
For QDCs defined above, how could we estimate the cost of time, hardware with a given requirement of error and privacy? Here we establish a general theory to estimate the hardware-time cost for QDCs, and determine the optimal parameters according to the cost function. 

In general, we define a cost function $F_{\text{cost}}^{{\text{QDC}}}$ for a given QDC architecture. The cost function could be written as
\begin{align}\label{costalgorithm}
 F_{\text{cost}}^{{\text{QDC}}}= F_{\text{cost}}^{{\text{QDC}}}({T_{{\rm{total}}}},{N_{{\rm{total}}}},P_{\text{total}})~.
\end{align}
Here, the cost function $F_{\text{cost}}^{{\text{QDC}}}$ includes the time cost $T_{\text{cost}}$, space (hardware) cost $N_{\text{cost}}$, and the privacy cost $P_{\text{total}}$ (The privacy cost here means a quantity the represents the level of consumption for the QDC users. We will provide an example in the situation of Section \ref{qpq}). For instance, one could simply assume that the above cost function is linear, 
\begin{align}
 F_{\text{cost}}^{{\text{QDC}}}= {\alpha _T}{T_{{\rm{total}}}} + {\alpha _N}{N_{{\rm{total}}}} + {\alpha _P}P_{\text{total}}~,
\end{align}
with fixed positive coefficients $\alpha_T$, $\alpha_N$ and $\alpha_P$. More generally, $ F_{\text{cost}}^{{\text{QDC}}}$ could be defined as a monotonic function of $T_{\text{cost}}$, $N_{\text{cost}}$ and $P_{\text{total}}$. Moreover, $T_{\text{cost}}$, $N_{\text{cost}}$ and $P_{\text{total}}$ are given by one collection of throughput parameters, and the other collection of hyperparameters (latency) $\theta$. The optimal hyperparameters could be determined by 
\begin{align}
{\theta ^*} = {{\mathop{\rm argmin}\nolimits} _\theta }F_{\text{cost}}^{{\text{QDC}}}~.
\end{align}
for given requirements of hardware. Similar analysis could be done for their counterparts without QDC, with the cost function $F_{\text{cost}}^{\overline {\text{QDC}} }$, and if QDCs have advantages, we want $F_{\text{cost}}^{\overline {\text{QDC}} }>F_{\text{cost}}^{\text{QDC}}$. 

The cost analysis examples discussed later could be understood as precise instances of the above framework. In Section \ref{comp}, we understand the hardware cost as the qubit cost, and we manifest the contribution of both  $T_{\text{total}}$ and $N_{\text{total}}$. In the $T$-gate example, we understand the entanglement cost as another form of the hardware cost, and we find significant advantages of QDCs in some cases. In Section \ref{qpq}, we emphasize  $T_{\text{total}}$, $N_{\text{total}}$ and $P_{\text{total}}$, the privacy cost in quantum communication. In general QDCs, all terms might be included based on the practical usage, time cost, and hardware. For a more systematic mathematical treatment of privacy in the complexity theory, see \emph{differential privacy} discussed in the machine learning community \cite{vadhan2017complexity}.

\subsection{Writing data in QRAM}\label{writing}
In this section, we more precisely define what it means to \emph{write} data to QRAM. The definition of writing depends on whether the data being written are classical or quantum, and also on whether the addressing scheme is classical or quantum. We elaborate on these four different situations below. 

% \begin{table}[ht]
% \begin{tabular}{ |c|c|c| } 
%  \hline
%   & Classical Pointer & Quantum Pointer \\ 
%  \hline
%  Classical Data & CC & CQ \\ 
%  \hline
%  Quantum Data & QC & QQ \\ 
%  \hline
% \end{tabular}
% \caption{A summary of four different situations in QRAM and QDCs: Classical Data, Classical Pointer (CC); Classical Data, Quantum Pointer (CQ); Quantum Data, Classical Pointer (QC); Quantum Data, Quantum Pointer (QQ).}
% \label{tableqdataaddress}
% \end{table}

\subsubsection{Classical data, classical addressing}
In this situation, the QRAM holds a classical data vector $x$, and the writing operation consists of specifying a classical address $i$ and a new classical value $y_i$, then overwriting the value  $i$-th cell of the QRAM's memory, $x_i\to y_i$. This writing process is entirely classical; it can be implemented simply by performing classical operations on the classical data. 

Even though this definition of writing to QRAM is completely classical, it is still useful in the context of quantum algorithms. In particular, after writing, the modified classical data in the QRAM can subsequently be read in superposition (i.e., with quantum addressing). For example, if each element in the database is replaced as $x_i\to y_i$, then reading the QRAM consists of the operation
\begin{align}
 %&\left| {{\psi _{{\rm{in }}}}} \right\rangle  =
 \sum\limits_{i = 0}^{N - 1} {{\alpha _i}} {\left| i \right\rangle ^{Q_1}}{\left| 0 \right\rangle ^{Q_2}} \to %\nonumber\\
 %& \left| {{\psi _{{\rm{out }}}}} \right\rangle  =
 \sum\limits_{i = 0}^{N - 1} {{\alpha _i}} {\left| i \right\rangle ^{Q_1}}{\left| {{y_i}} \right\rangle ^{Q_2}}~,
\end{align}  
Thus, the same QRAM can be re-used to perform a superposition of queries to a different data set. This is particularly useful in the context of QDCs, as multiple users could be running different algorithms that require access to different classical data sets. The QDC can cater to all of these users by overwriting the QRAM's classical data between queries from different users. 

\subsubsection{Classical data, quantum addressing}
In contrast to the previous definition, writing to QRAM for the case of classical data and quantum addressing is not well defined. To illustrate this, we propose a possible definition for writing in this situation, then show that it ultimately reduces to a probabilistic version of the classical writing procedure described above. 

We suppose that the QRAM's classical data is stored in a quantum memory, i.e.~each classical datum $x_i\in\{0,1\}$ is encoded in a qubit as $\ket{x_i}$, so that the full database consists of the product state $\bigotimes_i \ket{x_i}^{D_i}$, where $D_i$ denotes the $i$-th cell of the memory. The writing procedure consists of first specifying a quantum address $\sum_i \alpha_i \ket{i}^{Q_1}$. Then, coherently conditioned on the state of the $Q_1$ register, one prepares another qubit $Q_2$ in the state $\ket{y_i}$ with $y_i \in\{0,1\}$, then swaps this state with the $i$-th cell of the memory,
\begin{align}
&\sum_i {{\alpha _i}} {\left| i \right\rangle ^{Q_1}}{\left| y_i \right\rangle ^{Q_2}} \left[ \bigotimes_{j=1}^{N}\ket{x_j}^{D_j} \right]
\nonumber \\ 
\to & \sum_i {{\alpha _i}} {\left| i \right\rangle ^{Q_1}}{\left| x_i \right\rangle ^{Q_2}} \left[ \ket{y_i}^{D_{i}}  \bigotimes_{j\neq i}\ket{x_j}^{D_j} \right].
\end{align}
In general, this operation leaves the data registers entangled with the $Q_1$ and $Q_2$ registers. As such, tracing out the $Q_1$ and $Q_2$ registers leaves the database in a mixed state, wherewith probability $|\alpha_i|^2$ one finds that $i$-th entry has been overwritten as $x_i \to y_i$. To achieve the same result, one could instead simply have randomly chosen to overwrite the $i$-th element according to the distribution $|\alpha_i|^2$. Therefore, the use of quantum addressing and classical data does not confer an advantage over the case of classical addressing and classical data.

\subsubsection{Quantum data, classical addressing}
In this situation, the QRAM holds quantum data, i.e., an $N$-qubit quantum state. The writing operation consists of specifying a classical address $i$ and a new single-qubit state $\ket{\phi}^{Q_2}$, then swapping this state with the $i$-th qubit in the QRAM's memory. In particular, if the QRAM initially holds a product state $\bigotimes_j \ket{\psi_j}^{D_j}$, then this writing procedure enacts the operation
\begin{equation}
    \ket{\phi}^{Q_2}  \left[\bigotimes_j \ket{\psi_j}^{D_j}\right] 
    \to \ket{\psi_i}^{Q_2}  \left[ \ket{\phi}^{D_i}\bigotimes_{j\neq i} \ket{\psi_j}^{D_j}\right]. 
\end{equation}
Note that in the case where the quantum data consists of a product state, this operation does not entangle the $Q_2$ and $D_i$ registers. For general quantum data, however, this operation may leave these registers entangled, such that the data can be left in a mixed state when the $Q_2$ register is traced out.

\subsubsection{Quantum data, quantum addressing}
In this situation, the QRAM holds quantum data, i.e., an $N$-qubit quantum state. The writing operation consists of first specifying a quantum address $\sum_i\alpha_i\ket{i}$. Then, coherently conditioned on the state of the $Q_1$ register, one prepares another register $Q_2$ in the state $\ket{\phi_i}$, then swaps this state with the $i$-th cell of the memory. In the case where the QRAM initially holds a product state $\bigotimes_j \ket{\psi_j}^{D_j}$, then this writing procedure enacts the operation
\begin{align}
&\sum_i {{\alpha _i}} {\left| i \right\rangle ^{Q_1}}{\left| \phi_i \right\rangle ^{Q_2}} \left[ \bigotimes_j\ket{\psi_j}^{D_j} \right]
\nonumber \\ 
\to & \sum_i {{\alpha _i}} {\left| i \right\rangle ^{Q_1}}{\left| \psi_i \right\rangle ^{Q_2}} \left[ \ket{\phi_i}^{D_{i}}  \bigotimes_{j\neq i}\ket{\psi_j}^{D_j} \right].
\end{align}
We note that reading quantum data is a special instance of the above process where $\ket{\phi_i} = \ket{0}$ for all $i$. This operation generally leaves the $Q_1$ and $Q_2$ registers entangled with the data registers $D_i$.

\section{Computing}\label{comp}
\subsection{General discussions about oracles}\label{oraclegeneral}
One of the most important applications of QDC is quantum computation. In the minimal definition of QDCs, we could use the QRAM as a remote service center providing oracles for the user. Many famous quantum algorithms, such as Quantum Principle Component Analysis, require the construction of oracles to reach the quantum advantage \cite{lloyd2014quantum}. QRAM could provide substantial benefits regarding interfaces between the classical and the quantum world, serving as a natural hardware realization of the quantum oracle. Moreover, a hybrid QRAM/QROM construction will provide an optimal choice of the hardware-time overhead. Thus, one could imagine that the quantum computation is performed on the user side, and QDCs will serve as the source of the oracle. Connected by quantum networks, the user will call QDCs multiple times to complete the algorithm.

\begin{figure}
    \centering
    \includegraphics[width=0.45\textwidth]{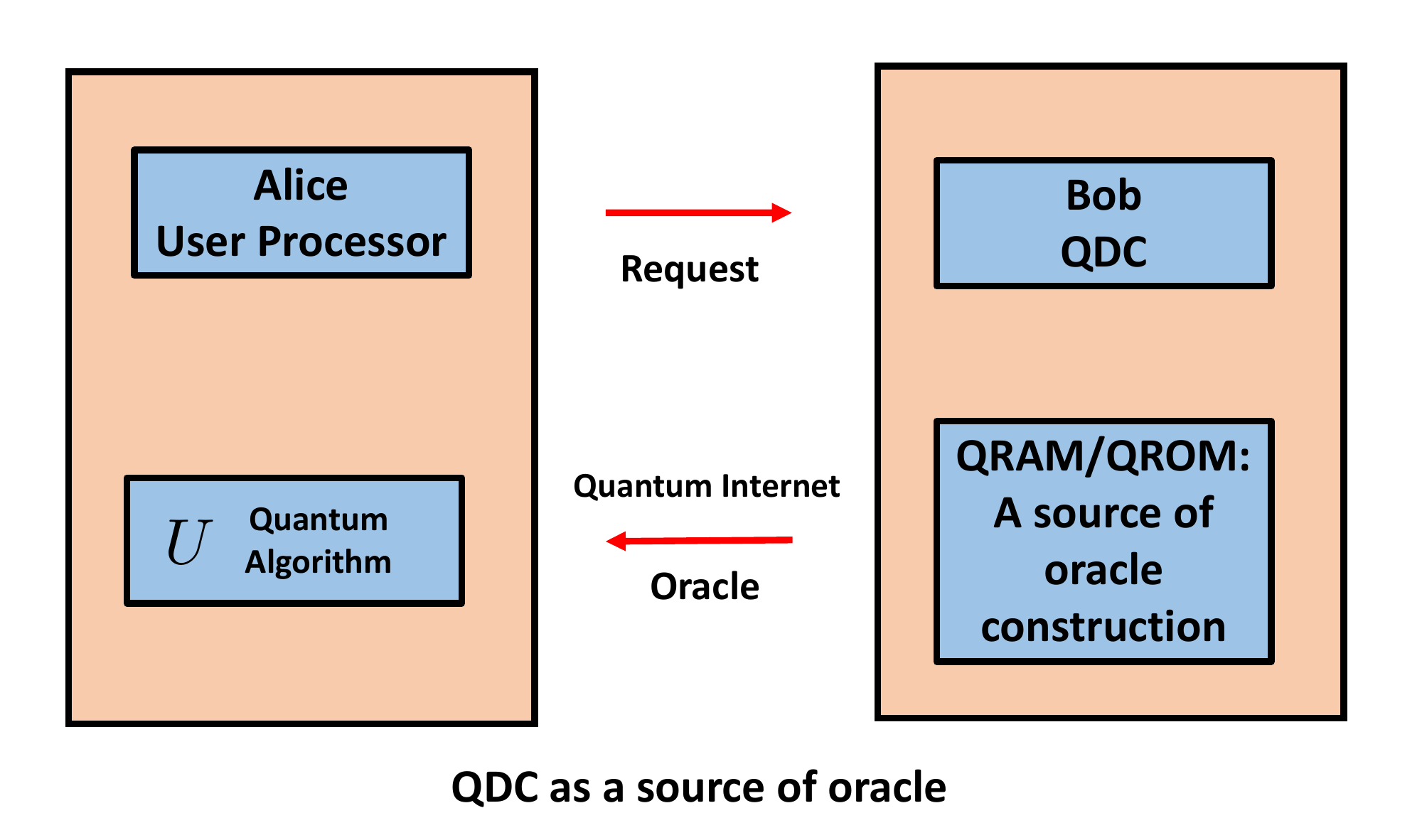}
    \caption{QDC for a general algorithm $U$. In this example, QDC could serve as a pool of the explicit oracle construction that is needed for the algorithm $U$.}
    \label{fig:eg1_qubi}
\end{figure}

Here, we will give a general discussion about the hardware-time cost of using QDC as a resource of oracle in a minimal setup. (see Figure \ref{fig:eg1_qubi}).

If we assume that a general quantum algorithm $U$ has the time cost and the qubit cost given by
\begin{align}
&{T_U} = {T_U}(L,\epsilon,\theta_0,\theta_U )~,\nonumber\\
&{N_U} = {N_U}(L,\epsilon,\theta_0,\theta_U )~,
\end{align}
depending on the problem size $L$, the precision $\epsilon$ and the collections of other hyperparameters $\theta_U$ and problem parameters $\theta_0$. Say that the time cost of the algorithm itself is expressed by the query complexity, and the corresponding oracle is prepared by QRAM itself multiple times. We assume the QRAM cost as 
\begin{align}
&{T_{Q}} = {T_{Q}}(L,\epsilon ,{\theta _{Q}})~,\nonumber\\
&{N_{Q}} = {N_{Q}}(L,\epsilon ,{\theta _{Q}})~,
\end{align}
with the QRAM parameter $\theta_{Q}$. Finally, we define the quantum network cost
\begin{align}
&{T_I} = {T_I}(L,\epsilon ,{L_{{\rm{tot}}}},{\theta _I})~,\nonumber\\
&{N_I} = {N_I}(L,\epsilon ,{L_{{\rm{tot}}}},{\theta _I})~,
\end{align}
with the total length $L_{\text{tot}}$ and the quantum network parameter $\theta_I$. Here, we are assuming that the oracles are prepared remotely with the total length $L_{\text{tot}}$, and transformed to the user with the quantum network. So the total cost is given by 
\begin{align}
&{T_{{\rm{total}}}}(L,\epsilon,\theta_0,{L_{{\rm{tot}}}},\theta_U,\theta_{Q},\theta_I) \nonumber\\
&={T_U}(L,\epsilon ,{\theta _U}) \times \left( {{T_{Q}}(L,\epsilon ,{\theta _{Q}}) + {T_I}(L,\epsilon ,{L_{{\rm{tot}}}},{\theta _I})} \right)~,\nonumber\\
&{N_{{\rm{total}}}}(L,\epsilon,\theta_0,{L_{{\rm{tot}}}},\theta_U,\theta_{Q},\theta_I)\nonumber\\
&={N_U}(L,\epsilon ,{\theta _U}) + {N_{Q}}(L,\epsilon ,{\theta _{Q}}) + {N_I}(L,\epsilon ,{L_{{\rm{tot}}}},{\theta _I})~.
\end{align}
Note that here $T_U$ is the query complexity for the quantum algorithm $U$. The time cost is a product, and the qubit cost is additive. Thus, for given $L$, $\epsilon$, and $\theta_0$, we could determine the optimal choice of QDC by
\begin{align}
&{(L_{{\rm{tot}}}^*,\theta _U^*,\theta _{Q}^*,\theta _I^*)_{L,\epsilon ,{\theta _0}}}\nonumber\\
&= \text{argmin} _{({L_{{\rm{tot}}}},{\theta _U},{\theta _{Q}},{\theta _I})}{F_{{\rm{cost}}}}~,
\end{align}
where 
\begin{align}\label{costcomputing}
F_{\text{cost}}=F_{\text{cost}}({T_{{\rm{total}}}},{N_{{\rm{total}}}})~,
\end{align}
is a given cost function based on the architecture of QDC. This is a specific example of the cost function algorithm Equation \ref{costalgorithm} for quantum computation. 

Here, we maintain the quantum algorithm mentioned here to be abstract. All quantum algorithms with the oracle required in the QRAM form could be adapted here. In Section \ref{oracleappend}, we discuss a specific quantum algorithm, quantum signal processing (QSP) for Hamiltonian simulation \cite{low2019hamiltonian,LowQSPprx,low2017optimal}, where quantum oracles are needed to address the information of the Hamiltonian.  According to Section \ref{oracleappend}, if we use the qubitization algorithm, and $L$ is the number of Pauli terms appearing in the Hamiltonian, we have
\begin{align}
&\frac{{{\operatorname{QDC}}{{}_{{\text{hardware cost}}}}}}{{{{\overline {{\rm{QDC}}} }_{{\text{hardware cost}}}}}} \nonumber\\
&= \frac{{{\cal O}(\log L) + {\cal O}\left( {\log {{\max }_i}\dim {\Pi _i}} \right)}}{{{\cal O}(\log L) + {\cal O}\left( {\log {{\max }_i}\dim {\Pi _i}} \right) + {\cal O}\left( {\frac{L}{M} + \log L} \right)}} \nonumber\\
&\approx \frac{{{\cal O}(M\log L)}}{{{\cal O}\left( L \right)}}~.
\end{align}
Here, we are comparing the hardware cost completely from the user side: in the $\overline{\text{QDC}}$ case, since the user does not have QDC, the user has to implement QRAM or QROM by himself or herself. $M$ is the parameter for the hybrid QRAM or QROM architecture. Moreover, we assume that $L$ is large (note that this will happen if we are assuming non-local Hamiltonians and the Hamiltonian might be dense, which is not always true in the quantum chemistry tasks). In this case, using QDC, we could provide a significant hardware cost saving from the user side: when $M$ does not scale with $L$, the saving could be even exponential. 

Moreover, we make some discussions about QDCs used for multiple users in Section \ref{multipleusers}. Furthermore, another potential saving of the hardware could come from the fact that the entanglement cost of accessing an $N$-element data set with a QDC is only $\log N$, where we have implicitly used in the above example, and it has been already manifest in the $T$-gate example.

\subsection{Quantum simulation and oracles from QDCs}\label{oracleappend}
A perfect example of running QDCs as oracle resources could be the quantum simulation algorithm, which has wide applications in quantum many-body physics, quantum field theory, and quantum computational chemistry with potential advantages compared to classical computers. Aside from the so-called Trotter simulation scheme \cite{uni,suzuki1976trotter,tro}, many quantum simulation algorithms are oracle-based, such as algorithms based on quantum walks \cite{childs2011walks,berry2015hamiltonian}, multiproduct formula~\cite{low2019multiproduct,childs2012multiproduct}, Taylor expansion~\cite{berry2012black,berry2015simulating}, fractional-query models~\cite{spar}, qubitization and quantum signal processing (QSP)~\cite{low2019hamiltonian,LowQSPprx,low2017optimal}. Those oracles could naturally be implemented by the QRAM model (see, for instance, \cite{connorthesis}).

We will give a short introduction to the qubitization and QSP algorithms here and discuss their costs. We will consider the linear combination of unitaries (LCU) decomposition as the input. We assume that the Hamiltonian is given by the following unitary strings
\begin{align}
H = \sum\limits_{i = 1}^L {{\alpha _i}} {\Pi _i}~.
\end{align}
For simplicity, we will assume that $\alpha_i>0$. This is called the LCU model, and $\Pi_i$s are usually the Pauli matrices. We introduce the ancilla states $\ket{i}: i=1,2,\cdots, L$ with the number of qubits $\log L$. Furthermore, we implement the following state $\ket{G}$
\begin{align}
\ket{G}=\sum_{i=1}^{L} g_{i}|i\rangle, \quad\left|g_{i}\right|^{2}=\frac{\alpha_{i}}{\lambda}, \quad \lambda=\sum_{i=1}^{L} \alpha_{i}~,
\end{align}
and
\begin{align}
&R = 2\left| G \right\rangle \left\langle G \right| - I~,\nonumber\\
&U = \sum\limits_{i = 1}^L {\left| i \right\rangle \left\langle i \right| \otimes {\Pi _i}}~, \nonumber\\
&W = RU~.
\end{align}
One could show that 
\begin{align}
&\frac{H}{\lambda } = \left( {\left\langle G \right| \otimes I} \right)U\left( {\left| G \right\rangle  \otimes I} \right)~,\nonumber\\
&\left\langle {G\left| {{W^n}} \right|G} \right\rangle  = {T_n}\left( {\frac{H}{\lambda }} \right)~,
\end{align}
where $T_n$ is the $n$-th Chebychev polynomial. The Hamiltonian evolution $e^{-i H t}$ could be given by
\begin{align}
{e^{ - iHt}} = {J_0}( - \lambda t) + 2\sum\limits_{n = 1}^{ + \infty } {{i^n}} {J_n}( - \lambda t){T_n}\left( {\frac{H}{\lambda }} \right)~.
\end{align}
Namely, one could separately add all terms together, and it requires 
\begin{align}
\mathcal{O}\left( {\lambda t \times \log \frac{1}{\epsilon }} \right)~,
\end{align}
number of queries to the operation $U$ and $\ket{G}$. Here, $t$ is the Hamiltonian evolution time, and $\epsilon$ is the error. Here, implementing $\ket{G}$ is a quantum oracle operation, which could be operated by QRAM or QROM. For simplicity, here we will mostly discuss the query complexity made by $\ket{G}$, and $U$ itself would cost $\mathcal{O}(L C_1)$ primitive gates, where $C_1$ is the maximal complexity of implementing a single Pauli term $\Pi_i$. In terms of gate counting, $G$ itself would cost $\mathcal{O}(L)$ primitive gates. A more complicated construction, which is called the quantum signal processing (QSP) \cite{low2019hamiltonian}, could reduce the above product in query complexity to addition
\begin{align}
\mathcal{O}\left( {\lambda t + \log \frac{1}{\epsilon }} \right)~.
\end{align}

Another ingredient of our analysis would combine the quantum network. Quantum network, based on hardware realizations of quantum teleportation and quantum cryptography, is expected to be efficient for transferring quantum states and their associated quantum data across long distances with guaranteed security \cite{kimble2008quantum}. Specifically, here we will discuss the quantum repeaters, architectures that could significantly overcome the loss errors and depolarization errors for quantum communication with photons (see, for instance, \cite{muralidharan2014ultrafast}). For cost estimation, we will follow the discussion in \cite{muralidharan2016optimal}. 
There are three different generations of quantum repeaters, and here we will, for simplicity, discuss them together. A universal measure of cost overhead for those quantum repeaters is the cost coefficient $C_2$ ($C'$ used in  \cite{muralidharan2014ultrafast}), which could be understood as the $\text{qubit}\times {\text{time}}$ cost for the transmission of one Bell pair per unit length. Now, we will assume that for the quantum teleportation task of the data center, we use $L_{\text{tot}}$ length. The characteristic time is given by $t_{\text{ch}}$ (which is different from three different generations of quantum repeaters). 

For our minimal definition of the quantum data center, with QDCs serving as the remote oracle resources, one could compute the total time cost $T_{\text{total}}$ and the qubit cost $N_{\text{total}}$ as the following
\begin{align}
&T_{\text{total}}=\mathcal{O}\left( {\lambda t + \log \frac{1}{\epsilon }} \right) \times ( \mathcal{O}(LC_1) + \mathcal{O}(M\log^2 L)  + t_{\text{ch}})~,\nonumber\\
&N_{\text{total}}=\mathcal{O}(\log L)+\mathcal{O}(\log \max_i \dim \Pi_i) + \mathcal{O}(\frac{L}{M}+ \log L) \nonumber\\
&+ \mathcal{O}(\log L \times \frac{{C_2}}{{{t_{{\rm{ch}}}}}}{L_{{\rm{tot}}}} )~.
\end{align}
We will give the following explanations to the above formula.
\begin{itemize}
\item The first term in the time cost, $\mathcal{O}\left( {\lambda t + \log \frac{1}{\epsilon }} \right) $, is exactly the query complexity of QSP. Based on our minimal definition of the quantum data center, the cost of each query, including the quantum communication cost and the QRAM/QROM cost.  
\item The term $\mathcal{O}(L C_1)$ in the time cost corresponds to the cost of each $U$ in the QSP algorithm. 
\item The parameter $M$ corresponds to the parameter of the hybrid QRAM/QROM construction \cite{hann2021resilience,connorthesis}, which is a way to unify the hardware-time cost. A pure QRAM would cost $\mathcal{O}(L)$ qubits in $\mathcal{O}(\log L)$ time, while QROM would cost $\mathcal{O}(\log L)$ qubits in $\mathcal{O}(L \log L)$ time. With the tunable parameter $M$, the hybrid construction would cost $\mathcal{O}(\log L+ L/M)$ qubits within $\mathcal{O}(M \log L)$ time, which could reduce to QRAM with $M=1$ and QROM with $M=L$. This is how the $ \mathcal{O}(\frac{L}{M}+ \log L) $ term comes in the second term.
\item The form of the oracle $\ket{G}$ is identical to the amplitude encoding oracle, which could reduce to the QRAM definition (the data-lookup oracle) with $\mathcal{O}(\log L)$ time cost overhead (without post-selection), or $\mathcal{O}(1)$ time cost overhead (with post-selection) \cite{connorthesis}. Here, for simplicity, we are using the case without post-selection. Thus, aside from the hybrid QRAM/QROM time cost $\mathcal{O}(M \log L)$, we have an extra $\mathcal{O}(\log L)$ factor, which gives the $\mathcal{O}(M \log^2 L)$ factor in $T_{\text{tot}}$.
\item The term $\mathcal{O}(\log L \times \frac{{C_2}}{{{t_{{\rm{ch}}}}}}{L_{{\rm{tot}}}} )$ comes from the definition of $C_2$ in quantum repeaters, followed by the actual qubits and the corresponding maximal possible Bell pairs we are using when doing teleportation \cite{muralidharan2016optimal}.  
\end{itemize}
As long as we know the exact setups of QDCs, we could decide the resources easily based on our requirement as described by the general setup. Assuming a cost function $F_{\text{cost}}$, one could determine the set of hyperparameters both in quantum communication and quantum simulation by
\begin{align}
M^*,{L_{{\rm{tot}}}^*},\text{other hyperparameters}\ldots  = \arg \min F_{\text{cost}}~.
\end{align}
Finally, we mention that QDCs could potentially provide transducers to transform different types of quantum data, for instance, from digital qubits to analog qubits. Since various different forms of qubits have their own advantages and challenges, it is necessary to consider hybrid quantum systems. For example, if we wish to combine quantum computation performed in the superconducting qubit systems, and quantum communication provided by transformations of optical photons across long distances in QDC and its users, quantum transducers might be necessary. See for instance \cite{zhang2018quantum}. Here in this example, since the quantum simulation algorithms could be performed by superconducting qubits, while the quantum network could be realized by optical photons, the quantum transducer is needed.

Finally, we consider the case where we only count the hardware cost from users. In the case where we do not have QDCs, the users have to implement QRAM or QROM by themselves in the quantum simulation algorithm. Thus, in the case where users have access to QDCs, we could subtract the hardware contribution from QDC. We could compute the hardware cost ratio between the case where we have QDCs, and the case where we do not have QDCs ($\overline{\operatorname{QDC}}$). The answer is
\begin{align}
&\frac{{{\operatorname{QDC}}{{}_{{\rm{hardware}}}}}}{{{{\overline {{\rm{QDC}}} }_{{\rm{hardware}}}}}} \nonumber\\
&= \frac{{{\cal O}(\log L) + {\cal O}\left( {\log {{\max }_i}\dim {\Pi _i}} \right)}}{{{\cal O}(\log L) + {\cal O}\left( {\log {{\max }_i}\dim {\Pi _i}} \right) + {\cal O}\left( {\frac{L}{M} + \log L} \right)}} \nonumber\\
&\approx \frac{{{\cal O}(M\log L)}}{{{\cal O}\left( L \right)}}~.
\end{align}
Here, we take the large $L$ limit. Thus the $L$-dependent term will be dominant. We could see that, especially when $M$ is not scaling with $L$, this will be an exponential saving of the hardware cost for QDC users.

\subsection{QDC for computing: multiple users}\label{multipleusers}
In this section, we discuss a simple situation where QDC has multiple users and discuss its usage. 

Consider the case where multiple users want the same answer of a quantum algorithm. For simplicity, we assume the answer should be classical such that it is able to be copied to multiple users (the result could also be quantum, but then we have to use approximate quantum cloning). We define the hardware cost of the quantum algorithm $U$ for a single user as ${f_U}({\theta _0},{\theta _U})$ where $\theta_0$ is the problem parameter, and $\theta_U$ is the hyperparameter of the algorithm. Say that we have $k$ users, and for each user, the network cost of the hardware is ${f_I}({\theta _0},{\theta _I})$ where $\theta_I$ is the hyperparameter of the algorithm. Thus, without QDC, calculations are performed independently from each user, and the total hardware cost scales as
\begin{align}
f({\theta _0},{\theta _U}) = k{f_U}({\theta _0},{\theta _U})~.
\end{align}
With the QDC, the hardware cost will scale as
\begin{align}
f({\theta _0},{\theta _U},\theta_I) =  {f_U}({\theta _0},{\theta _U}) + k{f_I}({\theta _0},{\theta _I})~.
\end{align}
Thus, the condition of the advantage of QDC is given by
\begin{align}
\frac{{{f_I}({\theta _0},{\theta _U})}}{{{f_U}({\theta _0},{\theta _I})}} \ll \frac{{k - 1}}{k}~.
\end{align}
Thus, we could define the ratio
\begin{align}
r = \frac{k}{{k - 1}}\frac{{{f_I}({\theta _0},{\theta _I})}}{{{f_U}({\theta _0},{\theta _U})}}~.
\end{align}
The smaller $r$ is, the more useful QDC should be. The optimal $r$ could be given by
\begin{align}
{r^*}({\theta _0},k) = \frac{k}{{k - 1}}{\min _{{\theta _U},{\theta _I}}}\frac{{{f_I}({\theta _0},{\theta _U})}}{{{f_U}({\theta _0},{\theta _I})}}~.
\end{align}

Here we make some comments about the above calculation. The observation of comparing the communication cost and the computational cost is one of the original motivations of QDCs: using teleportation, one could save computational costs for multiple users. The $r$ coefficient we defined here, and its possible variant, could serve as a generic measure for such observations. However, the task we described before is not using the full features of QDCs. If we teleport quantum states using the quantum network, the state itself is not copiable to multiple users (even though we could copy the state approximately, but the error might be significant). One could use classical networks instead, or encode the classical output to quantum repeaters and make use of quantum networks. Thus, in this case, the quantum network may not necessarily be needed. It could serve as a version of QDC where QRAM is used, but the quantum network is not (see another example where we use the quantum network but not QRAM in Section \ref{channelappend}). Moreover, we expect that the above generic protocol could be improved and extended to more practical applications in the real science or business situation, and the simple analysis presented here could be general guidance towards those applications.

\subsection{Surface code and the $T$-gate counting}\label{Tgatesec}
First, we give a brief comment on the alternative ``smart" usage of QDC assisted quantum computing. In fact, the native approach would incur a prohibitive $\mathcal{O}(\sqrt{N})$ communication cost per query. In contrast, by outsourcing entire queries to the QDC, one effectively funnels a large amount of ``magic'' (the $\mathcal{O}(\sqrt{N})$ magic states required to implement a query) into a very small number of transmitted qubits (the $\mathcal{O}(\log N)$ qubits comprising the query's output). This way, the user receives maximal assistance from the QDC at minimal communication cost. 

Moreover, we give a brief review of the $T$-gate counting techniques that are developed in \cite{litinski2019game} about surface-code quantum computation. Those techniques are based on a formalism of executions in a fault-tolerant surface-code architecture from a given quantum circuit (quantum algorithm). Estimations of hardware-time trade-off for given quantum algorithms, using this formalism, are based on the hardware and algorithm assumptions, which might be different compared to other protocols (see, for instance, \cite{herr2017lattice}). Further details could be found in \cite{litinski2019game}.

The formalism is established from making assumptions about basic qubit manipulations. Simple operations such as qubit initializations and single-patch measurements can be regarded as easy, and they will cost $0 \VarClock$, while operations like two or multiple-qubit measurements and patch deformation will cost $1 \VarClock$. Here, the time unit $1 \VarClock$ might be based on the real hardware. And in the examples of \cite{litinski2019game} we can set $1 \VarClock = 1\mu \text{s}$.

The procedure of estimating the hardware-time cost for a given quantum circuit is the following. Firstly, we decompose the target unitary operation as Clifford+$T$-gates. Usually, we assume that the Clifford gates are cheap and $T$-gates are expensive. In fact, $T$-gates could be regarded as classical operations, but a given $T$-gate will require consumption of a single magic state, $|0\rangle+e^{i \pi / 4}|1\rangle$. We need to use magic state distillation \cite{bravyi2005universal} to generate high-quality magic states in the large-scale quantum computation. 

Further treatment of a series of Clifford+$T$-gates will contain designing data blocks (blocks of tiles where the data qubits live), distillation blocks (blocks of tiles to distill magic states), and their combinations. In \cite{litinski2019game}, several protocols are concretely discussed for hardware-time costs. Finally, for given large-scale quantum algorithms, precise designs are presented to minimize the hardware-time cost, especially the costs from $T$-gates and magic state distillation, and the costs could be pinned down to the number of qubits, gates, and even hours of time costs from assumptions of $\VarClock$. In the $T$-gate example in the main text, we point out that QDCs could serve as a $T$-gate factory and could reduce the $T$-gate counts significantly. 

Moreover, we discuss some details about the calculation in the main text with the help of \cite{litinski2019game}. Based on the setup of qubit numbers and the required target failure probability, the main takeaways from this analysis are as follows. First, a 116-to-12 magic state distillation scheme is sufficient, as the probability of logical error in the distilled magic state is $< 10^{-10}$, hence the total probability that any of the $10^8$ magic states is faulty is $<1\%$. Once the distillation scheme is chosen, the total number of surface code tiles (210) and cycles ($11 d \times 10^8$) required by the algorithm can be determined, and hence the minimum code distance can be calculated. For the above parameters, a distance $d=27$ is required to keep the total logical error probability below $1\%$. This translates into a cost of $306,000$ physical qubits and a runtime of 7 hours (assuming each surface code cycle takes $1\mu s$). These costs constitute a baseline for our later comparisons. 
\begin{figure}[h!]
\begin{center}  
\includegraphics[width=0.8\textwidth]{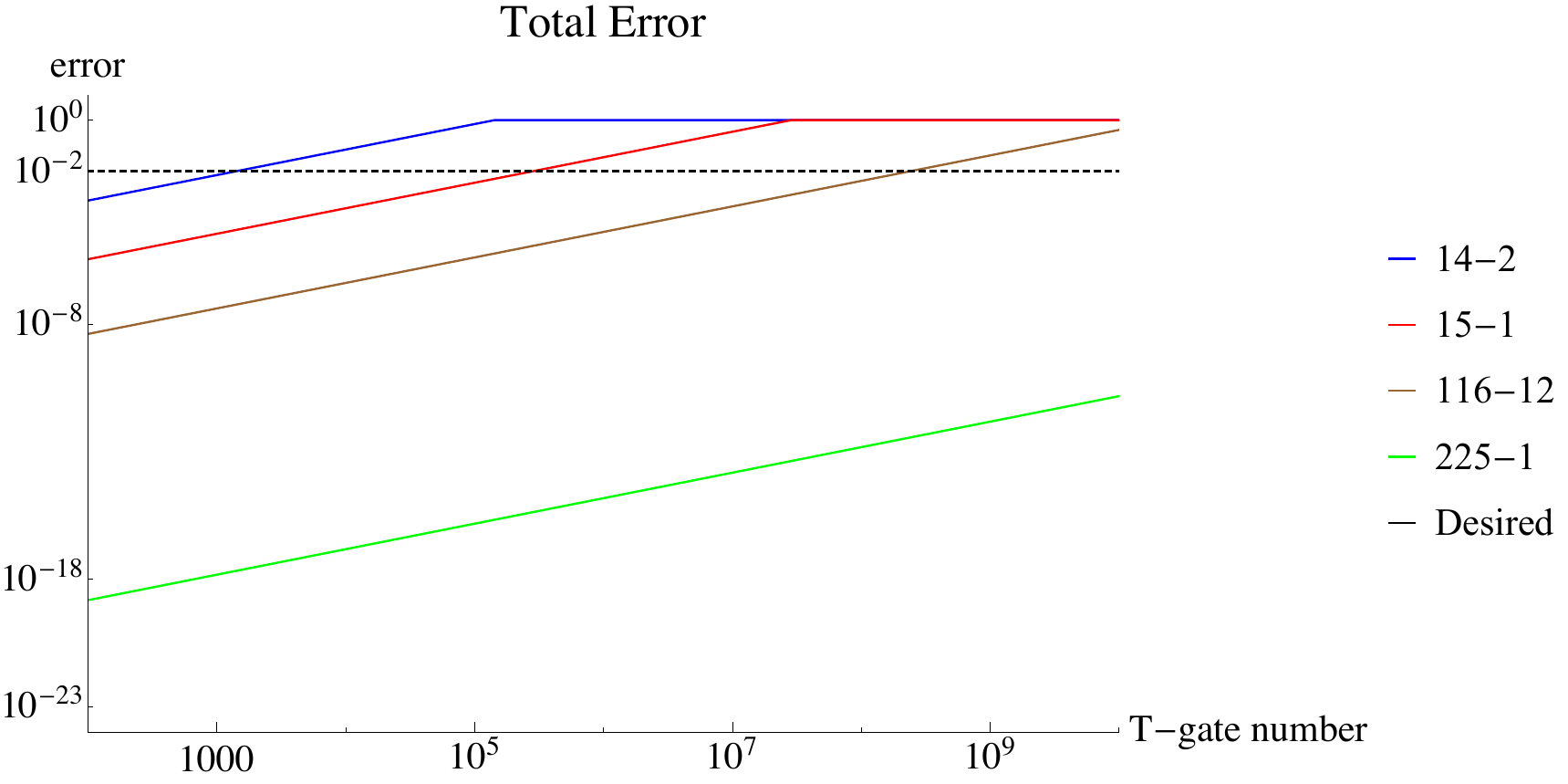}
\caption{The dependence of the total error on the physical error rate $p$ for different magic state distillation schemes according to the approximation in \cite{litinski2019game}. One could use this dependence to determine the optimal magic state distillation scheme for a given number of $T$-gate costs. In our example, we want the total error to be smaller than $10^{-2}$ (desired). Following the orders from up to down, we have 14-2 (blue), 15-1 (red), 116-12 (brown), and 225-1 (green) different distillation protocols showing in the plot as solid lines, compared to the dashed line $10^{-2}$. }
\label{fig:distillation}
\end{center}  
\end{figure}  
In Figure \ref{fig:distillation}, we are presenting the total failure probability of computations for different magic state distillation schemes, depending on different error rates of devices $p$. Following \cite{litinski2019game}, we are using the following formulas to estimate the total error,
\begin{align}
&{\text{total error}}{(p)_{14 - 2}} \approx 7{p^2}~,\nonumber\\
&{\text{total error}}{(p)_{15 - 1}} \approx 35{p^3}~,\nonumber\\
&{\text{total error}}{(p)_{116 - 12}} \approx 41.25{p^4}~,\nonumber\\
&{\text{total error}}{(p)_{225 - 1}} \approx 35{(35{p^3})^3}~,
\end{align}
before the total error meets 1. This figure could directly reveal the proper choice of the magic state distillation schemes. For instance, in our case, we are demanding the total error to be smaller than $10^{-2}$. When the number of $T$-gates is $10^8$, the 116-to-12 magic state distillation scheme is sufficient. 

\begin{figure}[h!]
\begin{center}  
$(a)$\includegraphics[width=0.8\textwidth]{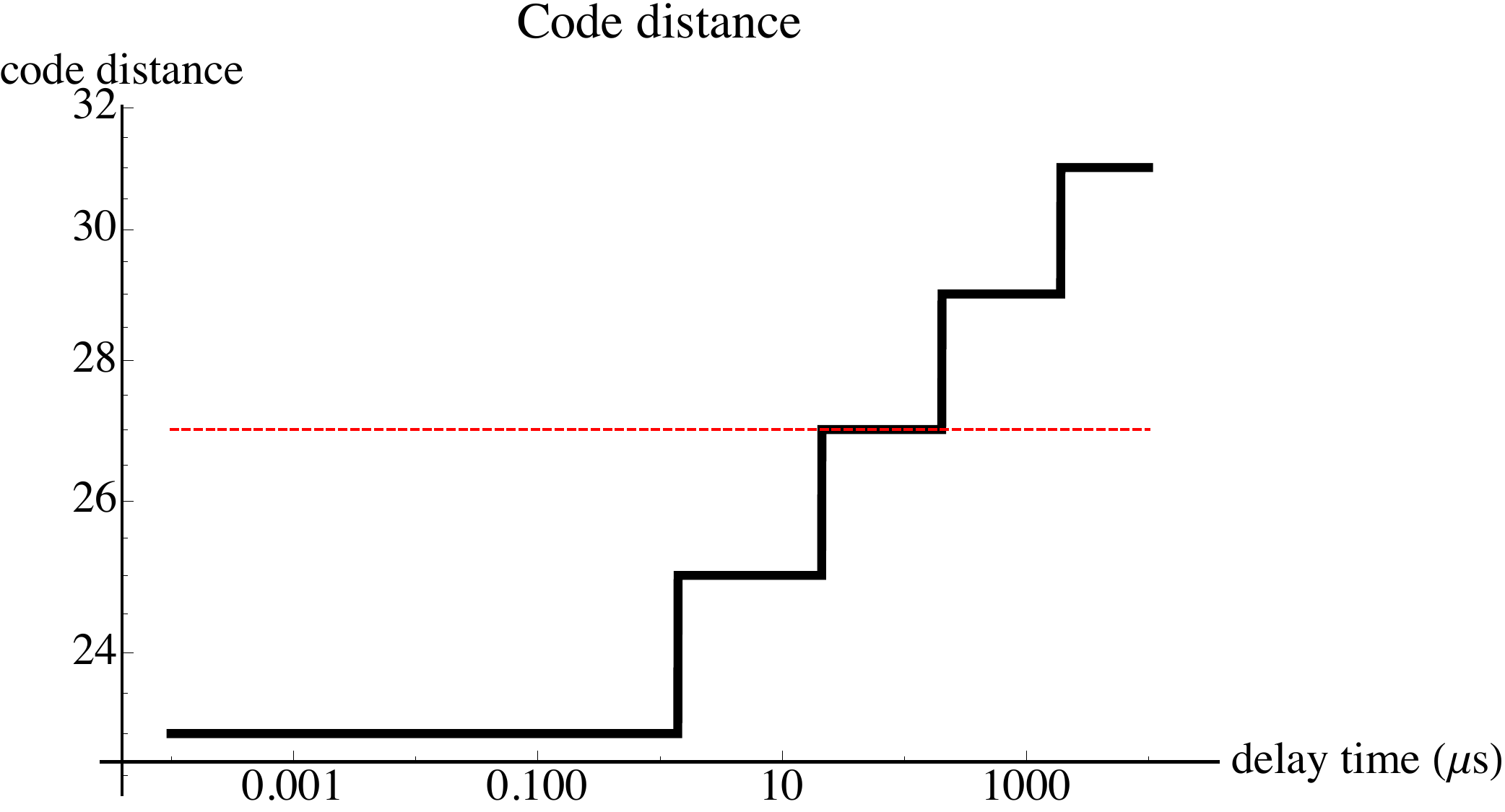}
$(b)$\includegraphics[width=0.8\textwidth]{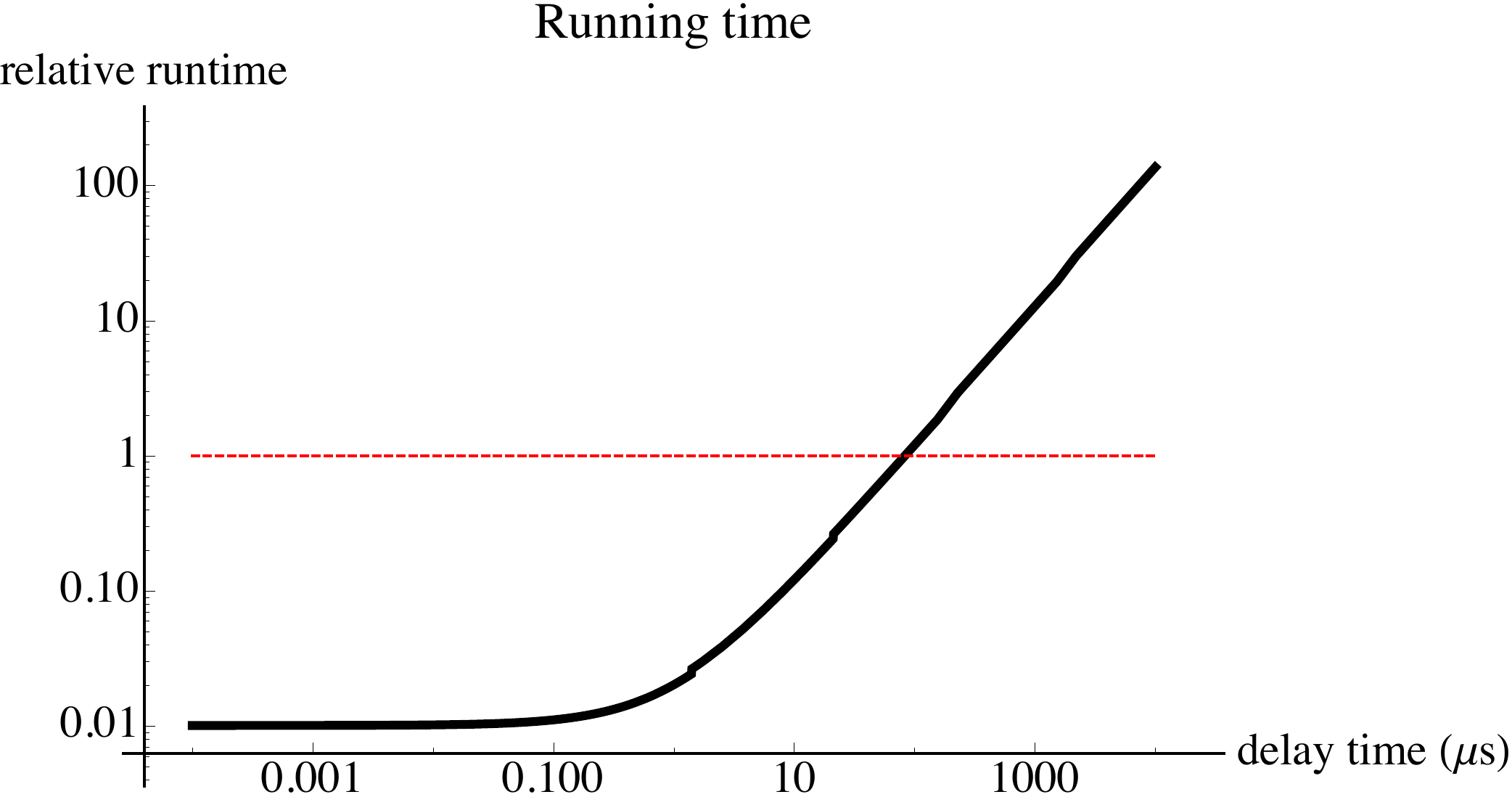}
\caption{The QDC-assisted code distance $(a)$ and the relative running time $(b)$ depending on the delay time. Here, the relative quantities are measured against the situation without QDC. The red dashed line represents the threshold where QDC has the same performance as the situation without QDC.}
\label{fig:magiccost}
\end{center}  
\end{figure}

Now, in the same situation of the main text, we make an analysis on the pure hardware savings depending on the delay factor and the delay time. In Figure \ref{fig:magiccost}, we investigate the code distance $d$, the hardware cost ratio (the number of qubits used in the QDC situation divided by the one without QDC), and the time cost ratio (the time cost used in the QDC situation divided by the one without QDC), depending on the delay factor or the actual delay time, assuming $1 \mu \text{s}$ per code cycle. Since the delay factor is not related to the choice of magic state distillation schemes, in the QDC situation, we keep the scheme to be the same (the 15-1 protocol). We could see that when the delay factor is high, namely, we have a relatively large waiting time from the quantum communication, we are not able to obtain a significant advantage from using QDC. However, if we assume that the quantum communication is fast and the delay factor is small, we are able to save more hardware and time by using QDC. For instance, when the delay factor $\le 21$ ($\le \mathcal{O}(10 \mu \text{s})$ for the delay time), we could set the code distance from $d=27$ to $d=25$. When the delay factor $\le 1910$ ($\le \mathcal{O}(1 \text{ms})$ for the delay time), the QDC could outperform the situation without QDC measured by hardware overhead. When the delay factor $\le 83$ ($\le \mathcal{O}(100 \mu \text{s})$ for the delay time), the QDC could outperform the situation without QDC measured by time overhead.

\section{Communication}\label{comm}

In the context of quantum communication, QDCs can be used to guarantee privacy, with a variety of potential applications. The essential feature of the QDC that enables this privacy is the ability of QRAM to perform queries to data in superposition. By secretly choosing to perform classical queries or superposed queries, then examining the results, users can determine whether other parties (including the QDC) may have tampered with the queries. 

This basic idea is operationalized in the Quantum Private Queries (QPQ) protocol of \cite{giovannetti2008queries}, which we describe below. This protocol allows users to access classical data with privacy guarantees, and this same idea can be applied to enable efficient blind quantum computation~\cite{giovannetti2013efficient} (also described below). Both of these protocols can be directly implemented using a QDC.

\subsection{Quantum Private Queries}\label{qpq}
A QDC can be directly used to implement the quantum private queries protocol of \cite{giovannetti2008queries}. In the protocol, a user (Alice) wants to access some classical data that is stored in a remote database (held by Bob). Alice wishes to access the data without revealing to Bob which data elements she has accessed. At the same time, Bob wants to maintain the privacy of his database, only sending Alice the information she requests. 

The protocol of \cite{giovannetti2008queries}, Quantum Private Query (QPQ), guarantees both user and database privacy by storing the data in QRAM. To access the $i$-th element of a length-$N$ database, Alice prepares a $\log N$-qubit register in the state $\ket{i}$ and transmits this state to Bob. Then, Bob uses this state as input to a QRAM query, so that the corresponding classical data, $x_i$, is encoded in an output qubit register. Both the input and output registers are then returned to Alice. As such, database privacy is guaranteed because Bob must only transmit one element of the data back to Alice. To guarantee user privacy, Alice randomly chooses to either send initial state $\ket{i}$ or a lure state $(\ket{i}+\ket{0})/\sqrt{2}$ to Bob (which she chooses is unknown to him). By performing measurements on the states Bob returns, Alice can ascertain whether or not Bob has attempted to learn the value of $i$. Thus Alice can guarantee her privacy.

The implementation of this protocol with a QDC is not hard. The QDC consists of a QRAM, so the QDC simply plays the role of Bob in the protocol. Moreover, QDC provides an application of the QPQ protocol through the quantum network. 

Now we quantify the protocol more precisely. In fact, instead of considering the states $\ket{i}$ and $(\ket{i}+\ket{0})/2$, we could consider more general states \cite{giovannetti2008queries}. Bob needs to make choices in one of the two following scenarios:
\begin{align}
&\left| {{S_A}} \right\rangle  = {\left| j \right\rangle _{{Q_1}}} \otimes \frac{1}{{\sqrt 2 }}\left( {{{\left| j \right\rangle }_{{Q_2}}} + {{\left| r \right\rangle }_{{Q_2}}}} \right)~,\nonumber\\
&\left| {{S_B}} \right\rangle  = \frac{1}{{\sqrt 2 }}\left( {{{\left| j \right\rangle }_{{Q_1}}} + {{\left| r \right\rangle }_{{Q_1}}}} \right) \otimes {\left| j \right\rangle _{{Q_2}}}~,
\end{align}
where $S_{A,B}$ are made by the joint states of two queries: $Q_1$ and $Q_2$. All possible operations from Bob could be summarized by two unitaries: $U_1$ and $U_2$. $U_1$ ($U_2$) acts on the query space $Q_1$ ($Q_2$), the associated register system $R_1$ ($R_2$), and Bob's ancillary system $B$ (now we could understand it as Bob's QDC). If Bob is honest, the algorithm of Bob is to make use of QRAM, uploading the information from $Q_2$ to registers, and the states in $Q_2$ will not be changed. If not, Bob's remaining system $Q_2$ will be entangled with the rest at the end. One could compute the final state of Alice:
\begin{align}
{\rho _\ell }(j) \equiv {{\mathop{\rm Tr}\nolimits} _B}\left[ {{U_2}{U_1}\left| {{\Psi _\ell }(j)} \right\rangle \left\langle {{\Psi _\ell }(j)} \right|U_1^\dag U_2^\dag } \right]~,
\end{align}
where $\ell=A,B$, and 
\begin{align}
\left| {{\Psi _\ell }(j)} \right\rangle  = {\left| {{S_\ell }} \right\rangle _{{Q_1}{Q_2}}}{\left| 0 \right\rangle _{RB}}~.
\end{align}
Moreover, the final state of $Q_2$ is given by
\begin{align}
{\sigma _\ell }(j) \equiv {{\mathop{\rm Tr}\nolimits} _{{Q_1}{Q_2}{R_1}{R_2}}}\left[ {{U_2}{U_1}\left| {{\Psi _\ell }(j)} \right\rangle \left\langle {{\Psi _\ell }(j)} \right|U_1^\dag U_2^\dag } \right]~.
\end{align}
One could quantify the amount of information Bob could obtain from Alice by the mutual information $I_B$. We will use the Holevo information associated with the ensemble $\left\{ {{p_j},\sigma (j)} \right\}$, where $p_j=1/N$ is the probability for choosing $j$, and $\sigma (j) = \left[ {{\sigma _A}(j) + {\sigma _B}(j)} \right]/2$ is the final state of $Q_2$, since Alice has an equal probability to choose $\ell=A,B$. Thus one could obtain \cite{giovannetti2010quantum}
\begin{align}
{I}_B \le c{\epsilon_p ^{1/4}}{\log _2}N ~.
\end{align}
Here $c$ is a constant, $c \leq 631$, and $\epsilon_p$ is the maximal probability where Alice finds that Bob is not cheating. Namely, if we use $1-P_\ell (j)$ to denote the probability where Bob will pass Alice's test, then $P_\ell (j) \le \epsilon_p$. As a summary, $I$ is a measure of how \emph{honest} Bob could actually have, and $\epsilon_p$ is a result of Alice's test. The above inequality is originated from the information-disturbance trade-off, and the Holevo bound \cite{nielsen2002quantum}. The $\epsilon_p^{1/4}$ dependence is coming from repetitively taking the square root between amplitudes and probabilities in quantum mechanics.  

Now we relate $\epsilon_p$ by the number of queries appearing in the QPQ protocol. Let us assume that Alice has $Q$ queries independently sent to Bob. Note that for multiple queries, if there is at least one time when Alice finds that Bob is cheating, Alice will know that privacy is not guaranteed. So the probability of \emph{Alice cannot find Bob is cheating among $Q_B$ times in all $Q$ times}, is given by $(1-\epsilon_p)^{Q_B}$, where $\epsilon_p$ is the maximal probability where Alice finds that Bob is cheating in a single time. When $Q_B$ increases, $(1-\epsilon_p)^{Q_B}=a$ will decay from 1 to an $\mathcal{O}(1)$ number $a$ where $1-a$ is not ignorable and we assume that $Q_B \ll 1/\epsilon_p$. In this case, we have
\begin{align}
Q_B = \frac{{\log a}}{{\log (1 - {\epsilon _p})}} \approx \frac{{\log \frac{1}{a}}}{{{\epsilon _p}}} = \mathcal{O}(\frac{1}{{{\epsilon _p}}})~.
\end{align}  
Now, we get
\begin{align}
I_B\le \mathcal{O}\left( {{Q_B^{ - 1/4}}\log N} \right)~.
\end{align}
Thus, we see that for larger $Q_B$, if Alice does not find Bob is cheating, then Alice could be more confident that Bob has less mutual information. One can also assume that Bob pick a cheating strategy by $Q_B \sim Q^{\alpha}$, where $0\le \alpha \le 1$ will imply how many times Bob is cheating during the whole process. So we have
\begin{align}
I_B\le \mathcal{O}\left( {{Q^{ - \alpha/4 }}\log N} \right)~.
\end{align}

When designing the QDC associated with QPQ, we could introduce a joint cost measurement among time, space and privacy. Similar to the analysis about quantum signal processing, we write down the costs for the QDC when implementing QPQ as
\begin{align}
&{T_{{\rm{total}}}} = \mathcal{O}(QM\log N) + \mathcal{O}(Q{t_{{\rm{ch}}}})~,\nonumber\\
&{N_{{\rm{total}}}} = \mathcal{O}\left( {\frac{N}{M} + \log N} \right) + \mathcal{O}\left( {\log N \times \frac{{C_2}}{{{t_{{\rm{ch}}}}}}{L_{{\rm{tot}}}}} \right)~.
\end{align}
Here, $N$ is the number of qubits, $Q$ is the total number of queries Alice has sent, $M$ is the parameter in the hybrid QRAM/QROM construction, $t_{\text{ch}}$ is the teleportation time per query, $C_2$ is the teleportation qubit $\times$ time cost for the transmission of one Bell pair per unit length, and $L_{\text{tot}}$ is the total length during teleportation. For the mutual information $I_B$, one could understand $I_B$ as the privacy cost, by defining $P_\text{total}$ as a monotonically decreasing function of $I_B$, since the smaller $I_B$, the larger privacy we are requiring. For simplicity, we could define $P_\text{total} = 1/I_B$. It does not matter how we choose the monotonic function, since a redefinition of the function could be absorbed to the definition of the cost function $F_{\text{cost}}$. Moreover, one could also understand $I_B$ as part of the hardware and the time cost, since we could write,
\begin{align}
Q = \mathcal{O}({I_B^{ - 4/\alpha}}{\log ^{4/\alpha}}N)~.
\end{align}
And if we demand a fixed value of $I_B$, we could adapt $Q$ into the hardware and the time cost. The larger $Q$ is, the higher costs are required
\begin{align}
&{T_{{\rm{total}}}} = {\cal O}(M \times I_B^{ - 4}{\log ^{4/\alpha+1}}N) + {\cal O}({t_{{\rm{ch}}}} \times I_B^{ - 4}{\log ^{4/\alpha}}N)~,\nonumber\\
&{N_{{\rm{total}}}} = \mathcal{O}\left( {\frac{N}{M} + \log N} \right) + \mathcal{O}\left( {\log N \times \frac{{C_2}}{{{t_{{\rm{ch}}}}}}{L_{{\rm{tot}}}}} \right)~.
\end{align}
The total cost estimation of QDC associated with QPQ will be a joint measurement among $T_{\text{total}}$, $N_\text{total}$ and $P_{\text{total}}$. 

Note that in the main text, we discuss the combination of QPQ with the quantum secret sharing protocol. The original proposal about quantum secret sharing given in \cite{hillery1999quantum} is based on the entanglement property of the Greenberger-Horne-Zeilinger (GHZ) state, and it is a quantum scheme for sharing classical data since it relies on the measurement result. Moreover, in \cite{cleve1999share}, a quantum scheme for sharing quantum data has been proposed, which is more suitable in our context. The paper \cite{cleve1999share} construct a $((k,n))$ threshold scheme, where a quantum state is divided into $n$ shares, while any $k$ of them could completely reconstruct the state, but any $k-1$ of them cannot. It was shown that as long as $n<2k$, the construction is possible, and the explicit scheme has been constructed. In our case, a single Alice could divide the information to $n$ QDCs. We could assume arbitrary $k$ such that $n<2k$. Based on the practical purposes, a more specific setup of $k$ might be used. A joint analysis of the multi-party quantum communication parameters might set the security standard for a given set of hardware, which we defer for future research.

Finally, we hope to mention that there are studies about limitations and insecurity concerns about concepts that are related to QPQs. In fact, there are no-go theorems \cite{mayers1997unconditionally,lo1997quantum,lo1997insecurity} about the imperfection of certain quantum computation and communication schemes. The QPQ protocol does not violate the no-go theorems \cite{lo1997insecurity} since the setup is different. In QPQ, we do not have the security requirement required for the no-theorem, where the user Alice cannot know the private key of QDCs, since QDCs do not have private keys. On the other hand, it will be interesting to understand better how those studies could potentially improve the capability of QDCs.

\subsection{Efficient Blind Quantum Computation}\label{blindQC}

A QDC can be directly used to implement the efficient blind quantum computation protocol of \cite{giovannetti2013efficient}. In blind quantum computation, a user, Alice, wants to perform a quantum computation using Bob's quantum computer without revealing to Bob what computation has been performed. \cite{giovannetti2013efficient} shows how this is possible through a simple application of the QPQ protocol. Bob holds a length-$N$ database stored in QRAM, where each entry in the database corresponds to a different unitary operation that he can perform on his quantum computer. Alice tells him which operation to perform by sending a $\log N$-qubit quantum state $\ket{i}$, indicating that Bob should apply the $i$-th unitary operation, $U_i$. Bob applies the operation without measuring the register (i.e., he applies a coherently-controlled operation $U_{\text{Bob}}=\sum_i \ket{i}\bra{i}\otimes U_i$) then sends the state back to Alice.  To protect her privacy, Alice periodically sends lure states and measures the states returned to her. If Bob attempts to cheat, Alice will be able to detect it. 

The implementation of this protocol with a QDC is simple, as we have already shown that a QDC can implement the QPQ protocol. In this case, though, the QDC also requires a universal quantum computer in order to perform the computation. Thus, the efficient blind quantum computation protocol could be understood as an extension of QPQ with extra powers in quantum computation. When estimating the computational cost for QDC, one should include the computational complexity of the quantum operation $U_i$, while other analysis stay the same as QDC associated with QPQ.

\subsection{Final comments on multi-party private quantum communication}
Private quantum communication refers to the possibility of transmitting quantum information without revealing this information to eavesdroppers. If multiple parties are communicating over a quantum network, eavesdroppers may nevertheless be able to learn who has sent information and who has received it, even if they cannot determine what that information was. Multi-party private quantum communication refers to a stronger notion, where eavesdroppers can neither learn what information was communicated \emph{nor which users were communicating to which others}. To our knowledge, this notion of \emph{multi-party private quantum communication} and the corresponding protocol is introduced for the first time in our paper.

This protocol constitutes private multi-party quantum communication because (1) the use of secret sharing means that no QDC can learn what information is being communicated, and (2) the use of Quantum Private Queries means that no QDC can learn which user $B_j$ is accessing the information transmitted by $A_i$. A crucial assumption in the protocol is that the QDCs are non-cooperating. If the QDCs cooperate, they could work together to reconstruct the secret. To mitigate this problem, the number of parts each secret is divided to can be increased (along with the number of QDCs). This way, revealing the secret would require cooperation between an increasingly large number of QDCs. We comment on the difference between multi-party private quantum communication and so-called covert quantum communication \cite{arrazola2016covert}. Covert quantum communication refers to a stronger notion, where eavesdroppers cannot even detect whether any information has been transmitted in the first place. In the protocol of \cite{arrazola2016covert}, Alice and Bob are assumed to share a random, secret key, and the quantum information is sent via optical photons from Alice to Bob at one of $N$ times specified by the key. The probability that an eavesdropper can distinguish between this situation and that where no information is communicated at all (i.e., when no photons are sent) is shown to decrease as $1/\sqrt{N}$. In the limit of large $N$, the eavesdropper cannot determine whether any information has been sent.

Note that, the multi-party private quantum communication scheme is teleporting quantum states, not classical information. Those quantum states are naturally merged with quantum private queries where the security is guaranteed quantumly, making the usage of superposition of addresses in QRAM. Moreover, an important technicality regarding the last step of the protocol is that the QDCs are storing quantum data, and, in general, the act of accessing this data can perturb the quantum database (a consequence of the no-cloning theorem). The QDC could, in principle, detect this perturbation, and use this information to infer which user $B_j$ is accessing the information transmitted by $A_i$. To prevent this, we suppose that the quantum data is accessed as follows. In addition to sending a state $\ket{i}$ specifying which element to access, each user $B_j$ also sends a quantum state $\rho_j$ to the QDC. The QDC then swaps $\rho_j$ with the state stored at location $i$ in the memory. If $\rho_j$ is chosen to be a maximally mixed state, this data access procedure has no backreaction on the database; from the perspective of each QDC, the states stored in the database always look maximally mixed.

\section{Sensing}\label{sensing}
\label{sec:sensing}

\subsection{Quantum data compression}\label{datacomp}
In this section, we describe how QRAM architectures can be used to implement the unary-to-binary compression described in our main text, which could be used for quantum sensing. In particular, we show that the compression operation can be implemented using a modified version of the bucket-brigade QRAM archiecture~\cite{giovannetti2008quantum,giovannetti2008architectures}. In the following, we assume familiarity with this QRAM architecture; we refer unfamiliar readers to Ref.~\cite{hann2021resilience}, which provides a recent, self-contained review.

\begin{figure*}[ht]
\centering
%\rule{\linewidth}{3cm}
\includegraphics[width=\textwidth]{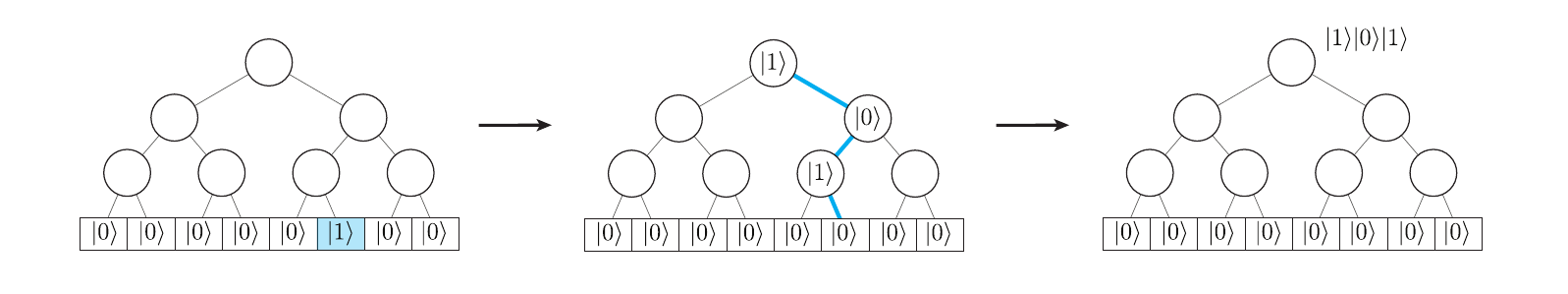}
\caption{Quantum compression with QRAM. (a) Compression procedure. The bucket-brigade QRAM consists of a binary tree of quantum routers~\cite{hann2021resilience}, with the $N$-qubit quantum memory located at the bottom of the tree. An excitation initially stored in the memory is routed up and out of the tree, such that its initial position is first encoded in the states of the quantum routers, then in the state of an external register. The excitation is routed up through the tree using the routing circuit shown in (b). The action of this circuit is shown diagrammatically in (c). We note that this compression procedure is a straightforward extension of the routing procedure described in Ref.~\cite{hann2021resilience}, which provides a more detailed description of QRAM and quantum routing. }
\label{fig:compression}
\end{figure*}

The basic compression scheme is illustrated in Figure~\ref{fig:compression}. Suppose a single excitation is stored in one of $N$ different cells in the QRAM's quantum memory (or in a superposition of multiple different cells). The procedure in the figure allows one to coherently extract the position of the excitation using a modified version of the bucket-brigade QRAM's binary-tree routing scheme. Specifically, the excitation is routed upward from the quantum memory at the bottom of the tree to the root node at the top.  As the excitation is routed upward, its original position is encoded into the states of the quantum routers comprising the tree. This encoding is accomplished using a simple modification to the quantum routing circuit of Ref.~\cite{hann2021resilience}, shown in Figure~\ref{fig:compression} (b,c). Subsequently, the position information is extracted from the routers and stored in an external $\log N$-qubit register, exactly as in the usual bucket-brigade approach~\cite{giovannetti2008quantum,giovannetti2008architectures,hann2021resilience}. 

In this way, QRAM enables us to convert the unary information of the photon's position into a more compact binary representation of $\log N$ qubits. That is, the scheme in Figure~\ref{fig:compression} implements the mapping $\ket{\psi_\text{unary}} \rightarrow \ket{\psi_\text{binary}}$ described in \Cref{sec:sensing}. 
As a result, the state of $N$-mode memory (assumed to lie within the single-excitation subspace) can be compressed to $\log N$ qubits. Generalization to multi-excitation subspaces is straightforward; the procedure can be repeated to extract multiple excitations from the memory, such that the $k$-excitation subspace can be compressed into $k\log N$ qubits. %More efficient compression schemes can be implemented using more sophisticated protocols~\cite{hann2022inprep}.

Additionally, the scheme in Figure~\ref{fig:compression} can also be used to implement the operation $U$ that coherently extracts the address of an excitation stored in the QRAM's quantum memory. To do so, first, the compression of procedure in Figure~\ref{fig:compression}(a) is applied,
\begin{align}
    &\sum_{i=0}^{N-1} \alpha_i  \ket{i}^{Q_1'}  \ket{0}^{Q_2} \left[\bigotimes_{j = 1}^{N} \ket{\delta_{ij}}^{D_j} \right]
    \nonumber \\
    \rightarrow& \sum_{i=0}^{N-1} \alpha_i  \ket{i}^{Q_1'}  \ket{1}^{Q_2} \left[\bigotimes_{j = 1}^{N} \ket{0}^{D_j} \right],
\end{align}
where here the labels $Q_1$, $Q_2$, and $D_j$ respectively denote the external $\log N$-qubit register, an external qubit used to hold the extracted excitation (not shown in Figure~\ref{fig:compression}), and the $j$-th cell of the quantum memory.
Then, a series of CNOT gates are used to copy the address information stored in register $Q_1$ into another external $\log N$-qubit register, denoted $Q_1$,
\begin{align}
    \rightarrow  \sum_{i=0}^{N-1} \alpha_i  \ket{i}^{Q_1}   \ket{i}^{Q_1'} \ket{1}^{Q_2} \left[\bigotimes_{j = 1}^{N} \ket{0}^{D_j} \right].
\end{align}
Finally, the compression procedure is run in reverse in order to return the excitation to its original location in memory 
\begin{align}
    \rightarrow \sum_{i=0}^{N-1} \alpha_i  \ket{0}^{Q_1}  \ket{i}^{Q_1'}  \ket{0}^{Q_2} \left[\bigotimes_{j = 1}^{N} \ket{\delta_{ij}}^{D_j} \right].
\end{align}
The $Q_1'$ and $Q_2$ registers can subsequently be discarded.
%For $k>1$ this compression rate is non-optimal, but better compression rates can be achieved with more sophisticated protocols, as described in Ref.~\cite{hann2022inprep}. 

% If the photon is in a superposition of different time bins, rather than a superposition of different modes, then a QROM-like scheme can similarly be used to achieve the same compression.

Finally, we comment on how QDC-assisted distributed sensing could provide benefits on the entanglement cost. Suppose that two or more physically separated users probe some system, such that a quantity of interest is encoded in an entangled state shared between them. If local operations do not suffice to measure this quantity, then quantum information must be transmitted between the users. If each user has $N$ qubits of quantum data, then $N$ entangled pairs will be required to transmit the information in general. In certain situations, this entanglement cost can be greatly reduced using a QDC. For example, if the $N$-qubit states are guaranteed to lie in the single-excitation subspace, then they can be transmitted using only $\log N$ entangled pairs using the unary-to-binary compression described above. This is the case for quantum-assisted telescope arrays~\cite{gottesman2012longer}, where quantum networks are used to enable optical interferometry. In this context, a single optical photon arrives at one of multiple telescopes in superposition, with its arrival time and frequency unknown. Supposing that the photon arrives at one of $N$ unknown time-frequency bins, Refs.~\cite{khabiboulline2019optical,khabiboulline2019quantum} show that unary-to-binary compression enables the optical phase difference to be extracted using only $\log N$ entangled pairs. QDCs could be directly used to implement this compression. In fact, using a QDC to implement the compression is more hardware efficient than the approach proposed in Ref.~\cite{khabiboulline2019quantum}. In that work, the authors consider the case where a photon arrives at one of $T_{\text{bin}}$ different time bins and in one of $R$ different frequency bands, and they describe a procedure that uses $\mathcal{O}(R\log T_{\text{bin}})$  qubits to compresses the photon's arrival time and frequency information. The same compression can be achieved with a QDC using only $\mathcal{O}(R) + \mathcal{O}(\log T_{\text{bin}})$ qubits as follows. At each time step, any incoming photon is stored in one of $R$ different memory qubits according to its frequency band. These $R$ qubits constitute the QDC's quantum memory, and the QDC performs unary-to-binary compression scheme described above to compress this which-frequency information. If the photon arrived at the present time step, it is now stored at a definite location (namely, register $Q_2$ in \Cref{eq:QRAM_compression}). The presence of this photon can then be used to control the binary encoding of the which-time information, as in Ref.~\cite{khabiboulline2019optical}. Altogether, this procedure requires $\mathcal{O}(R)$ qubits for the QDC and its quantum memory, plus an additional $\mathcal{O}(\log T_{\text{bin}})$ qubits to hold the compressed which-time information, hence the total hardware cost is  $\mathcal{O}(R) + \mathcal{O}(\log T_{\text{bin}})$ qubits. When counting the communication time cost, the savings will be more drastic. More sophisticated compression protocols can enable further reduction in entanglement cost. If each mode of an $M$-mode system is populated with a photon with probability $p$, then the quantum data can be transmitted as few as $M H(p)$ qubits, where $H(p)$ is the binary entropy, using a scheme for Schumacher coding~\cite{schumacher1995quantum}. Such schemes would further reduce the entanglement cost. Moreover, a QDC equipped with quantum sorting networks (a generalization of QRAM), can implement Schumacher coding in polylogarithmic time~\cite{hann2022inprep}, enabling improved detector bandwidth.

\subsection{Channel discrimination using QDCs}\label{channelappend}
In the main text, we discuss various aspects of QDC realizations with both QRAM and quantum communications. However, QDCs could still be made without either QRAM or quantum communications. Here we give a simple example from quantum sensing, where QRAM is not necessarily needed. 

The estimation and discrimination of quantum channels are natural problems in quantum sensing (see \cite{acin2001statistical,duan2007entanglement,duan2009perfect,zhou2021asymptotic,rossi2021quantum}). Following \cite{rossi2021quantum}, one of the simplest problems is the following channel discrimination problem: say that we have two distributions $\Theta_{b}$ with $b=0,1$. For a given $b$, we wish to find out the value of $b$ by accessing the quantum channel 
\begin{align}
\mathcal{E}_b = \exp (i \theta_b H)~~~~~~\theta_b \sim \Theta_b~,
\end{align}
with minimal numbers of times (for the qubit setup in this paper, we could specify $H$ as Pauli $X$. For higher dimensional channel discrimination, see \cite{duan2009perfect}). One could define the channel discrimination protocol by the following circuits. The quantum (coherent) protocol corresponds to the following circuit
\begin{align}
Q = {{\cal E}_{b,N}}\prod\limits_{\ell  = 1}^{N - 1} {\left( {{V_\ell }{{\cal E}_{b,\ell }}} \right)} ~.
\end{align}
Here, $N$ is the total number of queries, and $\mathcal{E}_{b,\ell}$ corresponds to $\ell$th copy of the channel $\mathcal{E}_b$. A series of unitaries $V_\ell$ will define the protocol (one could specify it by QSP angles, see \cite{rossi2021quantum}). Say that we define the input state to be $\ket{\psi_{\text{input}}}$ and the output state to be $\ket{\psi_{\text{output}}}$, the success probability is
\begin{align}
p = {\max _{V,{\psi _{{\rm{input}}}},{\psi _{{\rm{output}}}}}}{\left| {\left\langle {{\psi _{{\rm{output}}}}\left| Q \right|{\psi _{{\rm{input}}}}} \right\rangle } \right|^2}~.
\end{align}
For the incoherent (classical) protocol, we could perform $N$ different measurement. Say that for $\ell$th measurement, we expect the input $\ket{\psi_{\text{input},\ell}}$ and the output $\ket{\psi_{\text{output},\ell}}$, we will have the $\ell$th probability 
\begin{align}
{p_\ell } = {\left| {\left\langle {{\psi _{{\rm{output}},\ell }}\left| {{V_\ell }{{\cal E}_{b,\ell }}} \right|{\psi _{{\rm{input}},\ell }}} \right\rangle } \right|^2}~,
\end{align}
and the protocol could be specified by the majority vote \cite{rossi2021quantum}. Moreover, one could specify the $\xi$-hybrid protocol as performing a length-$\xi$ coherent protocol $N/\xi$ times, and the result could be determined by the majority vote of $N/\xi$ trials. It is shown that when the channel is noiseless ($\Theta_b$s are Dirac function distributions), coherent protocols always have the advantage over incoherent protocols. However, when the channel is too noisy, incoherent protocols might be better \cite{rossi2021quantum}. Thus the hybrid protocols might be useful when we increase $N$. 

If the user wants the quantum sensing to have high precision, they might have to go to the large $N$ regime. In the above example, precision $\delta$ is the difference between the mean of the distribution $\Theta_1$ and $\Theta_2$. It is proved that \cite{rossi2021quantum} in the noiseless case, the optimal $N$ scales as $1/\delta$. Moreover, the unitaries $V=\{V_\ell\}$ might be hard to construct and design. In those cases, QDC might be useful. 

The protocol with QDC could be defined as the following. On the user side, the user could generate the channel $\mathcal{E}_{b,j}$, and the user could also send the information about the distribution $\Theta_b$ to a QDC. The QDC could generate a series of unitaries $\{V_\ell\}$ and design the optimal hybrid $\xi$-protocol. Each time when the state passes through the channel $\mathcal{E}_{b,j}$, one could teleport the state by the quantum network to the QDC, and QDC will apply $V_{\ell}$ on the state and teleport it back. The measurement could be done either by the QDC or by the user. The majority vote could either be done classically with $\mathcal{O}(N/\xi)$ complexity, or quantumly by measuring $\mathcal{O}(N/\xi)$ times in the computational basis. The QDC-assisted channel discrimination protocols will have advantages when the user is hard to design the optimal circuits $V=\{V_\ell\}$.

\end{document}